\documentclass[12pt]{article}
\pdfoutput=1
\usepackage{latexsym,amsmath,amssymb, graphicx, subfigure, dsfont}
\usepackage[linktocpage]{hyperref}
\renewcommand{\thefootnote}{\fnsymbol{footnote}}

\usepackage{mathrsfs }
\usepackage{amsmath}
\usepackage{amsfonts}
\usepackage{tikz}

\numberwithin{equation}{section}

\def\doubleset#1#2{\bgroup%
\def\doit#1#2{%
\setbox\dblsetbox=\hbox{$\cstyle #1$}%
\raise#2\ht\dblsetbox\copy\dblsetbox%
\hskip-\wd\dblsetbox%
\raise-#2\ht\dblsetbox\box\dblsetbox}%
\mathchoice%
{\def\cstyle{\displaystyle}\doit#1#2}%
{\def\cstyle{\textstyle}\doit#1#2}%
{\def\cstyle{\scriptstyle}\doit#1#2}%
{\def\cstyle{\scriptscriptstyle}\doit#1#2}\egroup}
\def\underarrow#1{\vbox{\ialign{##\crcr$\hfil\displaystyle
 {#1}\hfil$\crcr\noalign{\kern1pt\nointerlineskip}$\longrightarrow$\crcr}}}

\newbox\dblsetbox

\newlength{\extraspace}
\newlength{\extraspaces}
\newcommand{\be}{\begin{equation}
\addtolength{\abovedisplayskip}{\extraspaces}
\addtolength{\belowdisplayskip}{\extraspaces}
\addtolength{\abovedisplayshortskip}{\extraspace}
\addtolength{\belowdisplayshortskip}{\extraspace}}
\newcommand{\ee}{\end{equation}}

\newcommand{\ba}{\begin{eqnarray}
\addtolength{\abovedisplayskip}{\extraspaces}
\addtolength{\belowdisplayskip}{\extraspaces}
\addtolength{\abovedisplayshortskip}{\extraspace}
\addtolength{\belowdisplayshortskip}{\extraspace}}
\newcommand{\ea}{\end{eqnarray}}

\newcommand{\bd}{\begin{displaymath}
\addtolength{\abovedisplayskip}{\extraspaces}
\addtolength{\belowdisplayskip}{\extraspaces}
\addtolength{\abovedisplayshortskip}{\extraspace}
\addtolength{\belowdisplayshortskip}{\extraspace}}
\newcommand{\ed}{\end{displaymath}}

\newcounter{saveeqn}

\newcommand{\newsection}[1]{
\vspace{12mm} \pagebreak[3] \addtocounter{section}{1}
\setcounter{equation}{0} \setcounter{subsection}{0}
\noindent{\bf \thesection. #1} \nopagebreak
\medskip
\nopagebreak
\addcontentsline{toc}{section}{\thesection. #1}}

\newcommand{\newsubsection}[1]{
\vspace{0.8cm} \pagebreak[3] \addtocounter{subsection}{1}
\setcounter{subsubsection}{0}
\noindent{ \it \thesubsection. #1} \nopagebreak \vspace{2mm}
\nopagebreak
\addcontentsline{toc}{subsection}{\thesubsection. #1}}

\makeatletter
\newcommand\xleftrightarrow[2][]{%
  \ext@arrow 9999{\longleftrightarrowfill@}{#1}{#2}}
\newcommand\longleftrightarrowfill@{%
  \arrowfill@\leftarrow\relbar\rightarrow}
\makeatother

\begin{document}
\addtolength{\baselineskip}{1.5mm}

\thispagestyle{empty}

\vbox{} \vspace{-0.0cm}

\begin{center}
\centerline{\Large{\bf Higher AGT Correspondences, ${\cal W}$-algebras, and}}
\medskip
\centerline{\Large{\bf Higher Quantum Geometric Langlands Duality}}
\medskip
\centerline{\Large{\bf from M-Theory}}

\vspace{1.4cm}

{\bf{Meng-Chwan~Tan}}
\\[0mm]
{ Department of Physics,
National University of Singapore}\\[0 mm]
mctan@nus.edu.sg
\end{center}

\vspace{0.3cm}

\centerline{\it{In loving memory of See-Hong Tan}}

\vspace{1.8 cm}

\centerline{\bf Abstract}\smallskip

We further explore the implications of our framework in~\cite{4d AGT, 5d AGT}, and physically derive, from the principle that the spacetime BPS spectra of string-dual M-theory compactifications ought to be equivalent, (i) a 5d AGT correspondence for any compact Lie group, (ii) a 5d and 6d AGT correspondence on ALE space of type $ADE$, and (iii) identities between the ordinary, $q$-deformed and elliptic affine $\cal W$-algebras associated with the 4d, 5d and 6d AGT correspondence, respectively, which also define a quantum geometric Langlands duality and its higher analogs formulated by Feigin-Frenkel-Reshetikhin in~\cite{FF, FR}. As an offshoot, we are led to the sought-after connection between the gauge-theoretic realization of the geometric Langlands correspondence by Kapustin-Witten~\cite{KW, K} and its algebraic CFT formulation by Beilinson-Drinfeld~\cite{BD}, where one can also understand Wilson and 't Hooft-Hecke line operators in 4d gauge theory as monodromy loop operators in 2d CFT, for example. In turn, this will allow us to argue that the higher 5d/6d analog of the geometric Langlands correspondence for simply-laced Lie (Kac-Moody) groups $G$ ($\widehat G$), ought to relate the quantization of circle (elliptic)-valued $G$ Hitchin systems to circle/elliptic-valued $^LG$ ($\widehat{^LG}$)-bundles over a complex curve on one hand, and the transfer matrices of a $G$ ($\widehat{G}$)-type XXZ/XYZ spin chain on the other, where $^LG$ is the Langlands dual of $G$. Incidentally, the latter relation also serves as an M-theoretic realization of Nekrasov-Pestun-Shatashvili's recent result in~\cite{NPS}, which relates the moduli space of 5d/6d supersymmetric $G$ ($\widehat{G}$)-quiver $SU(K_i)$ gauge theories to the representation theory of quantum/elliptic affine (toroidal) $G$-algebras.


\newpage

\renewcommand{\thefootnote}{\arabic{footnote}}
\setcounter{footnote}{0}

\tableofcontents

\newsection{Introduction, Summary and Acknowledgements}

In 2009, Alday-Gaiotto-Tachikawa computed in~\cite{AGT} that the Nekrasov instanton partition function of a 4d ${\cal N} = 2$ conformal $SU(2)$ quiver theory is equivalent to a conformal block of a 2d CFT with affine ${\cal W}(\frak {su}(2))$-algebra symmetry, i.e. Liouville theory. This celebrated 4d-2d correspondence, since then known as the 4d AGT correspondence, was expected to hold for other gauge theories as well. In particular, it was later proposed and checked that the correspondence ought to hold for asymptotically-free $SU(2)$ theories~\cite{irr}, and for conformal $SU(N)$ quiver theory whereby the corresponding 2d CFT is an $A_{N-1}$ conformal Toda field theory with affine ${\cal W}(\frak {su}(N))$-algebra symmetry~\cite{AGT-matter}. The correspondence for pure arbitrary $G$ theory whereby the corresponding 2d CFT has affine ${\cal W} (^L\frak g)$-algebra symmetry  ($^L \frak g$ being the Langlands dual of the Lie algebra of $G$), was also proposed and checked to hold up to the first instanton level~\cite{abcdefg}.  

The basis for the 4d AGT correspondence for $SU(N)$, as first pointed out in~\cite{Alday-Tachikawa}, is a conjectured relation between the equivariant cohomology of the moduli space of $SU(N)$-instantons on ${\bf R}^4$ and the integrable representations of an affine ${\cal W}(\frak {su}(N))$-algebra on a certain punctured Riemann surface. This conjectured relation, and its generalization to simply-laced $G$, were proved mathematically in~\cite{Vasserot, MO} and~\cite{NBF}, respectively, while its generalization to arbitrary $G$, and more, were physically derived in~\cite{4d AGT} via the principle that the spacetime BPS spectra  of {string-dual} M-theory compactifications ought to be equivalent.

The 4d AGT correspondence was subsequently generalized to 5d via a physical computation for pure $SU(2)$ theory~\cite{Awata-AGT} whereby the 2d CFT has $q$-deformed affine ${\cal W}(\frak {su}(2))$-algebra symmetry, and a mathematical conjecture for conformal $SU(N)$ linear quiver theory~\cite{Awata} whereby the corresponding 2d CFT has \emph{$q$-deformed} affine ${\cal W}(\frak {su}(N))$-algebra symmetry.  The 5d AGT correspondence for general $SU(N)$ theories, and more, were then physically derived in~\cite{5d AGT} via the principle that the spacetime BPS spectra  of {string-dual} M-theory compactifications ought to be equivalent.   

Exploiting the existence of an elliptic-deformation in~\cite{Saito} of the mathematical Ding-Iohara algebra in~\cite{DI}, a 6d generalization of the AGT correspondence for conformal $SU(N)$ linear quiver theory whereby the corresponding 2d CFT has \emph{elliptic} affine ${\cal W}(\frak {su}(N))$-algebra symmetry, and more, were also physically derived in~\cite{5d AGT}, again, via the principle that the spacetime BPS spectra  of {string-dual} M-theory compactifications ought to be equivalent.  Of late, there have also been efforts to physically derive the 6d AGT correspondence for the specific  case of $N=2$ via field and string-theoretic methods~\cite{N, Y}.

The main aim of this paper is to further explore the implications of our framework in~\cite{4d AGT, 5d AGT}, so as to physically derive, via M-theory, generalizations of the aforementioned 5d and 6d AGT correspondence, and more. 

One motivation for our effort is the recent work in~\cite{PK} which  furnishes a gauge-theoretic realization of the $q$-deformed affine $\cal W$-algebras constructed in~\cite{FR-T}. This work strongly suggests that we should be able to realize, in a unified manner through our M-theoretic framework, a quantum geometric Langlands duality and its higher analogs as defined in~\cite{FF, FR}, and more.

 Another motivation for our effort is the hitherto missing connection between the gauge-theoretic realization of the geometric Langlands correspondence by Kapustin-Witten in~\cite{KW, K} and its original algebraic CFT formulation by Beilinson-Drinfeld in~\cite{BD}. The fact that we can relate 4d supersymmetric gauge theory to ordinary affine $\cal W$-algebras which obey a geometric Langlands duality, suggests that the sought-after connection may actually reside within our M-theoretic framework.

Let us now give a brief plan and summary of the paper.

\bigskip\noindent{\it A Brief Plan and Summary of the Paper}

In $\S$2\ref{s2}, we will furnish an M-theoretic derivation of a 5d pure AGT correspondence for the  $A$, $B$, $C$, $D$, $G_2$ compact Lie groups, which relates the Nekrasov instanton partition function to the norm of a coherent state in a module of a $q$-deformed affine $\cal W$-algebra associated with the Langlands dual Lie algebra. By taking the topological string limit, we will be able to derive the correspondence for the $E_{6,7,8}$ and $F_4$ groups, too, although in this case, the Nekrasov instanton partition function would be related to the norm of a coherent state in a module of a Langlands dual\emph{ toroidal} Lie algebra. 

In $\S$3\ref{s3}, we will first derive an essential mathematical result of Nakajima's~\cite{Nak} which relates the equivariant K-theory of quiver varieties to quantum toroidal Lie algebras, and then proceed to furnish an M-theoretic derivation of a 5d and 6d AGT correspondence for $SU(N)$ on the smooth ALE space $\widetilde {{\bf R}^4 / {\Gamma}}$, where $\Gamma \subset SU(2)$ is a finite subgroup. We find that in the 5d and 6d case, the Nekrasov instanton partition function would be related to an $N$-tensor product of level 1 modules of a quantum and elliptic toroidal Lie algebra, respectively, that are associated with $\Gamma$ via the McKay correspondence~\cite{McKay}.

In $\S$4\ref{s4}, we will furnish an M-theoretic realization of various $\cal W$-algebras associated with the 4d, 5d and 6d AGT correspondence obtained hitherto, and derive respective identities which define a quantum geometric Langlands duality and its higher analogs. These identities, and their web of relations to one another, can be summarized as follows:
\tikzset{node distance=8.0cm, auto}
\begin{center}
\be
\label{dia 0}
 \begin{tikzpicture}
  \node (P) {$\boxed{{{\cal W}}_{{\rm aff}, k}({\frak g}) = {{\cal W}}_{{\rm aff}, ^Lk}({^L\frak g})}$};
  \node (B) [right of=P] {$\boxed{Z(U(\hat{\frak g})_{\rm crit}) = {\cal W}_{\rm cl}(^L\frak g)}
$};
  \node (A) [below of=P, yshift = 5.2cm] {$\boxed{{\cal W}^{q,t}_{{\rm aff}, k}({\frak g}_{ADE}) = {{\cal W}}^{t,q}_{{\rm aff}, ^Lk}({^L\frak g}_{ADE})}$};
  \node (C) [below of=B, yshift = 5.2cm] {$\boxed{Z(U_q(\hat{\frak g}_{ADE})_{\rm crit}) = {\cal W}^q_{\rm cl}(^L\frak g_{ADE})}
$};
\node (E) [below of=C, yshift = 5.2cm] {$\boxed{Z(U_{q,v}(\hat{\frak g}_{ADE})_{\rm crit}) = {\cal W}^{q,v}_{\rm cl}(^L\frak g_{ADE})}
$};
  \node (D) [below of=A, yshift = 5.2cm] {$\boxed{{\cal W}^{q,t, v}_{{\rm aff}, k}({\frak g}_{ADE}) = {{\cal W}}^{t,q, v}_{{\rm aff}, ^Lk}({^L\frak g}_{ADE})}$};
   \draw[transform canvas={yshift=0.5ex},->] (P) --(B) node[midway] {$\epsilon \to 0$};
\draw[transform canvas={yshift=-0.5ex},->](B) -- (P) node[midway] {$\epsilon \nrightarrow 0$}; 
 \draw[transform canvas={xshift=0.5ex},->] (P) --(A) node[midway] {$\beta \nrightarrow 0$};
\draw[transform canvas={xshift=-0.5ex},->](A) -- (P) node[midway] {$\beta \to 0$}; 
 \draw[transform canvas={yshift=0.5ex},->] (A) --(C) node[midway] {$\epsilon \to 0$};
\draw[transform canvas={yshift=-0.5ex},->](C) -- (A) node[midway] {$\epsilon \nrightarrow 0$}; 
\draw[transform canvas={xshift=0.5ex},->] (B) --(C) node[midway] {$\beta \nrightarrow 0$};
\draw[transform canvas={xshift=-0.5ex},->](C) -- (B) node[midway] {$\beta \to 0 $};   
 \draw[transform canvas={yshift=0.5ex},->] (D) --(E) node[midway] {$\epsilon \to 0$};
\draw[transform canvas={yshift=-0.5ex},->](E) -- (D) node[midway] {$\epsilon \nrightarrow 0$}; 
 \draw[transform canvas={xshift=-0.5ex},->] (D) --(A) node[midway] {$R_6 \to 0$};
  \draw[transform canvas={xshift=0.5ex},->] (A) --(D) node[midway] {$R_6 \nrightarrow 0$};
 \draw[transform canvas={xshift=-0.5ex},->] (E) --(C) node[midway] {$R_6 \to 0$};
  \draw[transform canvas={xshift=0.5ex},->] (C) --(E) node[midway] {$R_6 \nrightarrow 0$};
\end{tikzpicture}
\ee
 \end{center}
Here, ${\cal W}_{{\rm aff}, k}({\frak g})$, ${\cal W}^{q,t}_{{\rm aff}, k}({\frak g})$ and ${\cal W}^{q,t, v}_{{\rm aff}, k}({\frak g})$ are the ordinary, $q$-deformed and elliptic affine $\cal W$-algebras of level $k$ associated with the Lie algebra $\frak g$, where ${\cal W}_{\rm cl}(\frak g)$, ${\cal W}^{q}_{\rm cl}(\frak g)$ and ${\cal W}^{q,v}_{\rm cl}(\frak g)$ are their corresponding classical limits; $Z(U(\hat{\frak g})_{\rm crit})$, $Z(U_{q}(\hat{\frak g})_{\rm crit})$ and $Z(U_{q,v}(\hat{\frak g})_{\rm crit})$ are the centers, at critical level, of the (universal enveloping algebra of the) affine ${\frak g}$-algebra, quantum affine ${\frak g}$-algebra and elliptic affine ${\frak g}$-algebra; $\epsilon$ is one of the two Omega-deformation parameters of the Nekrasov instanton partition function; and $\beta$ and $R_6$ are the radii of the fifth and sixth circles associated with the 5d and 6d Nekrasov instanton partition functions, respectively. 

In $\S$5\ref{s5}, we will demonstrate the sought-after connection between the gauge-theoretic realization of the (quantum) geometric Langlands correspondence by Kapustin-Witten~\cite{KW, K} and its original algebraic CFT formulation by Beilinson-Drinfeld~\cite{BD}. We will explain how, in our M-theoretic framework which realizes a 4d AGT correspondence for \emph{massless} ${{\cal N} = 2}^\ast$ theories, 4d gauge-theoretic $S$-duality actually corresponds to a 2d conformal field-theoretic (quantum) $\cal W$-algebra duality in the (left) right box of the topmost relation in (\ref{dia 0}), whence we would be able to construct, out of pairs of dual M-theory compactifications which realize the correspondence for massless necklace quiver and pure ${\cal N}=2$ theories, an effective pair of dual M-theory compactifications which will naturally allow us to demonstrate the aforementioned connection for a complex curve of arbitrary genus greater than one. In our framework, the Wilson and 't Hooft-Hecke line operators of 4d gauge theory as realized by boundary M2-branes, will correspond to monodromy loop operators of 2d CFT as realized by M0-branes, whence the action of the former on the categories of sigma-model branes in the gauge theory picture, can be understood in terms of the action of the latter on modules of the classical affine $\cal W$-algebras in the right box of the topmost relation in (\ref{dia 0}) in the CFT picture.

And lastly in $\S$6\ref{s6}, we will generalize our derivation of the geometric Langlands correspondence in $\S$5\ref{s5}  to higher dimensions via a 5d/6d AGT correspondence. In doing so, we will find that the 5d/6d analog of the geometric Langlands correspondence for simply-laced Lie (Kac-Moody) groups $G$ ($\widehat G$), which one can associate with the $\cal W$-algebra duality in the right box of the middle/bottommost relation in (\ref{dia 0}), ought to relate the quantization of circle (elliptic)-valued $G$ Hitchin systems to circle/elliptic-valued $^LG$ ($\widehat{^LG}$)-bundles over a complex curve on one hand, and the transfer matrices of a $G$ ($\widehat{G}$)-type XXZ/XYZ spin chain on the other, where $^LG$ is the Langlands dual of $G$. As an offshoot, we would be able to furnish an M-theoretic realization of Nekrasov-Pestun-Shatashvili's recent result in~\cite{NPS}, which relates (a) the moduli space of 5d/6d, ${\cal N} =1$ $G$ ($\widehat{G}$)-quiver $SU(K_i)$ gauge theories captured by a classical integrable system of periodic $G$-monopoles (doubly-periodic $G$-instantons) to (b) the representation theory of quantum/elliptic affine (toroidal) $G$-algebras.

\bigskip\noindent{\it Acknowledgements} 

I would like to thank David Ben-Zvi, Sergey Cherkis, Naihuan Jing, Hiroaki Nakajima, Yosuke Saito and Richard S. Ward, for helpful exchanges. I would also like to thank the organizers of String-Math 2015 for the invitation to speak, as this work was motivated by a talk there. 

This work is supported in part by the NUS Tier 1 FRC Grant R-144-000-316-112.


\newpage

\newsection{A 5d AGT Correspondence for Compact Lie Groups} 
\label{s2}




\newsubsection{An M-Theoretic Derivation of a 5d Pure AGT Correspondence for the $A$ and $B$ Groups}
\label{s2.1}

We shall now furnish an M-theoretic derivation of a 5d pure AGT Correspondence for any compact Lie group. Let us start with the $A$ and $B$ groups. The derivation for the $A$ groups has already been carried out in~\cite{5d AGT}. Let us review the relevant results in \emph{loc.cit.} that will be useful here.  

\bigskip\noindent{\it A 5d Pure AGT Correspondence for the $A$ Groups}

From~\cite[$\S$3.3]{5d AGT}, the 5d pure AGT correspondence for the $A_{N-1}$ groups can be expressed as 
\be
\label{1-point to pure}
{Z^{{\rm pure}, \, {\rm 5d}}_{{\rm inst}, \, SU(N)} (\epsilon_1, \epsilon_2, \vec a, \beta, \Lambda)  =   \langle {\bf 0} |  {\Phi}^{\otimes N}_{A} (1) | {\bf 0} \rangle_{{\bf S}^2}}.
\ee 
On the LHS, $Z^{{\rm pure}, \, {\rm 5d}}_{{\rm inst}, \, SU(N)}$ is the 5d Nekrasov instanton partition function for pure $SU(N)$ gauge theory on $M_5 = {\bf S}^1 \times {\bf R}^4_{\epsilon_1, \epsilon_2}$ with corresponding Omega-deformation parameters $\epsilon_{1, 2}$;  the vector $\vec a = (a_1, \dots, a_N)$ is the Coloumb moduli; $\beta$ is the radius of ${\bf S}^1$, and; $\Lambda$ is the energy scale. On the RHS, ${\Phi}_{A}$ is a vertex operator of a level one module of the Ding-Iohara algebra~\cite{DI}, so ${\Phi}^{\otimes N}_{A}$ is a vertex operator of a level $N$ module of the Ding-Iohara algebra; also, the vacuum state $| {\bf 0} \rangle =  | \emptyset \rangle^{\otimes N}$, where $| \emptyset \rangle$ is the vacuum state associated with the level one module of the Ding-Iohara algebra.  In turn, as explained in~\cite[footnote 8]{5d AGT}, it would mean that we have the map
\be
\label{q-W-pure-SU(N)}
{\Phi}^{\otimes N}_{A}:  \, \widehat{{\cal W}^q}(\frak{su}(N)_{\textrm{aff}}) \to \widehat{{\cal W}^q}(\frak{su}(N)_{\textrm{aff}}),
\ee
where $\widehat{{\cal W}^q}(\frak{su}(N)_{\textrm{aff}})$ is a Verma module of ${{\cal W}^q}(\frak{su}(N)_{\textrm{aff}})$, a $q$-deformation of an affine ${\cal W}$-algebra ${{\cal W}}(\frak{su}(N)_{\textrm{aff}})$ associated with the affine Lie algebra $\frak{su}(N)_{\textrm{aff}}$.  

In fact, we can write (\ref{1-point to pure}) as
\be
\label{Z5d-pure-SU(N)}
\boxed{{Z^{{\rm pure}, \, {\rm 5d}}_{{\rm inst}, \, SU(N)} (\epsilon_1, \epsilon_2, \vec a, \beta, \Lambda)  = \langle G_{SU(N)}  | G_{SU(N)} \rangle}}
\ee
where
\be
\label{Z5d-pure-SU(N)-G}
{\hspace{-0.00cm}{| G_{SU(N)} \rangle =  (e^{-  \sum_{n_1 > 0} {1 \over n_1} { (\beta \Lambda)^{n_1} \over  1 - s^{n_1}} \, {a}_{-{n_1}}} \cdots e^{-  \sum_{n_N > 0} {1 \over n_N} {(\beta \Lambda)^{n_N} \over  1 - s^{n_N}} \, {a}_{-n_N}}) \cdot \, \, \left( | \emptyset \rangle_1 \otimes \cdots \otimes | \emptyset \rangle_N \right)}};
\ee
\be
\label{Z5d-pure-SU(N)-G-dual}
{\hspace{-0.02cm}{\langle G_{SU(N)} | = \left({_N\hspace{-0.02cm}\langle \emptyset |}\otimes \cdots \otimes{_1\hspace{-0.02cm}\langle \emptyset |} \right) \cdot (e^{\sum_{n_N > 0} {1 \over n_N} {(\beta \Lambda)^{n_N} \over  1 - s^{-n_N}} \, {a}_{n_N}} \cdots e^{\sum_{n_1 > 0} {1 \over n_1} {(\beta \Lambda)^{n_1} \over  1 - s^{-n_1}} \, {a}_{n_1}})}};
\ee
\be
{{{[{a}_{m_k}, {a}_{n_k}] = m_k {1- { s}^{|m_k|} \over 1- { r}^{|m_k|}} \delta_{m_k + n_k, 0}}}, \quad a_{m_k > 0} | \emptyset \rangle_k = 0};
\label{he-pure-SU(N)}
\ee
and
\be
\label{variables-pure-SU(N)}
{s = e^{- i \beta \sqrt{\epsilon_1 \epsilon_2}},  \quad r = e^{- i \beta ( \epsilon_1 + \epsilon_2 + \sqrt{\epsilon_1 \epsilon_2})}}.
\ee
From (\ref{q-W-pure-SU(N)}), and the fact that $| G_{SU(N)} \rangle $ is a sum over states of all possible energy levels (c.f.~(\ref{Z5d-pure-SU(N)-G})), it is clear that 
\be
\label{GSU(N)}
\boxed{| G_{SU(N)} \rangle \in {\widehat{{\cal W}^q}({^L\frak{su}}(N)_{\textrm{aff}})}}
\ee
is a \emph{coherent state}, where ${^L\frak{su}}(N)_{\textrm{aff}}$ is the Langlands dual affine Lie algebra.\footnote{Here, we have used the fact that ${\frak{g}}_{\textrm{aff}} \cong {{^L\frak{g}}}_{\textrm{aff}}$ for simply-laced $\frak g_{\textrm{aff}}$.\label{1}}  The relations (\ref{Z5d-pure-SU(N)}) and (\ref{GSU(N)}) define a 5d pure AGT correspondence for the $A_{N-1}$ groups.

\bigskip\noindent{\it Some Relevant Facts}

A mathematical fact that was not noted in~\cite{5d AGT} but will be relevant to our present discussion, is that the Ding-Iohara algebra is actually isomorphic to the quantum toroidal algebra ${\bf U}_q({\bf L}{\frak {gl}}(1)_{\textrm {aff}})$ associated with the Lie algebra ${\frak {gl}}(1)$~\cite{FT, SV}.\footnote{Specifically, ${\bf U}_q({\bf L}{\frak {gl}}(1)_{\textrm {aff}})$ is a $q$-deformation of the universal enveloping algebra of ${\bf L}{\frak {gl}}(1)_{\textrm {aff}}$, the loop algebra of ${\frak {gl}}(1)_{\textrm {aff}}$ that is hence a double loop or toroidal algebra of the Lie algebra ${\frak {gl}}(1)$.} From the results in~\cite{Miki} which tell us that the tensor product of $N$ level one modules of ${\bf U}_q({\bf L}{\frak {gl}}(1)_{\textrm {aff}})$ is isomorphic to $\widehat{{\cal W}^q}(\frak{su}(N)_{\textrm{aff}})$,\footnote{In \emph{loc.~cit.}, it was actually shown that the tensor product of $N$ level one modules of ${\bf U}_q({\bf L}{\frak {gl}}(1)_{\textrm {aff}})$ is isomorphic to a module of ${{\cal W}^q}(\frak{su}(N)_{\textrm{aff}}) \otimes {\frak u}(1)_{\rm aff}$. However, as explained in~\cite[footnote 8]{5d AGT}, the physics requires us to reduce away the ${\frak u}(1)_{\rm aff}$ factor whence in our context, this statement is consistent. \label{3}} we then arrive back at our conclusion about (\ref{Z5d-pure-SU(N)}). 

Yet another mathematical fact that was not noted in~\cite{4d AGT, 5d AGT} but will also be relevant to our present discussion, is that in the rational limit, i.e. when $\beta \to 0$, we have a reduction of ${\bf U}_q({\bf L}{\frak {gl}}(1)_{\textrm {aff}})$ to ${\bf Y}({\frak {gl}}(1)_{\textrm {aff}})$, the Yangian associated with the Heisenberg algebra ${\frak {gl}}(1)_{\textrm {aff}}$~\cite{GM}. From the results in~\cite[$\S$19.2]{MO} which tell us that the tensor product of $N$ level one modules of ${\bf Y}({\frak {gl}}(1)_{\textrm {aff}})$ is isomorphic to the Verma module $\widehat{{\cal W}}(\frak{su}(N)_{\textrm{aff}})$,\footnote{In \emph{loc.~cit.}, it was actually shown that the tensor product of $N$ level one modules of ${\bf Y}({\frak {gl}}(1)_{\textrm {aff}})$ is isomorphic to a module of ${{\cal W}}(\frak{su}(N)_{\textrm{aff}}) \otimes {\frak u}(1)_{\rm aff}$. However, as explained in~\cite[footnote 8]{5d AGT}, the physics requires us to reduce away the ${\frak u}(1)_{\rm aff}$ factor whence in our context, this statement is again consistent. \label{4}}  it would mean that in the rational limit, (\ref{Z5d-pure-SU(N)}) will reduce to the 4d pure AGT correspondence for the $A_{N-1}$ groups derived in~\cite[$\S$5.2]{4d AGT}.

In short, we have the following diagram
\tikzset{node distance=6.0cm, auto}
\begin{center}
\be
\label{dia 1}
 \begin{tikzpicture}
  \node (P) {$\underbrace{\widehat{{\bf Y}}({\frak {gl}}(1)_{\textrm {aff}, 1}) \otimes \dots \otimes \widehat{{\bf Y}}({\frak {gl}}(1)_{\textrm {aff}, 1})}_{N \,{\textrm {times}}}$};
  \node (B) [right of=P] {$\widehat {{\cal W}}({\frak{su}(N})_{\textrm{aff}, k})$};
  \node (A) [below of=P, yshift = 2.9cm] {$\underbrace{\widehat{{\bf U}_q}({\bf L}{\frak {gl}}(1)_{\textrm {aff}, 1}) \otimes \dots \otimes \widehat{{\bf U}_q}({\bf L}{\frak {gl}}(1)_{\textrm {aff}, 1})}_{N \,{\textrm {times}}}$};
  \node (C) [below of=B, yshift = 2.9cm] {$\widehat {{\cal W}^q}({\frak{su}(N})_{\textrm{aff}, k})$};
   \draw[transform canvas={yshift=0.0ex},->] (P) --(B) node[midway] {};
\draw[transform canvas={yshift=-0.0ex},->](B) -- (P) node[midway] {}; 
 \draw[transform canvas={xshift=0.5ex},->] (P) --(A) node[midway] {$\beta \nrightarrow 0$};
\draw[transform canvas={xshift=-0.5ex},->](A) -- (P) node[midway] {$\beta \to 0$}; 
 \draw[transform canvas={yshift=0.0ex},->] (A) --(C) node[midway] {};
\draw[transform canvas={yshift=-0.0ex},->](C) -- (A) node[midway] {}; 
\draw[transform canvas={xshift=0.5ex},->] (B) --(C) node[midway] {$\beta \nrightarrow 0$};
\draw[transform canvas={xshift=-0.5ex},->](C) -- (B) node[midway] {$\beta \to 0 $};   
\end{tikzpicture}
\ee
 \end{center}
where $\widehat{{\bf Y}}({\frak {gl}}(1)_{\textrm {aff}, 1})$ and $\widehat{{\bf U}_q}({\bf L}{\frak {gl}}(1)_{\textrm {aff}, 1})$ are level one modules of the defining algebras, while $\widehat {{\cal W}}({\frak{su}(N})_{\textrm{aff}, k})$ and $\widehat {{\cal W}^q}({\frak{su}(N})_{\textrm{aff}, k})$ are modules of ${{\cal W}}(\frak{su}(N)_{\textrm{aff}})$ and ${{\cal W}^q}(\frak{su}(N)_{\textrm{aff}})$ associated with ${\frak{su}(N})_{\textrm{aff}, k}$ of some level $k(N, \epsilon_{1,2})$. 

Also, as explained in~\cite[$\S$3]{5d AGT}, the $N$-tensor products in the above diagram arise because along an ${\bf S}^2$ over which the 2d theory of the 5d pure AGT correspondence lives, there are, in the dual type IIA frame, $N$ coincident D6 branes (intersecting a single D4-brane) whereby on each brane, one has a Ding-Iohara module $\widehat{{\bf U}_q}({\bf L}{\frak {gl}}(1)_{\textrm {aff}, 1})$ associated with an underlying Heisenberg algebra ${\frak {gl}}(1)_{\textrm {aff}}$.    

Last but not least, note that in the topological string limit where $\epsilon_1 + \epsilon_2 = 0$, according to~\cite[$\S$2.2]{5d AGT}, we have to replace, in (\ref{dia 1}), $ {{\cal W}}({\frak{su}(N})_{\textrm{aff},k})$ with  ${{\frak{su}(N})}_{\textrm{aff},1}$ and $ {{\cal W}^q}({\frak{su}(N})_{\textrm{aff},k})$  with $ {{\bf L}{\frak{su}(N})}_{\textrm{aff},1}$, the loop algebra of ${{\frak{su}(N})}_{\textrm{aff},1}$. Together with (i) the (conformal) equivalence of the $N$-tensor product ${\frak {gl}}(1)_{\textrm {aff}, 1} \otimes \dots \otimes {\frak {gl}}(1)_{\textrm {aff}, 1}$ with ${\frak{su}(N})_{\textrm{aff}, 1} \otimes {\frak{u}(1})_{\textrm{aff}, 1}$; (ii)  the fact that looping an affine Lie algebra does not modify the underlying central charge; and (iii) the comments in footnotes~\ref{3} and~\ref{4}; it would mean that in the topological string limit, the diagram (\ref{dia 1}) ought to become 
\tikzset{node distance=5.0cm, auto}
\begin{center}
\be
\label{dia 2}
 \begin{tikzpicture}
  \node (P) {$\underbrace{\widehat{{\frak {gl}}(1)}_{\textrm {aff}, 1} \otimes \dots \otimes \widehat{{\frak {gl}}(1)}_{\textrm {aff}, 1}}_{N \,{\textrm {times}}}$};
  \node (B) [right of=P] {$\widehat {{\frak{su}(N})}_{\textrm{aff},1}$};
  \node (A) [below of=P, yshift = 1.9cm] {$\underbrace{\widehat{{\bf L}{\frak {gl}}(1)}_{\textrm {aff}, 1} \otimes \dots \otimes \widehat{{\bf L}{\frak {gl}}(1)}_{\textrm {aff}, 1})}_{N \,{\textrm {times}}}$};
  \node (C) [below of=B, yshift = 1.9cm] {$\widehat {{\bf L}{\frak{su}(N})}_{\textrm{aff},1}$};
   \draw[transform canvas={yshift=0.0ex},->] (P) --(B) node[midway] {};
\draw[transform canvas={yshift=-0.0ex},->](B) -- (P) node[midway] {}; 
 \draw[transform canvas={xshift=0.5ex},->] (P) --(A) node[midway] {$\beta \nrightarrow 0$};
\draw[transform canvas={xshift=-0.5ex},->](A) -- (P) node[midway] {$\beta \to 0$}; 
 \draw[transform canvas={yshift=0.0ex},->] (A) --(C) node[midway] {};
\draw[transform canvas={yshift=-0.0ex},->](C) -- (A) node[midway] {}; 
\draw[transform canvas={xshift=0.5ex},->] (B) --(C) node[midway] {$\beta \nrightarrow 0$};
\draw[transform canvas={xshift=-0.5ex},->](C) -- (B) node[midway] {$\beta \to 0 $};   
\end{tikzpicture}
\ee
 \end{center}
Since the limit $\epsilon_1 + \epsilon_2 = 0$ is tantamount to turning off Omega-deformation on the 2d side of the AGT correspondence~\cite[$\S$2.2]{5d AGT}, the diagrams (\ref{dia 1}) and (\ref{dia 2}) tell us that turning on Omega-deformation on the 2d side effects, when $\beta = 0$, the transformation ${\frak {gl}}(1)_{\textrm {aff},1} \to {\bf Y}({\frak {gl}}(1)_{\textrm {aff},1})$ and ${{\frak{su}(N})}_{\textrm{aff},1} \to {\cal W}({{\frak{su}(N})}_{\textrm{aff},k})$, and when $\beta \neq 0$, the transformation ${\bf L}{\frak {gl}}(1)_{\textrm {aff},1} \to {\bf U}_q({\bf L}{\frak {gl}}(1)_{\textrm {aff},1})$  and  ${{\bf L}{\frak{su}(N})}_{\textrm{aff},1} \to {\cal W}^q({{\frak{su}(N})}_{\textrm{aff},k})$. 

\bigskip\noindent{\it A 5d Pure AGT Correspondence for the $B$ Groups}

Let us now derive a 5d pure AGT correspondence for the $B_{N/2}$ groups (with $N$ even). To this end, note that according to~\cite[$\S$5.2]{4d AGT}, in order to get, on the gauge theory side, a $B_{N/2} = SO(N+1)$ group when $\beta =0$, one must $\mathbb Z_2$-twist the affine Lie algebra underlying the 2d theory. In particular, this would mean that when $\epsilon_1 + \epsilon_2 = 0$, one would have to replace, in the upper line of (\ref{dia 2}), ${\frak{su}(N})_{\textrm{aff}, 1}$ and ${\frak {gl}}(1)_{\textrm {aff},1}$ with their $\mathbb Z_2$-twisted versions ${\frak{su}(N})^{(2)}_{\textrm{aff}, 1}$ and ${\frak {gl}}(1)^{(2)}_{\textrm {aff},1}$. As this $\mathbb Z_2$-twist is independent of the value of $\beta$, and looping a twisted affine Lie algebra does not alter its twist characteristics, it would mean that for $B$ groups, we have, instead of (\ref{dia 2}), the diagram
\tikzset{node distance=5.0cm, auto}
\begin{center}
\be
\label{dia 3}
 \begin{tikzpicture}
  \node (P) {$\underbrace{\widehat{{\frak {gl}}(1)}^{(2)}_{\textrm {aff}, 1} \otimes \dots \otimes \widehat{{\frak {gl}}(1)}^{(2)}_{\textrm {aff}, 1}}_{N \,{\textrm {times}}}$};
  \node (B) [right of=P] {$\widehat {{\frak{su}(N})}^{(2)}_{\textrm{aff},1}$};
  \node (A) [below of=P, yshift = 1.7cm] {$\underbrace{\widehat{{\bf L}{\frak {gl}}(1)}^{(2)}_{\textrm {aff}, 1} \otimes \dots \otimes \widehat{{\bf L}{\frak {gl}}(1)}^{(2)}_{\textrm {aff}, 1})}_{N \,{\textrm {times}}}$};
  \node (C) [below of=B, yshift = 1.7cm] {$\widehat {{\bf L}{\frak{su}(N})}^{(2)}_{\textrm{aff},1}$};
   \draw[transform canvas={yshift=0.0ex},->] (P) --(B) node[midway] {};
\draw[transform canvas={yshift=-0.0ex},->](B) -- (P) node[midway] {}; 
 \draw[transform canvas={xshift=0.5ex},->] (P) --(A) node[midway] {$\beta \nrightarrow 0$};
\draw[transform canvas={xshift=-0.5ex},->](A) -- (P) node[midway] {$\beta \to 0$}; 
 \draw[transform canvas={yshift=0.0ex},->] (A) --(C) node[midway] {};
\draw[transform canvas={yshift=-0.0ex},->](C) -- (A) node[midway] {}; 
\draw[transform canvas={xshift=0.5ex},->] (B) --(C) node[midway] {$\beta \nrightarrow 0$};
\draw[transform canvas={xshift=-0.5ex},->](C) -- (B) node[midway] {$\beta \to 0 $};   
\end{tikzpicture}
\ee
 \end{center}
 
 Turning on Omega-deformation on the 2d side of the AGT correspondence does not affect the $\mathbb Z_2$-twist. Hence, from (\ref{dia 3}), and the paragraph before last, it would mean that for general $\epsilon_{1,2}$, we have, for the $B$ groups, instead of (\ref{dia 1}), the diagram
\tikzset{node distance=5.9cm, auto}
\begin{center}
\be
\label{dia 4}
 \begin{tikzpicture}
  \node (P) {$\underbrace{\widehat{{\bf Y}}({\frak {gl}}(1)^{(2)}_{\textrm {aff}, 1}) \otimes \dots \otimes \widehat{{\bf Y}}({\frak {gl}}(1)^{(2)}_{\textrm {aff}, 1})}_{N \,{\textrm {times}}}$};
  \node (B) [right of=P] {$\widehat {{\cal W}}({\frak{su}(N})^{(2)}_{\textrm{aff}, k})$};
  \node (A) [below of=P, yshift = 2.7cm] {$\underbrace{\widehat{{\bf U}_q}({\bf L}{\frak {gl}}(1)^{(2)}_{\textrm {aff}, 1}) \otimes \dots \otimes \widehat{{\bf U}_q}({\bf L}{\frak {gl}}(1)^{(2)}_{\textrm {aff}, 1})}_{N \,{\textrm {times}}}$};
  \node (C) [below of=B, yshift = 2.7cm] {$\widehat {{\cal W}^q}({\frak{su}(N})^{(2)}_{\textrm{aff}, k})$};
   \draw[transform canvas={yshift=0.0ex},->] (P) --(B) node[midway] {};
\draw[transform canvas={yshift=-0.0ex},->](B) -- (P) node[midway] {}; 
 \draw[transform canvas={xshift=0.5ex},->] (P) --(A) node[midway] {$\beta \nrightarrow 0$};
\draw[transform canvas={xshift=-0.5ex},->](A) -- (P) node[midway] {$\beta \to 0$}; 
 \draw[transform canvas={yshift=0.0ex},->] (A) --(C) node[midway] {};
\draw[transform canvas={yshift=-0.0ex},->](C) -- (A) node[midway] {}; 
\draw[transform canvas={xshift=0.5ex},->] (B) --(C) node[midway] {$\beta \nrightarrow 0$};
\draw[transform canvas={xshift=-0.5ex},->](C) -- (B) node[midway] {$\beta \to 0 $};   
\end{tikzpicture}
\ee
 \end{center}

Comparing the bottom right-hand corner of (\ref{dia 4}) with the bottom right-hand corner of (\ref{dia 1}) for the $A$ groups, and bearing in mind the isomorphism $\frak{su}(N)^{(2)}_{\textrm{aff}} \cong {{^L\frak{so}}(N+1)_{\textrm{aff}}}$, it would mean that in place of (\ref{Z5d-pure-SU(N)}), we ought to have
\be
\label{Z5d-pure-SO(N+1)}
\boxed{{Z^{{\rm pure}, \, {\rm 5d}}_{{\rm inst}, \, SO(N+1)} (\epsilon_1, \epsilon_2, \vec a, \beta, \Lambda)  = \langle G_{SO(N+1)}  | G_{SO(N+1)} \rangle}}
\ee
where the \emph{coherent state}
\be
\label{GSO(N+1)}
\boxed{| G_{SO(N+1)} \rangle \in {\widehat{{\cal W}^q} ({{^L\frak{so}}(N+1)_{\textrm{aff}}})}}
\ee
In arriving at (\ref{Z5d-pure-SO(N+1)})--(\ref{GSO(N+1)}), we have just derived a 5d pure AGT correspondence for the $B_{N/2}$ groups.

In the limit $\beta \to 0$, the top right-hand side of (\ref{dia 4}) tells us that (\ref{Z5d-pure-SO(N+1)})--(\ref{GSO(N+1)}) would reduce to the result for the 4d case in~\cite[$\S$5.2]{4d AGT}, as they should.

\newsubsection{An M-Theoretic Derivation of a 5d Pure AGT Correspondence for the $C$, $D$ and $G_2$ Groups}
\label{s2.2}

Let us now proceed to furnish an M-theoretic derivation of a 5d pure AGT Correspondence for the $C$, $D$ and $G_2$ groups. The derivation is similar to that for the $A$ and $B$ groups in the last subsection, except for a few modifications.

\bigskip\noindent{\it A 5d Pure AGT Correspondence for the $D$ Groups}

Let us now derive a 5d pure AGT correspondence for the $D_N$ groups. To this end, note that according to~\cite[$\S$5.3, $\S$3.2]{4d AGT}, if we have, on the gauge theory side, a $D_{N} = SO(2N)$ group when $\beta =0$ and $\epsilon_1 + \epsilon_2 = 0$ (i.e. when Omega-deformation on the 2d side of the AGT correspondence is turned off), we ought to replace, in the upper line of (\ref{dia 2}),  ${\frak{su}}(N)_{\textrm{aff},1}$ and ${\frak {gl}}(1)_{\textrm {aff}, 1}$ with ${\frak{so}}(2N)_{\textrm{aff},1}$ and ${\frak{so}}(2)_{\textrm{aff},1}$, respectively. Also, according to~\cite[$\S$2.2]{5d AGT}, the looping of the underlying affine Lie algebra when $\beta \neq 0$ is independent of the Lie algebra type, and as mentioned in the last subsection, looping does not modify its central charge either. Altogether, this means that for the $D$ groups, in place of (\ref{dia 2}), we ought to have
 \tikzset{node distance=5.0cm, auto}
\begin{center}
\be
\label{dia 5}
 \begin{tikzpicture}
  \node (P) {$\underbrace{\widehat{{\frak {so}}(2)}_{\textrm {aff}, 1} \otimes \dots \otimes \widehat{{\frak {so}}(2)}_{\textrm {aff}, 1}}_{N \,{\textrm {times}}}$};
  \node (B) [right of=P] {$\widehat {{\frak{so}(2N})}_{\textrm{aff},1}$};
  \node (A) [below of=P, yshift = 2cm] {$\underbrace{\widehat{{\bf L}{\frak {so}}(2)}_{\textrm {aff}, 1} \otimes \dots \otimes \widehat{{\bf L}{\frak {so}}(2)}_{\textrm {aff}, 1})}_{N \,{\textrm {times}}}$};
  \node (C) [below of=B, yshift = 2cm] {$\widehat {{\bf L}{\frak{so}(2N})}_{\textrm{aff},1}$};
   \draw[transform canvas={yshift=0.0ex},->] (P) --(B) node[midway] {};
\draw[transform canvas={yshift=-0.0ex},->](B) -- (P) node[midway] {}; 
 \draw[transform canvas={xshift=0.5ex},->] (P) --(A) node[midway] {$\beta \nrightarrow 0$};
\draw[transform canvas={xshift=-0.5ex},->](A) -- (P) node[midway] {$\beta \to 0$}; 
 \draw[transform canvas={yshift=0.0ex},->] (A) --(C) node[midway] {};
\draw[transform canvas={yshift=-0.0ex},->](C) -- (A) node[midway] {}; 
\draw[transform canvas={xshift=0.5ex},->] (B) --(C) node[midway] {$\beta \nrightarrow 0$};
\draw[transform canvas={xshift=-0.5ex},->](C) -- (B) node[midway] {$\beta \to 0 $};   
\end{tikzpicture}
\ee
 \end{center}

If $\epsilon_1 + \epsilon_2  \neq 0$ (i.e. when Omega-deformation on the 2d side of the AGT correspondence is turned on), (\ref{dia 5}), and the explanation below (\ref{dia 2}), would mean that we ought to have
\tikzset{node distance=6.0cm, auto}
\begin{center}
\be
\label{dia 6}
 \begin{tikzpicture}
  \node (P) {$\underbrace{\widehat{{\bf Y}}({\frak {so}}(2)_{\textrm {aff}, 1}) \otimes \dots \otimes \widehat{{\bf Y}}({\frak {so}}(2)_{\textrm {aff}, 1})}_{N \,{\textrm {times}}}$};
  \node (B) [right of=P] {$\widehat {{\cal W}}({\frak{so}(2N})_{\textrm{aff}, k'})$};
  \node (A) [below of=P, yshift = 2.9cm] {$\underbrace{\widehat{{\bf U}_q}({\bf L}{\frak {so}}(2)_{\textrm {aff}, 1}) \otimes \dots \otimes \widehat{{\bf U}_q}({\bf L}{\frak {so}}(2)_{\textrm {aff}, 1})}_{N \,{\textrm {times}}}$};
  \node (C) [below of=B, yshift = 2.9cm] {$\widehat {{\cal W}^q}({\frak{so}(2N})_{\textrm{aff}, k'})$};
   \draw[transform canvas={yshift=0.0ex},->] (P) --(B) node[midway] {};
\draw[transform canvas={yshift=-0.0ex},->](B) -- (P) node[midway] {}; 
 \draw[transform canvas={xshift=0.5ex},->] (P) --(A) node[midway] {$\beta \nrightarrow 0$};
\draw[transform canvas={xshift=-0.5ex},->](A) -- (P) node[midway] {$\beta \to 0$}; 
 \draw[transform canvas={yshift=0.0ex},->] (A) --(C) node[midway] {};
\draw[transform canvas={yshift=-0.0ex},->](C) -- (A) node[midway] {}; 
\draw[transform canvas={xshift=0.5ex},->] (B) --(C) node[midway] {$\beta \nrightarrow 0$};
\draw[transform canvas={xshift=-0.5ex},->](C) -- (B) node[midway] {$\beta \to 0 $};   
\end{tikzpicture}
\ee
 \end{center}
Note that the upper horizontal relation in (\ref{dia 6}) is consistent with the mathematical results of~\cite[$\S$19.2]{MO},\footnote{This claim can be justified as follows. First, note that  $SO(2) \cong U(1)$ whence we can relate ${{\bf Y}}({\frak {so}}(2)_{\textrm {aff}})$ to ${{\bf Y}}({\frak {gl}}(1)_{\textrm {aff}})$. Second, note that in this case, the additional O6-plane below the stack of $N$ coincident D6-branes wrapping the ${\bf S}^2$ would result in a mirror image of the $N$ D6-branes. Hence,  to each factor of $\widehat{{\bf Y}}({\frak {so}}(2)_{\textrm {aff}})$, one must associate \emph{two} factors of ${\widehat{\bf Y}}({\frak {gl}}(1)_{\textrm {aff}})$, i.e.  we effectively have $2N$ factors of ${\widehat{\bf Y}}({\frak {gl}}(1)_{\textrm {aff}})$ whose corresponding tensor product, according to \emph{loc.~cit}, should be identified with $\widehat{\cal W}({\frak {gl}}(2N)_{\textrm {aff}})$. However, because of the reality condition of the chiral fermions on the ${\bf S}^2$ which generate the underlying affine Lie algebra, we necessarily have a special orthogonal group~\cite[$\S$3.2]{4d AGT}, i.e. we ought to have $\widehat{\cal W}({\frak {so}}(2N)_{\textrm {aff}})$ instead. In other words, we have our claim. \label{so(2) = u(1)}} while the left vertical relations are consistent with the mathematical results of~\cite{GM}.\footnote{In \emph{loc~cit.}, the correspondence between ${\bf U}_q({\bf L}{\frak g}_{\textrm {aff}})$ and ${\bf Y}({\frak g}_{\textrm {aff}})$ holds for $\frak g = {\frak {so}} (2N)$, too.} Also, the right vertical relation going upwards is consistent with the physical results of~\cite[$\S$5.3]{4d AGT}, while the right vertical relation going downwards is consistent with the corresponding relation in diagrams (\ref{dia 1}) and (\ref{dia 4}) (where it is clear that the $\beta \nrightarrow 0$ limit results in a $q$-deformation of the relevant affine $\cal W$-algebra). Altogether, this implies that the lower horizontal relation in (\ref{dia 6}) is also consistent, as expected.

By comparing the bottom right-hand corner of (\ref{dia 6}) with the bottom right-hand corner of (\ref{dia 1}) for the $A$ groups, bearing in mind footnote~\ref{1}, we find that in place of (\ref{Z5d-pure-SU(N)}), we ought to have
\be
\label{Z5d-pure-SO(2N)}
\boxed{{Z^{{\rm pure}, \, {\rm 5d}}_{{\rm inst}, \, SO(2N)} (\epsilon_1, \epsilon_2, \vec a, \beta, \Lambda)  = \langle G_{SO(2N)}  | G_{SO(2N)} \rangle}}
\ee
where the \emph{coherent state}
\be
\label{GSO(2N)}
\boxed{| G_{SO(2N)} \rangle \in {\widehat{{\cal W}^q} ({{^L\frak{so}}(2N)_{\textrm{aff}}})}}
\ee
In arriving at (\ref{Z5d-pure-SO(2N)})--(\ref{GSO(2N)}), we have just derived a 5d pure AGT correspondence for the $D_{N}$ groups.

In the limit $\beta \to 0$, the top right-hand side of (\ref{dia 6}) tells us that (\ref{Z5d-pure-SO(2N)})--(\ref{GSO(2N)}) would reduce to the result for the 4d case in~\cite[$\S$5.3]{4d AGT}, as they should.

\bigskip\noindent{\it A 5d Pure AGT Correspondence for the $C$ and $G_2$ Groups}

Last but not least, let us now derive a 5d pure AGT correspondence for the $C_{N-1}$ and $G_2$ groups. To this end, note that according to~\cite[$\S$5.3]{4d AGT}, in order to get, on the gauge theory side, a $C_{N-1} = USp(2N-2)$ or $G_2$ group when $\beta =0$, one must $\mathbb Z_n$-twist the affine Lie algebra underlying the 2d theory, where $n=2$ or $3$ (with $N=4$). In particular, this would mean that when $\epsilon_1 + \epsilon_2 = 0$, one would have to replace, in the upper line of (\ref{dia 5}), ${\frak{so}(2N})_{\textrm{aff}, 1}$ and ${\frak {so}}(2)_{\textrm {aff},1}$ with their $\mathbb Z_n$-twisted versions ${\frak{so}(2N})^{(n)}_{\textrm{aff}, 1}$ and ${\frak {so}}(2)^{(n)}_{\textrm {aff},1}$. As this $\mathbb Z_n$-twist is independent of the value of $\beta$, and looping a twisted affine Lie algebra does not alter its twist characteristics, it would mean that for $C$ and $G_2$ groups, we ought to have, instead of (\ref{dia 5}), the diagram
\tikzset{node distance=5.0cm, auto}
\begin{center}
\be
\label{dia 7}
 \begin{tikzpicture}
  \node (P) {$\underbrace{\widehat{{\frak {so}}(2)}^{(n)}_{\textrm {aff}, 1} \otimes \dots \otimes \widehat{{\frak {so}}(2)}^{(n)}_{\textrm {aff}, 1}}_{N \,{\textrm {times}}}$};
  \node (B) [right of=P] {$\widehat {{\frak{so}(2N})}^{(n)}_{\textrm{aff},1}$};
  \node (A) [below of=P, yshift = 1.8cm] {$\underbrace{\widehat{{\bf L}{\frak {so}}(2)}^{(n)}_{\textrm {aff}, 1} \otimes \dots \otimes \widehat{{\bf L}{\frak {so}}(2)}^{(n)}_{\textrm {aff}, 1})}_{N \,{\textrm {times}}}$};
  \node (C) [below of=B, yshift = 1.8cm] {$\widehat {{\bf L}{\frak{so}(2N})}^{(n)}_{\textrm{aff},1}$};
   \draw[transform canvas={yshift=0.0ex},->] (P) --(B) node[midway] {};
\draw[transform canvas={yshift=-0.0ex},->](B) -- (P) node[midway] {}; 
 \draw[transform canvas={xshift=0.5ex},->] (P) --(A) node[midway] {$\beta \nrightarrow 0$};
\draw[transform canvas={xshift=-0.5ex},->](A) -- (P) node[midway] {$\beta \to 0$}; 
 \draw[transform canvas={yshift=0.0ex},->] (A) --(C) node[midway] {};
\draw[transform canvas={yshift=-0.0ex},->](C) -- (A) node[midway] {}; 
\draw[transform canvas={xshift=0.5ex},->] (B) --(C) node[midway] {$\beta \nrightarrow 0$};
\draw[transform canvas={xshift=-0.5ex},->](C) -- (B) node[midway] {$\beta \to 0 $};   
\end{tikzpicture}
\ee
 \end{center}

Turning on Omega-deformation on the 2d side of the AGT correspondence does not affect the $\mathbb Z_n$-twist.  Hence, from (\ref{dia 7}), and the explanation below (\ref{dia 2}), it would mean that for general $\epsilon_{1,2}$, we have, for the $C$ and $G_2$ groups, instead of (\ref{dia 6}), the diagram
\tikzset{node distance=6.2cm, auto}
\begin{center}
\be
\label{dia 8}
 \begin{tikzpicture}
  \node (P) {$\underbrace{\widehat{{\bf Y}}({\frak {so}}(2)^{(n)}_{\textrm {aff}, 1}) \otimes \dots \otimes \widehat{{\bf Y}}({\frak {so}}(2)^{(n)}_{\textrm {aff}, 1})}_{N \,{\textrm {times}}}$};
  \node (B) [right of=P] {$\widehat {{\cal W}}({\frak{so}(2N})^{(n)}_{\textrm{aff}, k'})$};
  \node (A) [below of=P, yshift = 3.0cm] {$\underbrace{\widehat{{\bf U}_q}({\bf L}{\frak {so}}(2)^{(n)}_{\textrm {aff}, 1}) \otimes \dots \otimes \widehat{{\bf U}_q}({\bf L}{\frak {so}}(2)^{(n)}_{\textrm {aff}, 1})}_{N \,{\textrm {times}}}$};
  \node (C) [below of=B, yshift = 3.0cm] {$\widehat {{\cal W}^q}({\frak{so}(2N})^{(n)}_{\textrm{aff}, k'})$};
   \draw[transform canvas={yshift=0.0ex},->] (P) --(B) node[midway] {};
\draw[transform canvas={yshift=-0.0ex},->](B) -- (P) node[midway] {}; 
 \draw[transform canvas={xshift=0.5ex},->] (P) --(A) node[midway] {$\beta \nrightarrow 0$};
\draw[transform canvas={xshift=-0.5ex},->](A) -- (P) node[midway] {$\beta \to 0$}; 
 \draw[transform canvas={yshift=0.0ex},->] (A) --(C) node[midway] {};
\draw[transform canvas={yshift=-0.0ex},->](C) -- (A) node[midway] {}; 
\draw[transform canvas={xshift=0.5ex},->] (B) --(C) node[midway] {$\beta \nrightarrow 0$};
\draw[transform canvas={xshift=-0.5ex},->](C) -- (B) node[midway] {$\beta \to 0 $};   
\end{tikzpicture}
\ee
 \end{center}

Comparing the bottom right-hand corner of (\ref{dia 8}) with the bottom right-hand corner of (\ref{dia 1}) for the $A$ groups, and bearing in mind the isomorphisms $\frak{so}(2N)^{(2)}_{\textrm{aff}} \cong {{^L\frak{usp}}(2N-2)_{\textrm{aff}}}$ and $\frak{so}(2N)^{(3)}_{\textrm{aff}} \cong {{^L\frak{g}_2}_{ \, \textrm{aff}}}$ (when $N=4)$, it would mean that in place of (\ref{Z5d-pure-SU(N)}), we ought to have
\be
\label{Z5d-pure-USp(2N-2)}
\boxed{{Z^{{\rm pure}, \, {\rm 5d}}_{{\rm inst}, \, USp(2N-2)} (\epsilon_1, \epsilon_2, \vec a, \beta, \Lambda)  = \langle G_{USp(2N-2)}  | G_{USp(2N-2)} \rangle}}
\ee
and
\be
\label{Z5d-pure-G2}
\boxed{{Z^{{\rm pure}, \, {\rm 5d}}_{{\rm inst}, \, G_2} (\epsilon_1, \epsilon_2, \vec a, \beta, \Lambda)  = \langle G_{G_2}  | G_{G_2} \rangle}}
\ee
where the \emph{coherent states}
\be
\label{GUSp(2N-2)}
\boxed{| G_{USp(2N-2)} \rangle \in \widehat{{\cal W}^q} ({^L\frak{usp}}(2N-2)_{\textrm{aff}})}
\ee
and
\be
\label{GG2}
\boxed{| G_{G_2} \rangle \in \widehat{{\cal W}^q} ({^L\frak{g}_2}_{\, \textrm{aff}})}
\ee
In arriving at (\ref{Z5d-pure-USp(2N-2)})--(\ref{GG2}), we have just derived a 5d pure AGT correspondence for the $C_{N -1}$ and $G_2$ groups.

In the limit $\beta \to 0$, the top right-hand side of (\ref{dia 8}) tells us that (\ref{Z5d-pure-USp(2N-2)})--(\ref{GG2}) would reduce to the result for the 4d case in~\cite[$\S$5.3]{4d AGT}, as they should.

%
%
%

\newsubsection{An M-Theoretic Derivation of a 5d Pure AGT Correspondence for the $E_{6,7,8}$ and $F_4$ Groups in the Topological String Limit}

Lastly, let us now proceed to furnish an M-theoretic derivation of a 5d pure AGT Correspondence for the $E_{6,7,8}$ and $F_4$ groups. 

\bigskip\noindent{\it A 4d Pure AGT Correspondence for $E_{6,7,8}$ and $F_4$ Groups in the Topological String Limit}

Consider the discussion behind~\cite[eqns.~(3.137)--(3.148) and (3.149)--(3.152)]{4d AGT} with $k=1$ and $E_6$ generalized to $E_{6,7,8}$ therein. Let us turn on Omega-deformation via a fluxbrane as indicated in~\cite[eqn.~(5.4)]{4d AGT}, where the $x_{2,3,4,5}$ directions span the $\mathbb R^4$ in~\cite[eqn.~(3.137) and (3.149)]{4d AGT}. Let us also take the topological string limit and set $\epsilon_3 = \epsilon_1 + \epsilon_2 =0$ and $\epsilon_1 = h = - \epsilon_2$. Then, if we were to repeat the discussion there,\footnote{In particular, since $\epsilon_3 = 0$, there would be no effect on the space of gauge fields $A$ as we go around the circle ${\bf S}^1_r$ in~\cite[eqn.~(3.147)]{4d AGT} whence the analysis which follows remains the same, unlike in the case where $\epsilon_3 \neq 0$. \label{intricate}} we find that in place of~\cite[eqn.~(3.148) and (3.152)]{4d AGT}, we would have
\be
\label{4d AGT-TS-EF-identity}
 {\rm H}^\ast_{U(1)_h \times U(1)_{-h} \times T}({\cal M}^{G}_{{\bf R}^4}) = \widehat {^L\frak g}_{{\rm aff}, 1},
\ee
where ${\rm H}^\ast_{U(1)_h \times U(1)_{-h} \times T}({\cal M}^{G}_{{\bf R}^4})$ is the $U(1)_h \times U(1)_{-h} \times T$-equivariant cohomology of ${\cal M}^{G}_{{\bf R}^4}$, the (compactified) moduli space of $G$-instantons on $\bf R^4$, while $G = E_{6,7,8}$ or $F_4$ with Cartan subgroup $T$ and Lie algebra $\frak g$. Here, the RHS is a module of a 2d CFT on a cylinder, and the instanton number on the LHS corresponds to the (holomorphic) conformal weight eigenvalue on the RHS. 

Repeating the arguments in~\cite[eqn.~(5.27)--(5.36)]{4d AGT} with (\ref{4d AGT-TS-EF-identity}) in mind,  we find that we can express the 4d instanton partition function for a pure $G$ theory as
\be
\label{q | q}
{Z^{{\rm pure}, \, {\rm 4d}}_{{\rm inst}, \, G} (h, \vec a, \Lambda) =  \langle {\rm coh}_{h}  | {\rm coh}_{h} \rangle},
\ee
where $| {\rm coh}_{h} \rangle \in \widehat {^L\frak g}_{{\rm aff}, 1}$ is a \emph{coherent} state. Note that this is just an $E$-$F$ group generalization of~\cite[eqn.~(2.32)]{5d AGT}, and it agrees with the Nekrasov-Okounkov conjecture in~\cite[$\S$5.4]{NO}, as explained below~\cite[eqn.~(7.9)]{4d AGT}.

%
%

\bigskip\noindent{\it A 5d Pure AGT Correspondence for $E_{6,7,8}$ and $F_4$ Groups in the Topological String Limit}

We would now like to generalize (\ref{q | q}) to 5d. To this end, first note that according to the explanations above~\cite[eqn.~(3.57)]{5d AGT}, the 4d and 5d instanton partition function can be associated with a gauged supersymmetric quantum mechanics and cylindrical sigma-model with target ${\cal M}^{G}_{{\bf R}^4}$, respectively; in other words, the 4d instanton partition function is determined by points in ${\cal M}^{G}_{{\bf R}^4}$, while the 5d instanton partition function is determined by loops in ${\cal M}^{G}_{{\bf R}^4}$. 

Second, note that like in~\cite[eqn.~(2.32)]{5d AGT}, one can interpret the RHS of (\ref{q | q}) as a correlation function of two coherent state vertex operators inserted at the points $0$ and $\infty$ of ${\bf S}^2$, i.e. its poles. 

Therefore, it would mean that in order to go from 4d to 5d, one must replace the states $ \langle {\rm coh}_{\hbar} |$  and $|{\rm coh}_{\hbar} \rangle$ defined at the points $0$ and $\infty$, with the states $ \langle {\rm cir}_{\hbar} |$  and $|{\rm cir}_{\hbar} \rangle$ which are their projections onto a loop. Indeed, as $\beta \nrightarrow 0$, i.e. as we go from 4d to 5d, the higher KK modes which were previously decoupled will now contribute to the partition function of the gauge theory. This is consistent with the fact that when a quantum state  $|{\rm coh}_{\hbar} \rangle$ is no longer confined to a space of infinitesimal size, i.e. a point, but is projected onto a loop to become $|{\rm cir}_{\hbar} \rangle$, we will get contributions from higher energy modes.    

To understand the state $|{\rm cir}_{\hbar} \rangle$ (and its dual  $ \langle {\rm cir}_{\hbar} |$), firstly, recall that the (chiral) WZW model which underlies $^L{\frak g}_{{\rm aff}, 1} = {^L\frak e}_{6,7,8 \, {\rm aff}, 1} = {\frak e}_{6,7,8 \, {\rm aff}, 1}$ or $^L{\frak f}_{4 \, {\rm aff}, 1} = {\frak e}^{(2)}_{6 \, {\rm aff}, 1}$ on ${\bf S}^2$, can be regarded as a (twisted) bosonic sigma-model with worldsheet ${\bf S}^2$ and target a $G = {E}_{6,7,8}$ or $E_6$ group manifold. Thus, $|{\rm coh}_{\hbar} \rangle$, which is defined over a point in ${\bf S}^2$, would be associated with a point in the space of all points into the target, i.e. a point in the $G$ group itself. Similarly, $|{\rm cir}_{\hbar} \rangle$, which is defined over a loop in ${\bf S}^2$, would be associated with a point in the space of all loops into the target, i.e. a point in the loop group of $G$.  Hence, since $| {\rm coh}_{\hbar} \rangle \in \widehat {^L\frak g}_{{\rm aff}, 1}$, we ought to have  
\be
\label{cir | cir}
\boxed{{Z^{{\rm pure}, \, {\rm 5d}}_{{\rm inst}, \, G} (h, \vec a, \Lambda) =  \langle {\rm cir}_{h}  | {\rm cir}_{h} \rangle}}
\ee
where
\be
\label{state double loop-EF}
\boxed{|{\rm cir}_{h} \rangle \in \widehat{{{\bf L} {^L\frak g}}}_{\rm aff,1}} 
\ee
and ${{{\bf L} {^L\frak g}}}_{\rm aff,1}$ is a Langlands dual {toroidal} Lie algebra given by the loop algebra of $ {^L\frak g}_{{\rm aff},1}$. 

In arriving at (\ref{cir | cir}) and (\ref{state double loop-EF}), we have just derived a 5d pure AGT correspondence for the $E_{6,7,8}$ and $F_4$ groups in the topological string limit.

Note that (\ref{cir | cir}) and (\ref{state double loop-EF}) are consistent with the (lower horizontal relations of the) diagrams (\ref{dia 2}), (\ref{dia 3}), (\ref{dia 5}) and (\ref{dia 7}) that are associated with a 5d pure AGT correspondence for the $A$, $B$, $C$, $D$ and $G_2$ groups in the topological string limit, as they should.  

\bigskip\noindent{\it Beyond the Topological String Limit}

To go beyond the topological string limit such that $\epsilon_3 \neq 0$ requires a more intricate analysis, as mentioned in footnote~\ref{intricate}. As such, we shall leave the discussion for another occasion.

\newsection{A 5d and 6d AGT Correspondence on ALE Space of Type $ADE$}
\label{s3}


\newsubsection{An M-Theoretic Derivation of Nakajima's Result Relating Equivariant K-Theory of Quiver Varieties to Quantum Toroidal Algebras}
\label{s3.1}

Before we proceed to furnish an M-theoretic derivation of a 5d and 6d AGT correspondence for $SU(N)$ on the smooth ALE space $\widetilde {{\bf R}^4 / {\Gamma}}$, where the finite subgroup $\Gamma \subset SU(2)$ is, via the McKay correspondence~\cite{McKay}, related to the Lie algebra ${\frak g}_\Gamma$ of $ADE$ groups, it will be useful to first derive a mathematical result of Nakajima's~\cite{Nak} which relates the equivariant K-theory of quiver varieties to quantum toroidal algebras. Specifically, his result means that the $U(1)^2 \times \mathscr T$-equivariant K-cohomology of the (compactified) moduli space ${\cal M}^{U(1)}_{\Gamma}$ of framed $U(1)$-instantons on $\widetilde {{\bf R}^4 / \Gamma}$, where $\mathscr T$ (related to $\Gamma$) acts to change the framing at infinity while $U(1)^2$ corresponds to the rotation of the two ${\bf R}^2 \subset {\bf R}^4$ planes, is related  to a module of the quantum toroidal algebra of  $\frak g_\Gamma$ of type $ADE$ at level $1$.

\bigskip\noindent{\it Deriving Nakajima's Result for Type $A$}

Let $\Gamma$ be the cyclic group $\mathbb Z_p$, so that $\widetilde {{\bf R}^4 / \Gamma}$ is an ALE space of type $A_{p-1}$. Also, let the group be $U(1)$ on the pure gauge theory side with $\epsilon_1 = \epsilon_2 = \beta = 0$, i.e. we turn off Omega-deformation in the 4d pure $U(1)$ case. Then, according to our discussion in~\cite[$\S$5.2]{4d AGT} (adapted to $\widetilde {{\bf R}^4 / \mathbb Z_p}\vert_{0,0}$ instead of ${\bf R}^4\vert_{\epsilon_1, \epsilon_2}$), and the fact that an ${\cal N} = (2,0)$ theory on $\Sigma_t  \times M_4$ is topological along $M_4$ if it is a hyperk\"ahler four-manifold~\cite{junya}, we find that the relation underlying the correspondence would be
\be
{\rm H}^\ast_{{\bf L}^2}({\cal M}^{U(1)}_{\mathbb Z_p}) = \widehat {{\frak{u}(1})}_{\textrm{aff},p},
\ee
where the ${\bf L}^2$-cohomology classes on the left are associated with the 4d instanton partition function on $\widetilde {{\bf R}^4 / {\mathbb Z}_p}$, while the module of the Heisenberg algebra on the right is associated with the 2d theory on a cylinder. 

Let us now turn on Omega-deformation but with $\epsilon_1 = - \epsilon_2 = h$, i.e. we turn on Omega-deformation only on the gauge theory side. Then, according to~\cite[$\S$5.2]{4d AGT}, the relation underlying the correspondence would be\footnote{In \emph{loc~cit.}, what was discussed is actually an $SU(N)$ gauge theory on ${\bf R}^4$ whose equivariance group is $U(1)^2 \times T$, where the Cartan subgroup $T \subset SU(N)$ acts to change the framing at infinity. In order to deduce our following relation, we just have to note that when one replaces ${\bf R}^4$ with $\widetilde {{\bf R}^4 / {\mathbb Z}_p}$ and $SU(N)$ with $U(1)$, the group that acts to change the framing at infinity would be $\mathscr T$ instead of $T$~\cite[$\S$6]{Nak}.}
\be
\label{nak-1}
{\rm H}^\ast_{U(1)_h \times U(1)_{-h} \times \mathscr T}({\cal M}^{U(1)}_{\mathbb Z_p}) = \widehat {{\frak{u}(1})}_{\textrm{aff},p},
\ee 
where the (square-integrable) $U(1)_h \times U(1)_{-h} \times \mathscr T$-equivariant cohomology classes on the left are associated with the 4d instanton partition function on $\widetilde {{\bf R}^4 / {\mathbb Z}_p}$ with the two ${\bf R}^2 \subset {\bf R}^4$ plane rotation parameters being $h$ and $-h$, respectively, while the module of the Heisenberg algebra on the right is again associated with the 2d theory on a cylinder. 

As such, when $\beta \neq 0$, the corresponding relation in the 5d pure case for $\epsilon_1 = - \epsilon_2 = h$, is, according to~\cite[$\S$2.2]{5d AGT}, 
\be
\label{above}
{\rm K}^\ast_{U(1)_h \times U(1)_{-h} \times \mathscr T}({\cal M}^{U(1)}_{\mathbb Z_p}) = \widehat {{\bf L}{\frak{u}(1})}_{\textrm{aff},p},
\ee 
where the (square-integrable) $U(1)_h \times U(1)_{-h} \times \mathscr T$-equivariant K-cohomology classes on the left are associated with the 5d instanton partition function on ${\bf S}^1 \times \widetilde {{\bf R}^4 / {\mathbb Z}_p}$ with the two ${\bf R}^2 \subset {\bf R}^4$ plane rotation parameters being $h$ and $-h$, respectively, while the module of the loop algebra on the right is associated with the 2d theory on a cylinder. 

Notice that (\ref{above}) (at $p=1$) is indeed consistent with (\ref{Z5d-pure-SU(N)})--(\ref{GSU(N)}) (at $N=1$) -- when $\epsilon_1 = - \epsilon_2 = h$, one can see from (\ref{variables-pure-SU(N)}) that $s =  e^{\beta h} = r$, i.e. (\ref{he-pure-SU(N)}) reduces to a Heisenberg algebra, whence (\ref{Z5d-pure-SU(N)-G})--(\ref{Z5d-pure-SU(N)-G-dual}) would mean that we have, on the 2d side, a module of the loop algebra in (\ref{above}).   

Now, if $\epsilon_1 \neq \epsilon_2$, i.e. Omega-deformation on the 2d side is also turned on, according to the explanations below (\ref{dia 2}) regarding the Heisenberg algebra,\footnote{The aforementioned explanations involve the Heisenberg algebra at level 1 and not $p$. However, this disparity is inconsequential at the level of modules.} and a derived level-rank duality of the 2d theory (see~\cite[eqn.~(3.158)]{4d AGT}), it would mean that  in place of (\ref{above}), we would have
\be
\label{Nak}
\boxed{{\rm K}^\ast_{U(1)_{\epsilon_1} \times U(1)_{\epsilon_2} \times \mathscr T}({\cal M}^{U(1)}_{\mathbb Z_p}) = \widehat {{\bf U}_q}({\bf L}{\frak{su}(p})_{\textrm{aff},1})}
\ee 
This is just Nakajima's result for type $A$ in~\cite[$\S$6]{Nak}.

\bigskip\noindent{\it Deriving Nakajima's Result for Type $D$}

Let the group be $SO(2p)$ on the pure gauge theory side with $\epsilon_1 = \epsilon_2 = \beta = 0$, i.e. we turn off Omega-deformation in the 4d pure $SO(2p)$ case. Then, according to our discussion in~\cite[$\S$5.3]{4d AGT}, and the fact that the identity~\cite[eqn.~(3.162)]{4d AGT} also holds for resolved orbifolds (because the ${\cal N} = (2,0)$ theory on ${\bf S}^1 \times \mathbb R_t  \times M_4$ in~\cite[eqn.~(3.154)]{4d AGT} is topological along hyperk\"ahler $M_4$), we find that the relation underlying the correspondence would be
\be
{\rm H}^\ast_{{\bf L}^2}({\cal M}^{U(1)}_{\mathbb D_p}) = \widehat {{\frak{so}(2p})}_{\textrm{aff},1},
\ee
where $\mathbb D_p$ is a dihedral group, the ${\bf L}^2$-cohomology classes on the left are associated with the 4d instanton partition function, while the module of the affine algebra on the right is associated with the 2d theory on a cylinder. 

If we now repeat the arguments which took us to (\ref{nak-1}) and beyond to (\ref{Nak}) (omitting the last step of using a level-rank duality), we find that
\be
\label{Nak-D}
\boxed{{\rm K}^\ast_{U(1)_{\epsilon_1} \times U(1)_{\epsilon_2} \times \mathscr T}({\cal M}^{U(1)}_{\mathbb D_p}) = \widehat {{\bf U}_q}({\bf L}{\frak{so}(2p})_{\textrm{aff},1})}
\ee 
This is just Nakajima's result for type $D$ in~\cite[$\S$6]{Nak}.

\bigskip\noindent{\it A Useful Excursion}

We now make a useful excursion before we proceed to derive Nakajima's result for type $E$. 

Consider the following type IIB compactification:
\be
{\textrm {Type IIB}}: \, {\rm K3}_{E_{6,7,8}} \times {\bf S}^1 \times {\bf R}_t \times {\rm K3}_{A_{p-1}},
 \ee
where ${\rm K3}_{\mathscr G}$ denotes a singular K3 manifold with a singularity at the origin of type $\mathscr G$. 

If we scale ${\rm K3}_{E_{6,7,8}} $ to be much smaller than ${\bf S}^1 \times {\bf R}_t  \times {\rm K3}_{A_{p-1}}$, we effectively have, in the low energy limit, a six-dimensional spacetime theory that is ${\cal N} = (2,0)$ $E_{6,7,8}$ theory on ${\bf S}^1 \times {\bf R}_t  \times {\rm K3}_{A_{p-1}}$. As mentioned, an ${\cal N} = (2,0)$ theory on ${\bf S}^1 \times {\bf R}_t  \times M_4$, where $M_4$ is any hyperk\"ahler four-manifold, is topological along $M_4$ (and conformal along ${\bf S}^1 \times \mathbb R_t$). In particular, this means that the BPS spectrum of minimal energy states of the  ${\cal N} = (2,0)$ $E_{6,7,8}$ theory on ${\bf S}^1 \times {\bf R}_t  \times {\rm K3}_{A_{p-1}}$ -- which are states annihilated by all eight unbroken supercharges whence they satisfy $H=P$, where $H$ and $P$ are the Hamiltonian and momentum operators which generate translations along ${\bf R}_t$ and ${\bf S}^1$, respectively -- is invariant under topological deformations of ${\rm K3}_{A_{p-1}}$.  As such, according to~\cite[$\S$3.1]{4d AGT}, the Hilbert space of such BPS states would be spanned by
\be
 {\rm H}^\ast_{{\bf L}^2}({\cal M}^{E_{6,7,8}}_{\mathbb Z_p}),
\ee
and by scaling ${\rm K3}_{A_{p-1}}$ to be much smaller than ${\bf S}^1 \times {\bf R}_t$, we find that they live along ${\bf S}^1 \times {\bf R}_t$.

On the other hand, if  we scale ${\rm K3}_{A_{p-1}}$ to be much smaller than ${\bf S}^1 \times {\bf R}_t  \times {\rm K3}_{E_{6,7,8}}$, we effectively have, in the low energy limit, a six-dimensional spacetime theory that is ${\cal N} = (2,0)$ $A_{p-1}$ theory on ${\bf S}^1 \times {\bf R}_t  \times {\rm K3}_{E_{6,7,8}}$. Then, the Hilbert space of BPS states would be spanned by
\be
 {\rm H}^\ast_{{\bf L}^2}({\cal M}^{SU(p)}_{\Gamma_E}),
\ee
where ${\cal M}^{SU(p)}_{\Gamma_E}$ is the moduli space of $SU(p)$-instantons on $\widetilde {{\bf R}^4 / \Gamma_E}$, and $\Gamma_E$ is either a tetrahedral, octahedral or icosahedral group related to the $\frak e_{6,7,8}$ Lie algebra. By scaling ${\rm K3}_{E_{6,7,8}}$ to be much smaller than ${\bf S}^1 \times {\bf R}_t$, we again find that these states live along  ${\bf S}^1 \times {\bf R}_t$.

Therefore, if we were to simultaneously scale  ${\rm K3}_{E_{6,7,8}} $ and  ${\rm K3}_{A_{p-1}}$ to be much smaller than ${\bf S}^1 \times {\bf R}_t$, we would get a 2d ${\cal N} = (8,0)$ theory along ${\bf S}^1 \times {\bf R}_t$ whose minimal energy states with $H=P$ are spanned  by $ {\rm H}^\ast_{{\bf L}^2}({\cal M}^{E_{6,7,8}}_{\mathbb Z_p})$ and $ {\rm H}^\ast_{{\bf L}^2}({\cal M}^{SU(p)}_{\Gamma_E})$ at the \emph{same} time.  In other words, we ought to have
\be
\label{Nak-E-help}
{\rm H}^\ast_{{\bf L}^2}({\cal M}^{E_{6,7,8}}_{\mathbb Z_p}) = {\rm H}^\ast_{{\bf L}^2}({\cal M}^{SU(p)}_{\Gamma_E}).
\ee

\bigskip\noindent{\it Deriving Nakajima's Result for Type $E$}

We are now ready to derive Nakajima's result for type $E$. To this end, consider the discussion behind~\cite[eqns.~(3.137)--(3.148)]{4d AGT} with $k=1$ and $E_6$ generalized to $E_{6,7,8}$ therein, from which we then have
\be
\label{E-duality}
 {\rm H}^\ast_{{\bf L}^2}({\cal M}^{E_{6,7,8}}_{\mathbb Z_1}) = \widehat {{\frak e}}_{6,7,8 \,{\rm aff}, 1}.
\ee
Via (\ref{Nak-E-help}), we can also write this as
\be
\label{E-duality-relevant}
{\rm H}^\ast_{{\bf L}^2}({\cal M}^{U(1)}_{\Gamma_E}) = \widehat {{\frak e}}_{6,7,8 \,{\rm aff}, 1}.
\ee

Let us turn on Omega-deformation via a fluxbrane as indicated in~\cite[eqn.~(5.4)]{4d AGT}, where the $x_{2,3,4,5}$ directions span the $\mathbb R^4$ in~\cite[eqn.~(3.137)]{4d AGT}. Let us also set $\epsilon_3 = \epsilon_1 + \epsilon_2 =0$ and $\epsilon_1 = h = - \epsilon_2$. Then, if we were to repeat the discussion there, we find that only the LHS of (\ref{E-duality}) and therefore (\ref{E-duality-relevant}), would be Omega-deformed whence we would have
\be
\label{Nak-E-TS-4d}
 {\rm H}^\ast_{U(1)_h \times U(1)_{-h} \times \mathscr T}({\cal M}^{U(1)}_{\Gamma_E}) = \widehat {{\frak e}}_{6,7,8 \,{\rm aff}, 1},
\ee
where the (square-integrable) $U(1)_h \times U(1)_{-h} \times \mathscr T$-equivariant cohomology classes on the left can be associated with a 4d $U(1)$ instanton partition function on $\widetilde {{\bf R}^4 / \Gamma_E}$ with the two ${\bf R}^2 \subset {\bf R}^4$ plane rotation parameters being $h$ and $-h$, respectively, while the module of the affine algebra on the right is associated with a 2d theory on a cylinder. 

Next, applying the arguments in~\cite[$\S$2.2]{5d AGT} here, it would mean that from (\ref{Nak-E-TS-4d}), we can write 
\be
\label{Nak-E-TS-5d}
 {\rm K}^\ast_{U(1)_h \times U(1)_{-h} \times \mathscr T}({\cal M}^{U(1)}_{\Gamma_E}) = \widehat {{\bf L}{\frak e}}_{6,7,8 \,{\rm aff}, 1},
\ee
where the (square-integrable) $U(1)_h \times U(1)_{-h} \times \mathscr T$-equivariant K-cohomology classes on the left can be associated with a 5d  $U(1)$ instanton partition function on ${\bf S}^1 \times \widetilde {{\bf R}^4 / \Gamma_E}$ with the two ${\bf R}^2 \subset {\bf R}^4$ plane rotation parameters being $h$ and $-h$, respectively, while the module of the loop algebra on the right is associated with a 2d theory on a cylinder.

Consequently, if $\epsilon_1 + \epsilon_2 \neq 0$, according to our explanations below (\ref{dia 2}) for when $\beta \neq 0$,\footnote{This explanation -- that when $\epsilon_1 + \epsilon_2 \neq 0$, we ought to have $ {{\bf L}{\frak g}}_{{\rm aff}, 1} \to {\bf U_q}({\bf L}{\frak g}_{{\rm aff}, 1})$ -- should also hold for any other \emph{unique} Lie algebra $\frak g$ aside from $\frak {gl}(1)$ or $\frak {so}(2)$, such as ${\frak e}_{6,7,8}$.} we eventually have 
\be
\label{Nak-E-final}
\boxed{ {\rm K}^\ast_{U(1)_{\epsilon_1} \times U(1)_{\epsilon_2} \times \mathscr T}({\cal M}^{U(1)}_{\Gamma_E}) = \widehat {\bf U_q}({\bf L}{\frak e}_{6,7,8 \,{\rm aff}, 1})}
\ee
This is just Nakajima's result for type $E$ in~\cite[$\S$6]{Nak}.

\newsubsection{An M-Theoretic Derivation of a 5d AGT Correspondence for $SU(N)$ on ALE Space of Type $ADE$}


Armed with relations (\ref{Nak}), (\ref{Nak-D}) and (\ref{Nak-E-final}), we are now ready to derive a 5d AGT correspondence for $SU(N)$ on $\widetilde {{\bf R}^4 / \Gamma}$. For brevity, we shall limit ourselves to the pure case. 

From the K-theoretic expression of the relevant 5d instanton partition function on the LHS of~\cite[eqn.~(3.57)]{5d AGT} (which also holds for other groups), and the relation~(\ref{Nak}), (\ref{Nak-D}) and (\ref{Nak-E-final}),  we find that for a 5d pure $U(1)$ gauge theory on ${\bf S}^1 \times \widetilde {{\bf R}^4 / \Gamma}$, we have, in place of (\ref{Z5d-pure-SU(N)}), 
\be
\label{U(1)}
{\tilde Z}^{{\rm pure}, \, {\rm 5d}}_{{\rm inst}, \, U(1)} (\epsilon_1, \epsilon_2, \beta, \Lambda, \Gamma) =  \langle \widetilde{G}_{U(1)}  | \widetilde {G}_{U(1)} \rangle,
\ee
where $ | \widetilde {G}_{U(1)} \rangle$ is a \emph{coherent state} in $\widehat {{\bf U}_q}({\bf L}{\frak g}_{\Gamma  \, \textrm{aff},1})$. 

To obtain the result for the $SU(N)$ case, note that according to the lower horizontal relation in diagram (\ref{dia 1}), we ought to replace the states on the RHS of (\ref{U(1)}) with those in the $N$-tensor product of $\widehat {{\bf U}_q}({\bf L}{\frak g}_{\Gamma  \, \textrm{aff},1})$.  In other words, for a 5d pure $SU(N)$ gauge theory on ${\bf S}^1 \times  \widetilde {{\bf R}^4 / \Gamma}$, we have, in place of (\ref{U(1)}), 
\be
\label{SU(N) on Zp}
\boxed{{\tilde Z}^{{\rm pure}, \, {\rm 5d}}_{{\rm inst}, \, SU(N)} (\epsilon_1, \epsilon_2, \vec a, \beta, \Lambda, \Gamma) =  \langle \widetilde{G}_{SU(N)}  | \widetilde {G}_{SU(N)} \rangle}
\ee
where the \emph{coherent state}
\be
\label{G on Zp}
\boxed{| \widetilde {G}_{SU(N)} \rangle \in  \widehat{{\bf U}_q}({\bf L}{\frak g}_{\Gamma  \, \textrm{aff},1})_1 \otimes \dots \otimes \widehat{{\bf U}_q}({\bf L}{\frak g}_{\Gamma  \, \textrm{aff},1})_N}
\ee
and $\widehat{{\bf U}_q}({\bf L}{\frak g}_{\Gamma  \, \textrm{aff},1})_i$ is the $i$-th module. 

In arriving at (\ref{SU(N) on Zp})--(\ref{G on Zp}), we have just derived a 5d pure AGT correspondence for $SU(N)$ on the smooth ALE space $\widetilde {{\bf R}^4 / \Gamma}$!

When $\Gamma = \mathbb Z_p$ or $\mathbb D_p$ with $p=1$, (\ref{SU(N) on Zp}) and (\ref{G on Zp}), via the lower horizontal relation in (\ref{dia 1}) or (\ref{dia 6}), reduce to (\ref{Z5d-pure-SU(N)}) and (\ref{GSU(N)}) or (\ref{Z5d-pure-SO(2N)}) and (\ref{GSO(2N)}), respectively, as they should.

\newsubsection{An M-Theoretic Derivation of a 6d AGT Correspondence for $SU(N)$ on ALE Space of Type $ADE$}
\label{s3.3}


Let us now proceed to derive a 6d AGT correspondence for $SU(N)$ on $\widetilde {{\bf R}^4 / \Gamma}$. For brevity, we shall limit ourselves to the case of a conformal linear quiver theory which thus has $N_f = 2N$ fundamental matter. 

When $\Gamma = \mathbb Z_p$ and $p=1$, according to~\cite[$\S$5.1]{5d AGT}, the 6d Nekrasov instanton partition function on $\bar{\bf S}^1 \times {\bf S}^1 \times {\bf R}^4$ would be given by
\be
\label{Z6d-SU(N)}
{{Z^{\rm lin, \, 6d}_{{\rm inst}, \, SU(N)}({q_1}, \epsilon_1, \epsilon_2, {\vec a}, {\vec m}, \beta, R_6) =  \langle  {\bar \Phi}^{\bf w}_{\bf v} (z_1) \bar\Phi^{\bf v}_{\bf u} (z_2)   \rangle_{{\bf T}^2}}}.
\ee
Here, $q_1 = e^{2 \pi i \tau}$ (and $\tau$ is the complexified gauge coupling); ${\vec m} = (m_1, m_2, \dots, m_{2N})$ are the $2N$ masses; $R_6$ is the radius of the sixth circle $\bar{\bf S}^1$; the 2d vertex operators\footnote{Here, we have made use of the fact that the Ding-Iohara algebra is actually isomorphic to ${\bf U}_q({\bf L}{\frak {gl}}(1)_{\textrm {aff}})$~\cite{FT, SV}.}
\be
\label{Z6d-SU(N)-matter-vertex operator-map}
{{\bar\Phi}^{\bf c}_{\bf d}: \widehat{{\bf U}_{q, v}}({\bf L}{\frak {gl}}(1)_{\textrm {aff}, 1})_{d_1} \otimes \cdots \otimes \widehat{{\bf U}_{q, v}}({\bf L}{\frak {gl}}(1)_{\textrm {aff},1})_{d_N} \longrightarrow \widehat{{\bf U}_{q, v}}({\bf L}{\frak {gl}}(1)_{\textrm {aff},1})_{c_1} \otimes \cdots \otimes \widehat{{\bf U}_{q, v}}({\bf L}{\frak {gl}}(1)_{\textrm {aff},1})_{c_N}},
\ee
where  $\widehat{{\bf U}_{q, v}}({\bf L}{\frak {gl}}(1)_{\textrm {aff},1})_{\gamma}$ is a $\gamma$-dependent module of an elliptic deformation of the quantum toroidal algebra ${{\bf U}_{q}}({\bf L}{\frak {gl}}(1)_{\textrm {aff},1})$ with deformation parameter $v = e^{-1 / R_6}$, that is supposed to be isomorphic to the elliptic Ding-Iohara algebra at level 1 constructed in~\cite{Saito}; and the parameters ${\bf v} = (v_1, \dots, v_N)$, ${\bf w} = (w_1, \dots, w_N)$ and ${\bf u} = (u_1, \dots, u_N)$ are such that
\be
\label{Z6d-SU(N)-variables}
{w_i = e^{- \beta m_i},  \quad v_i = e^{-\beta a_i}, \quad u_i = e^{- \beta m_{N+i}}}.
\ee

When $R_6 \to 0$, i.e. in the 5d case, according to~\cite[$\S$3.2]{5d AGT}, the algebra ${{\bf U}_{q, v}}({\bf L}{\frak {gl}}(1)_{\textrm {aff},1})$ in (\ref{Z6d-SU(N)-matter-vertex operator-map}) ought to be replaced by the algebra ${{\bf U}_{q}}({\bf L}{\frak {gl}}(1)_{\textrm {aff},1})$, which is isomorphic to the Ding-Iohara algebra at level 1 (as mentioned earlier in $\S$2.1\ref{s2.1}). In other words, we have to replace the above 2d vertex operators with
\be
\label{Z5d-SU(N)-matter-vertex operator-map}
{{\Phi}^{\bf c}_{\bf d}: \widehat{{\bf U}_{q}}({\bf L}{\frak {gl}}(1)_{\textrm {aff}, 1})_{d_1} \otimes \cdots \otimes \widehat{{\bf U}_{q}}({\bf L}{\frak {gl}}(1)_{\textrm {aff},1})_{d_N} \longrightarrow \widehat{{\bf U}_{q}}({\bf L}{\frak {gl}}(1)_{\textrm {aff},1})_{c_1} \otimes \cdots \otimes \widehat{{\bf U}_{q}}({\bf L}{\frak {gl}}(1)_{\textrm {aff},1})_{c_N}}.
\ee

Now, notice that  (\ref{GSU(N)}), the lower horizontal relation in (\ref{dia 1}), and (\ref{G on Zp}), mean that in going from $\Gamma = \mathbb Z_1 = \mathds{1}$ to general $\Gamma$ in the 5d case, one ought to replace ${\bf L}{\frak {gl}}(1)_{\textrm {aff},1}$ with ${\bf L}{\frak g}_{\Gamma  \, \textrm{aff},1}$ in the formulas. In turn, by reversing the discussion about the $R_6 \to 0$ limit, and from (\ref{Z6d-SU(N)})--(\ref{Z6d-SU(N)-matter-vertex operator-map}), one can readily see that we can express the 6d Nekrasov instanton partition function on $\bar{\bf S}^1 \times {\bf S}^1 \times \widetilde {{\bf R}^4 / \Gamma}$ as
\be
\label{Z6d-SU(N)-Zp}
\boxed{{{{\tilde Z}^{\rm lin, \, 6d}_{{\rm inst}, \, SU(N)}({q_1}, \epsilon_1, \epsilon_2, {\vec a}, {\vec m}, \beta, R_6, \Gamma) =  \langle  {\bar \Psi}^{\bf w}_{\bf v} (z_1) \bar\Psi^{\bf v}_{\bf u} (z_2)   \rangle_{{\bf T}^2}}}}
\ee
where
 \be
\label{Z6d-SU(N)-matter-vertex operator-map-Zp}
\boxed{{{\bar\Psi}^{\bf c}_{\bf d}: \widehat{{\bf U}_{q, v}}({\bf L}{\frak g}_{\Gamma  \, \textrm{aff},1})_{d_1} \otimes \cdots \otimes \widehat{{\bf U}_{q, v}}({\bf L}{\frak g}_{\Gamma  \, \textrm{aff},1})_{d_N} \longrightarrow \widehat{{\bf U}_{q, v}}({\bf L}{\frak g}_{\Gamma  \, \textrm{aff},1})_{c_1} \otimes \cdots \otimes \widehat{{\bf U}_{q, v}}({\bf L}{\frak g}_{\Gamma  \, \textrm{aff},1})_{c_N}}}
\ee
and $\widehat{{\bf U}_{q, v}}({\bf L}{\frak g}_{\Gamma  \, \textrm{aff},1})_{\gamma}$ is a $\gamma$-dependent module of an algebra that can be obtained by the method of elliptic deformation applied to the free-field realization of the quantum toroidal algebra ${{\bf U}_{q}}({\bf L}{\frak g}_{\Gamma  \, \textrm{aff},1})$ constructed in~\cite{Saito-elliptic}.\footnote{I would like to thank Y.~Saito for his expertise on this point.}

In arriving at (\ref{Z6d-SU(N)-Zp})--(\ref{Z6d-SU(N)-matter-vertex operator-map-Zp}), we have just derived a 6d AGT correspondence for $SU(N)$ with $N_f =2N$ fundamental matter on the smooth ALE space $\widetilde {{\bf R}^4 / \Gamma}$!

\newsection{$\cal W$-algebras and Higher Quantum Geometric Langlands Duality}
\label{s4}


\newsubsection{An M-Theoretic Realization of Affine $\cal W$-algebras and a Quantum Geometric Langlands Duality}
\label{s4.1}

Let us now furnish an M-theoretic realization of affine $W$-algebras associated with the compact $A$, $B$, $C$, $D$ and $G_2$ Lie groups which appear in the context of the 4d AGT correspondence in~\cite[Part II]{4d AGT}, and show that they obey identities obtained earlier by mathematicians Feigin-Frenkel in~\cite{FF} which underlie a quantum geometric Langlands duality.  

\bigskip\noindent{\it Affine $\cal W$-algebras Associated with the Compact $A$, $B$, $C$, $D$ and $G_2$ Lie Groups} 

Since our ultimate aim is to derive identities which underlie a Langlands duality, let us specialize our discussion of the 4d AGT correspondence in~\cite[Part II]{4d AGT} to the ${\cal N} = 4$ (or massless ${\cal N} = 2^\ast$) case so that we can utilize $S$-duality. 

 According to ~\cite[eqns.~(5.5) and (5.8)]{4d AGT}, the string-dual M-theory compactifications which are relevant in this case would be\footnote{In the following, we have, for our convenience and purpose, replaced ${\bf R}_t$ in \cite[eqns.~(5.5) and (5.8)]{4d AGT} with ${\bf S}^1_t$. This replacement is inconsequential, as the sought-after quantities in \emph{loc.~cit.} are the relevant BPS states furnished by the spectrum of a topological sigma-model on ${\bf S}^1_n \times {\bf R}_t$ defined by target space differential forms which are independent of the global topology of its worldsheet.} 
\be
\label{4d dual AB}
\underbrace{{\bf R}^4\vert_{\epsilon_1, \epsilon_2}  \times {{\bf S}^1_n} \times {\bf S}^1_t}_{\textrm{$N$ M5-branes}}\times {\bf R}^{5}\vert_{\epsilon_3; \,  x_{6,7}} \Longleftrightarrow {{\bf R}^{5}}\vert_{\epsilon_3; \, x_{4,5}} \times \underbrace{ {\bf S}^1_t \times {{\bf S}^1_n}  \times TN_N^{R\to 0}\vert_{\epsilon_3; \, x_{6,7}}}_{\textrm{$1$ M5-branes}},
\ee
where $n=1$ or $2$ in (the $\mathbb Z_n$-twisted circle) ${\bf S}^1_n$ if we are considering the $G = SU(N)$ or $SO(N+1)$ (with $N$ even) case, and 
\be
\label{4d dual CDG}
\underbrace{{\bf R}^4\vert_{\epsilon_1, \epsilon_2}  \times {{\bf S}^1_n} \times {\bf S}^1_t}_{\textrm{$N$ M5-branes + OM5-plane}}\times {\bf R}^{5}\vert_{\epsilon_3; \,  x_{6,7}} \Longleftrightarrow {{\bf R}^{5}}\vert_{\epsilon_3; \, x_{4,5}} \times \underbrace{{\bf S}^1_t \times {{\bf S}^1_n}  \times SN_N^{R\to 0}\vert_{\epsilon_3; \, x_{6,7}}}_{\textrm{$1$ M5-branes}},
\ee
where $n=1$, $2$ or $3$ in (the $\mathbb Z_n$-twisted circle) ${\bf S}^1_n$ if we are considering the $G = SO(2N)$, $USp(2N-2)$ or $G_2$ (with $N=4$) case. Here, $\epsilon_3 = \epsilon_1 + \epsilon_2$, and  $TN_N^{R}$ and $ SN_N^{R}$ are an $N$-centered Taub-NUT and Sen's four-manifold with asymptotic radius $R$, respectively. 

According to~\cite[$\S$5.2, $\S$5.3]{4d AGT}, the Hilbert space of spacetime BPS states on the LHS of (\ref{4d dual AB}) and (\ref{4d dual CDG}) which underlie the gauge theory side of the correspondence is
\be
{\cal H}^\Omega_{\rm BPS} = \bigoplus_{m} {\cal H}^\Omega_{{\rm BPS}, m}  =  \bigoplus_{m} ~{\rm IH}^\ast_{U(1)^2 \times T} \, {\cal U}({\cal M}_{G, m}), 
\label{BPS-AGT-AB}
\ee
where $T \subset G$ is a Cartan subgroup, and ${\rm IH}^\ast_{U(1)^2 \times T} \, {\cal U}({\cal M}_{G, m})$ is the ($\mathbb Z_n$-invariant) $U(1)^2 \times T$-equivariant intersection cohomology of the Uhlenbeck compactification ${\cal U}({\cal M}_{G, m})$ of the (singular) moduli space ${\cal M}_{G, m}$ of $G$-instantons on $\mathbb R^4$ with instanton number $m$. 

On the other hand, according to~\cite[$\S$5.2, $\S$5.3]{4d AGT}, the Hilbert space of spacetime BPS states on the RHS of (\ref{4d dual AB}) and (\ref{4d dual CDG}) which underlie the 2d theory side of the correspondence is
\be
{\cal H}^{\Omega'}_{\rm BPS}  = \widehat{{\cal W}_{^L\kappa}}({^L \frak g}_{\rm aff}),
\label{AGT-AB-H=W}
\ee
where $ \widehat{{\cal W}_{^L\kappa}}({\frak g}^\vee_{\rm aff})$ is a module of an affine ${\cal W}$-algebra ${\cal W}_{^L\kappa}({^L\frak g}_{\textrm{aff}})$  associated with the Langlands dual affine Lie algebra ${^L\frak g}_{\textrm{aff}}$ of level $^L\kappa$.

From the principle that the spacetime BPS spectra of string-dual M-theory compactifications ought to be equivalent, we obtain, from (\ref{BPS-AGT-AB}) and (\ref{AGT-AB-H=W}), the relation
\be
{\bigoplus_{m} ~{\rm IH}^\ast_{U(1)^2 \times T} \, {\cal U}({\cal M}_{G, m}) = \widehat{{\cal W}_{^L\kappa}}({^L\frak g}_{\rm aff})},
 \label{AGT-duality}
\ee
where $G = A_{N-1}$, $B_{N/2}$, $D_N$, $C_{N-1}$ or $G_2$ groups, and $\frak g_{\textrm{aff}} = {\frak {su}}(N)_{\textrm{aff}}$, ${\frak {so}}(N+1)_{\textrm{aff}}$, ${\frak {so}}(2N)_{\textrm {aff}}$, ${\frak {usp}}(2N-2)_{\textrm{aff}}$ or ${{\frak g}_2}_{\, \textrm{aff}}$, accordingly.

\bigskip\noindent{\it Affine $\cal W$-algebra Identities and a Quantum Geometric Langlands Duality for the Simply-Laced $A$ and $D$ Groups} 

Let us now focus on the case where $G = SU(N)$ or $SO(2N)$, i.e. $n=1$ in (\ref{4d dual AB}) or (\ref{4d dual CDG}). In this case, according to~\cite[eqns.~(5.20) or (5.57)]{4d AGT}, the levels $^L\kappa_A$ or $^L\kappa_D$ of the respective affine $\cal W$-algebras on the RHS of (\ref{AGT-duality}) obey
\be
\label{level AD}
^L\kappa_{A,D} + {^Lh}_{A, D} = - {\epsilon_2 \over \epsilon_1},
\ee
where $^Lh_A =h (^L\frak {su}(N))$ or $^Lh_D  = h(^L\frak {so}(2N))$ are Coxeter numbers.  

Notice that the M-theory compactifications (\ref{4d dual AB}) and (\ref{4d dual CDG}) are invariant under the exchange $\epsilon_1 \leftrightarrow \epsilon_2$. Consequently, the relation (\ref{AGT-duality}) would also be invariant under the exchange $\epsilon_1 \leftrightarrow \epsilon_2$. In particular, the invariance of the RHS of (\ref{AGT-duality}) under the exchange $\epsilon_1 \leftrightarrow \epsilon_2$, (\ref{level AD}), and the identification ${^L\frak g}_{\rm aff} \cong {\frak g}_{\rm aff}$ for simply-laced ${\frak g}_{\rm aff}$, together mean that if we regard ${{\cal W}}_{\kappa}({\frak g}_{\rm aff})$ as an affine $\cal W$-algebra of level $\kappa$ labeled by the Lie algebra $\frak g$, i.e. ${{\cal W}}_{{\rm aff}, \kappa}({\frak g})$, we can write  
\be
\label{GL-AD}
\boxed{{{\cal W}}_{{\rm aff}, k}({\frak g}) = {{\cal W}}_{{\rm aff}, ^Lk}({^L\frak g}), \quad {\textrm {where}} \quad r^\vee(k + h^\vee) =  (^Lk + {^Lh}^\vee)^{-1}}
\ee
$r^\vee$ is the lacing number of $\frak g$; $h^\vee = h^\vee (\frak g)$ and $^Lh^\vee = h^\vee (^L \frak g)$ are dual Coxeter numbers, and $\frak g = {\frak {su}}(N)$ or ${\frak {so}}(2N)$.\footnote{To arrive at this statement, we made use of the fact that   $r^\vee =1$ and $h(^L\frak g) = h^\vee(^L\frak g) = h^\vee(\frak g) = h(\frak g)$ for simply-laced $\frak g$.}

In arriving at the identity (\ref{GL-AD}), we have just derived Feigin-Frenkel's result in~\cite{FF} which defines a quantum geometric Langlands duality for the $A$ and $D$ groups.

\bigskip\noindent{\it A Quantum Geometric Langlands Duality for the Nonsimply-Laced $B$ and $C$ Groups} 

Let us now set $n=2$ and $N = 2M$ in the dual compactifications in (\ref{4d dual AB}). According to~\cite[$\S$5.2]{4d AGT}, the Hilbert space of spacetime BPS states in (\ref{BPS-AGT-AB}) that appear on the LHS of (\ref{AGT-duality}), is furnished by the spectrum of a {topological} gauged sigma-model over the two-torus ${\bf T}^2_\sigma = {\bf S}^1_n \times {\bf S}^1_t$ on the LHS of (\ref{4d dual AB}). As the sigma-model is topological, this Hilbert space of spacetime BPS states would be invariant under a modular transformation of ${\bf T}^2_\sigma$ which maps its complex structure $\tau \to - 1/n \tau$.   

That said, this modular transformation will effect an $S$-duality of the relevant ${\cal N} =4$ gauge theory along the ${\bf R}^4\vert_{\epsilon_1, \epsilon_2}$ on the LHS of (\ref{4d dual AB}),\footnote{Note that we are working with the low energy regime of the M5-brane worldvolume theory in (\ref{4d dual AB}), which can then be regarded as a 6d ${\cal N} = (2,0)$ SCFT on ${\bf T}^2_\sigma  \times {\bf R}^4\vert_{\epsilon_1, \epsilon_2}$, whence we would have an ${\cal N} = 4$ gauge theory along ${\bf R}^4\vert_{\epsilon_1, \epsilon_2}$ with $S$-duality effected by the aforementioned modular transformation of ${\bf T}^2_\sigma $.} transforming the gauge group $SO(2M +1)$ into its Langlands dual $USp(2M)$. In particular, this means that the instantons on ${\bf R}^4\vert_{\epsilon_1, \epsilon_2}$ are, after the modular transformation, $USp(2M)$-instantons.  

The preceding two paragraphs, and (\ref{AGT-duality}), together mean that we have the diagram 
\tikzset{node distance=6.2cm, auto}
\begin{center}
\be
\label{dia qGL-BC}
 \begin{tikzpicture}
  \node (P) {$\bigoplus_{m} {\rm IH}^\ast_{U(1)^2 \times T} \, {\cal U}({\cal M}_{SO(2M+1), m}) $};
  \node (B) [right of=P] {$\widehat{{\cal W}_{^L\kappa_B}}({^L\frak {so}}(2M+1)_{\rm aff})$};
  \node (A) [below of=P, yshift = 4.0cm] {$\bigoplus_{m} {\rm IH}^\ast_{U(1)^2 \times T} \, {\cal U}({\cal M}_{USp(2M), m})$};
  \node (C) [below of=B, yshift = 4.0cm] {$\widehat{{\cal W}_{^L\kappa_C}}({^L\frak {usp}}(2M)_{\rm aff})$};
   \draw[transform canvas={yshift=0.0ex},->] (P) --(B) node[] {};
\draw[transform canvas={yshift=-0.0ex},->](B) -- (P) node[] {}; 
 \draw[transform canvas={xshift=0.0ex},->] (P) --(A) node[] {};
\draw[transform canvas={xshift=-0.0ex},->](A) -- (P) node[] {}; 
 \draw[transform canvas={yshift=0.0ex},->] (A) --(C) node[] {};
\draw[transform canvas={yshift=-0.0ex},->](C) -- (A) node[] {}; 
\draw[transform canvas={xshift=0.0ex},->] (B) --(C) node[] {};
\draw[transform canvas={xshift=-0.0ex},->](C) -- (B) node[] {};   
\end{tikzpicture}
\ee
 \end{center}
In particular, the right vertical relation means that 
\be
\label{WW-BC}
{{\cal W}_{^L\kappa_B}}({^L\frak {so}}(2M+1)_{\rm aff}) = {{\cal W}_{^L\kappa_C}}({^L\frak {usp}}(2M)_{\rm aff}),
\ee
where the RHS of (\ref{WW-BC}) can also be regarded as the 2d algebra of the AGT correspondence realized in (\ref{4d dual CDG}) with $N = M +1$ and $n=2$, albeit defined on a two-surface that is a modular transformation of ${\bf T}^2_\sigma $. 

What about the relationship between the levels $^L\kappa_B$ and $^L\kappa_C$ in (\ref{WW-BC})?  According to~\cite[eqns.~(5.20) and (5.57)]{4d AGT}, the levels $^L\kappa_B$ and $^L\kappa_C$ obey
\be
\label{level BC}
^L\kappa_{B,C} + {^Lh}_{B, C} = - {\epsilon_2 \over  \epsilon_1}, 
\ee
where $^Lh_B  = h({^L\frak {so}}(2M+1))$ and $^Lh_C = h({^L\frak {usp}}(2M))$ are Coxeter numbers.  Also, notice that  (\ref{4d dual CDG}), from which the RHS of (\ref{WW-BC}) is derived, is invariant under the exchange $\epsilon_1 \leftrightarrow \epsilon_2$.  Altogether, this means that we can write  
\be
\label{prelim BC}
(^L\kappa_{B} + {^Lh}_B)  =  (^L\kappa_{C} + {^Lh}_C)^{-1}.
\ee
However, (\ref{prelim BC}) cannot be exact, as we didn't take into account the fact that the algebra ${{\cal W}_{^L\kappa_C}}({^L\frak {usp}}(2M)_{\rm aff})$, associated with its RHS, is actually realized not on ${\bf T}^2_\sigma $ but a modular transformation thereof.


To ascertain the required modification of the RHS of (\ref{prelim BC}), first, note that $\tau = i R_{{\bf S}^1_t} / R_{{\bf S}^1_n}$, whence the modular transformation $\tau \to -1 /n\tau :  i R_{{\bf S}^1_t} / R_{{\bf S}^1_n} \to  i R_{{\bf S}^1_n} / nR_{{\bf S}^1_t}$ means that we have a switch $R_{{\bf S}^1_t} \rightarrow R_{{\bf S}^1_n}$ and $R_{{\bf S}^1_n} \rightarrow n R_{{\bf S}^1_t}$ in the radii of the circles  as we go down along the left vertical relation in (\ref{dia qGL-BC}). Second, since there are no \emph{a priori} restrictions on the radii, let us, for ease of illustration, set $R_{{\bf S}^1_t} = R_{{\bf S}^1_n} =1$; then, under $R_{{\bf S}^1_t} \rightarrow R_{{\bf S}^1_n}$ and $R_{{\bf S}^1_n} \rightarrow n R_{{\bf S}^1_t}$, we have $(R_{{\bf S}^1_t}, R_{{\bf S}^1_n}) = (1, 1) \to (1,n)$; in particular,  $R_{{\bf S}^1_n}$ effectively goes from $1 \to n$ as $\tau \to -1 /n\tau$. Third, recall from~\cite[Fig.~8]{4d AGT} (where $n=1$ therein in our case) that the building blocks of our theory involve a two-sphere $C = {\bf S}^2$, where $C$ can be understood as a fibration of ${\bf S}^1_n$ over an interval ${\textrm I}_t$; hence, when $R_{{\bf S}^1_n}$ goes from $1 \to n$, the scalar curvature $\cal R$ of $C$ will go from ${\cal R} \to {\cal R} / n$.\footnote{One can understand this as follows. In two dimensions, the scalar curvature is given by $S = 2 / \rho_1 \rho_2$, where $\rho_{1,2}$ are the principal radii of the surface. Hence, by scaling up one of the principal radii of $C$, i.e. $R_{{\bf S}^1_n}$, by a factor of $n$, we effectively scale down its scalar curvature $\cal R$ by a factor of $n$.}  Fourth, according to the explanations leading to~\cite[eqn.~(5.59)]{4d AGT}, the Omega-deformation-induced component $c_\Omega = f(M) \, (b  + 1/b)^2$ of the total central charge of ${{\cal W}_{^L\kappa_C}}({^L\frak {usp}}(2M)_{\rm aff})$, is defined via the expectation value of the trace of the stress tensor as $  c_\Omega \sim \langle T_{z \bar z} \rangle /{\cal R}$; this means that the change ${\cal R} \to  {\cal R} / n$ can also be interpreted as a change $c_\Omega \to  n c_\Omega$ in the \emph{original} underlying 2d theory on $C$ with scalar curvature $\cal R$. Lastly, since $c_\Omega = f(M) \, (b  + 1/b)^2$, the previous point would mean that we must rescale expressions equal to $b$ or $1/b$ by a factor of $\sqrt n$; in particular, from the original expression $   1/b = \sqrt{- (^L\kappa_{C} + {^Lh}_C)}$ (see~\cite[eqn.~5.57]{4d AGT}), it would mean that we ought to replace the RHS of (\ref{prelim BC}) with $[n(^L\kappa_{C} + {^Lh}_C)]^{-1}$. Thus, we ought to replace (\ref{prelim BC}) with
 \be
\label{prelim BC-corrected}
 (^L\kappa_{B} + {^Lh}_B)  =  n^{-1}(^L\kappa_{C} + {^Lh}_C)^{-1} ,
\ee
where $n =2$.

In other words, if we regard ${{\cal W}}_{\kappa}({\frak g}_{\rm aff})$ as an affine $\cal W$-algebra of level $\kappa$ labeled by the Lie algebra $\frak g$, i.e. ${{\cal W}}_{{\rm aff}, \kappa}({\frak g})$, from (\ref{WW-BC}) and (\ref{prelim BC-corrected}), we have
\be
\label{GL-LB}
{{ {{\cal W}_{{\rm aff}, k}}(\frak g) = {{\cal W}_{{\rm aff}, ^Lk}}( ^L{\frak g}), \quad \textrm{where} \quad r^\vee (k + h) =  (^Lk + {^Lh})^{-1}}};
\ee
$r^\vee$ is the lacing number of  $\frak g$;  $h =  h(\frak g)$ and $^Lh = h (^L\frak g)$ are Coxeter numbers; and ${\frak g} = {^L\frak {so}}(2M +1)$.\footnote{To arrive at this statement, we made use of the fact that the lacing number of  ${^L\frak {so}}(2M+1)$ is 2, and recalled that $ {^Lh}_B = h({^L\frak {so}}(2M+1)) $ and ${^Lh}_C = h(^L\frak {usp}(2M)) = h({\frak {so}}(2M+1))$.}

If we now start by setting $n=2$ and $N = M +1$ in the dual compactifications in (\ref{4d dual CDG}), and repeat the above arguments which led us to (\ref{GL-LB}), we would have
\be
\label{GL-LC}
{{ {{\cal W}_{{\rm aff}, k}}( {\frak g}) = {{\cal W}_{{\rm aff}, ^Lk}}({^L\frak g}), \quad \textrm{where} \quad  r^\vee (k + {h}) = (^Lk + {^Lh})^{-1}}};
\ee
 and ${\frak g} = {^L\frak {usp}}(2M)$.\footnote{To arrive at this statement, we made use of the fact that the lacing number of  ${^L\frak {usp}}(2M)$ is also 2, and in this case, $ {^Lh}_B = h({^L\frak {usp}}(2M))$ and ${^Lh}_C  = h(^L\frak {so}(2M+1)) = h({\frak {usp}}(2M))$.}

Clearly, (\ref{GL-LB}) and (\ref{GL-LC}) mean that  
\be
\label{GL-LBC}
{{ {{\cal W}_{{\rm aff}, k}}( {\frak g}) = {{\cal W}_{{\rm aff}, ^Lk}}({^L\frak g}), \quad \textrm{where} \quad r^\vee (k + {h}) = (^Lk + {^Lh})^{-1}}};
\ee
$r^\vee$ is the lacing number of  $\frak g$; $h = h (\frak g)$ and $^Lh = h (^L \frak g)$ are Coxeter numbers; and ${\frak g} =  {^L\frak {so}}(2M +1)$ or ${^L\frak {usp}}(2M)$. 

Notice that in arriving at (\ref{GL-LBC}), we have just derived an identity of affine $\cal W$-algebras which defines a quantum geometric Langlands duality for the $^LB$ and $^LC$ Langlands dual groups.

In order to obtain an identity for ${\frak g} =  {\frak {so}}(2M +1)$ or ${\frak {usp}}(2M)$, i.e. the Langlands dual of (\ref{GL-LBC}), one must exchange the roots and coroots of the Lie algebra underlying (\ref{GL-LBC}). In particular, this means that $h$ must be replaced by its dual $h^\vee$. In other words, from (\ref{GL-LBC}), one  also has
\be
\label{qGL-BC}
\boxed{{ {{\cal W}_{{\rm aff}, k}}( {\frak g}) = {{\cal W}_{{\rm aff}, ^Lk}}({^L\frak g}), \quad \textrm{where} \quad r^\vee (k + {h^\vee}) = (^Lk + {^Lh^\vee})^{-1}}}
\ee
$r^\vee$ is the lacing number of  $\frak g$; $h^\vee = h^\vee (\frak g)$ and $^Lh^\vee = h^\vee (^L \frak g)$ are dual Coxeter numbers; and ${\frak g} =  {\frak {so}}(2M +1)$ or ${\frak {usp}}(2M)$. 
 
In arriving at the identity (\ref{qGL-BC}), we have just derived Feigin-Frenkel's result in~\cite{FF} which defines a quantum geometric Langlands duality for the $B$ and $C$ groups.

\bigskip\noindent{\it A Quantum Geometric Langlands Duality for the Nonsimply-Laced $G_2$ Group} 

Last but not least, let us set $n=3$ and $N = 4$ in the dual compactifications in (\ref{4d dual CDG}). Then, by repeating the arguments which led us to (\ref{dia qGL-BC}) whilst noting that the Langlands dual of $G_2$ is itself, we would arrive at the diagram
\tikzset{node distance=5.5cm, auto}
\begin{center}
\be
\label{dia qGL-G2}
 \begin{tikzpicture}
  \node (P) {$\bigoplus_{m} {\rm IH}^\ast_{U(1)^2 \times T} \, {\cal U}({\cal M}_{G_2, m}) $};
  \node (B) [right of=P] {$\widehat{{\cal W}_{^L\kappa_G}}({^L\frak g}_{2 \, \rm aff})$};
  \node (A) [below of=P, yshift = 3.2cm] {$\bigoplus_{m} {\rm IH}^\ast_{U(1)^2 \times T} \, {\cal U}({\cal M}_{G_2, m})$};
  \node (C) [below of=B, yshift = 3.2cm] {$\widehat{{\cal W}_{^L\kappa'_G}}({^L\frak g}_{2 \,\rm aff})$};
   \draw[transform canvas={yshift=0.0ex},->] (P) --(B) node[] {};
\draw[transform canvas={yshift=-0.0ex},->](B) -- (P) node[] {}; 
 \draw[transform canvas={xshift=0.0ex},->] (P) --(A) node[] {};
\draw[transform canvas={xshift=-0.0ex},->](A) -- (P) node[] {}; 
 \draw[transform canvas={yshift=0.0ex},->] (A) --(C) node[] {};
\draw[transform canvas={yshift=-0.0ex},->](C) -- (A) node[] {}; 
\draw[transform canvas={xshift=0.0ex},->] (B) --(C) node[] {};
\draw[transform canvas={xshift=-0.0ex},->](C) -- (B) node[] {};   
\end{tikzpicture}
\ee
 \end{center}
In particular, the right vertical relation means that 
\be
\label{WW-G2}
{{\cal W}_{^L\kappa_G}}({^L\frak g}_{2 \, \rm aff}) = {{\cal W}_{^L\kappa_G'}}({^L\frak g}_{2 \, \rm aff}),
\ee
where the RHS of (\ref{WW-G2}) can also be regarded as the 2d algebra of the AGT correspondence realized in (\ref{4d dual CDG}) with $N = 4$ and $n=3$, albeit defined on a two-surface that is a modular transformation of ${\bf T}^2_\sigma $. 

Repeating the arguments which led us to (\ref{prelim BC-corrected}), we get
\be
\label{prelim G2-corrected}
(^L\kappa_G + {^Lh_G})  =  n^{-1} (^L\kappa'_G + {^Lh'_G})^{-1},
\ee
where $ {^Lh} = {^Lh'_G} = h(^L \frak g_2)$, and $ n =3$. 

In other words, if we regard ${{\cal W}}_{\kappa}({\frak g}_{\rm aff})$ as an affine $\cal W$-algebra of level $\kappa$ labeled by the Lie algebra $\frak g$, i.e. ${{\cal W}}_{{\rm aff}, \kappa}({\frak g})$, from (\ref{WW-G2}) and (\ref{prelim G2-corrected}), and the identification $^L\frak g_2 \cong \frak g_2$,  we can write
\be
\label{GL-LG}
{\cal W}_{{\rm aff}, k}( \frak g) = {\cal W}_{{\rm aff}, ^Lk}(^L\frak g), \quad \textrm{where} \quad r^\vee (k + h) = (^Lk + {^Lh})^{-1};
\ee
$r^\vee$ is the lacing number of  $\frak g$; $h= h(\frak g)$ and $^Lh = h(^L \frak g)$ are Coxeter numbers; and ${\frak g} = {^L\frak g}_2$.\footnote{To arrive at this statement, we made use of the fact that the lacing number of  $^L{\frak g}_2$ is 3, and the Coxeter numbers $h(^L{\frak g}_2) = h({\frak g}_2)$.} 
 
Notice that in arriving at (\ref{GL-LG}), we have just derived an identity of affine $\cal W$-algebras which defines a quantum geometric Langlands duality for the $^LG_2$ Langlands dual group.

In order to obtain an identity for ${\frak g} =  {\frak g}_2$, i.e. the Langlands dual of (\ref{GL-LG}), one must exchange the roots and coroots of the Lie algebra underlying (\ref{GL-LG}). In particular, this means that $h$ must be replaced by its dual $h^\vee$. In other words, from (\ref{GL-LG}), one also has
\be
\label{qGL-G}
\boxed{{ {{\cal W}_{{\rm aff}, k}}( {\frak g}) = {{\cal W}_{{\rm aff}, ^Lk}}(^L{\frak g}), \quad \textrm{where} \quad r^\vee (k + {h}^\vee) = (^Lk + {^Lh}^{\vee})^{-1}}}
\ee
$r^\vee$ is the lacing number of  $\frak g$; $h^\vee =  h ^\vee(\frak g)$  and $^Lh^\vee = h^\vee (^L \frak g)$ are dual Coxeter numbers of the Lie algebras; and ${\frak g} = {\frak g}_2$.

In arriving at the identity (\ref{qGL-G}), we have just derived Feigin-Frenkel's result in~\cite{FF} which defines a quantum geometric Langlands duality for the $G_2$ group. 

\bigskip\noindent{\it An M-theoretic Proof of a Corollary by Braverman-Finkelberg-Nakajima}

A corollary of~\cite[Conjecture 1.8]{NBF} states that there should be an isomorphism between ${\rm IH}^\ast_G$ and ${\rm IH}^\ast_{^LG}$ which sends $ \epsilon_2 / \epsilon_1$ to $r^\vee \epsilon_1 / \epsilon_2$, where ${\rm IH}^\ast_{\cal G} = \bigoplus_{m}~{\rm IH}^\ast_{U(1)^2 \times T} \, {\cal U}({\cal M}_{{\cal G}, m})$. 

For the $A$-$D$, $B$-$C$ and $G_2$ groups, one can see from (\ref{AGT-duality}), respectively (\ref{GL-AD}), (\ref{qGL-BC}) and (\ref{qGL-G}), and (\ref{level AD}) (bearing in mind that it also holds for all groups and that we have the physical symmetry $\epsilon_1 \leftrightarrow \epsilon_2$), that the above statement ought to be true. Thus, we have a purely physical M-theoretic proof of this corollary by Braverman-Finkelberg-Nakajima for these compact Lie groups. 

\newsubsection{An M-Theoretic Realization of $q$-deformed Affine $\cal W$-algebras and a Quantum $q$-Geometric Langlands Duality}

Let us now analyze, within our M-theoretic framework, the properties of $q$-deformed affine $\cal W$-algebras associated with the 5d AGT correspondence, and derive various identities conjectured by mathematicians Frenkel-Reshetikhin in~\cite{FR} which define a quantum $q$-geometric Langlands duality and its variants. For brevity, we shall only consider the case of the simply-laced $A$ and $D$ groups. The case of the nonsimply-laced $B$, $C$ and $G_2$ groups is somewhat more intricate, and we shall leave it for future discussion. 

\bigskip\noindent{\it A $q$-deformed Affine $\cal W$-algebra and a Quantum $q$-Geometric Langlands Duality for the $A$ Groups}

Notice that (\ref{Z5d-pure-SU(N)-G}), (\ref{he-pure-SU(N)}), (\ref{variables-pure-SU(N)}) and (\ref{GSU(N)}) mean that the $q$-deformation in the module $\widehat {{\cal W}}^q({\frak{su}(N})_{\textrm{aff},k})$ of (\ref{dia 1}) actually depends on the two parameter $s$ and $r$ given in (\ref{variables-pure-SU(N)}). As such, one can regard its underlying $q$-deformed algebra $ {{\cal W}}^q({\frak{su}(N})_{\textrm{aff},k})$ as a two-parameter generalization of an affine $\cal W$-algebra of level $k$ associated with the Lie algebra $\frak {su}(N)$, i.e. ${{\cal W}}^{r,s}_{{\rm aff}, k}({\frak {su}(N)})$.

 Note that reversing the overall orientation of the M-theory compactification (\ref{4d dual AB}) sends $\epsilon_{1,2} \to - \epsilon_{1,2}$ but leaves the physics invariant; indeed, the central charge and level of the affine $\cal W$-algebra underlying ${{\cal W}}^{r,s}_{{\rm aff}, k}({\frak {su}(N)})$, given by~\cite[eqn. (5.19) and (5.20)]{4d AGT}, \emph{are} invariant under $\epsilon_{1,2} \to - \epsilon_{1,2}$. Also, recall that the M-theory compactification (\ref{4d dual AB}) is invariant under the exchange $\epsilon_1 \leftrightarrow \epsilon_2$. In sum, this means that the change $(\epsilon_1, \epsilon_2) \to (-\epsilon_2, -\epsilon_1)$ is a symmetry of our physical setup. 

Now, let $p = r/s$. From (\ref{variables-pure-SU(N)}), it would mean that $p = e^{-i \beta (\epsilon_1 + \epsilon_2)}$. Then, according to the previous paragraph, the change $p \to p^{-1}$, i.e. $r/s \to s/r$, is also a symmetry of our physical setup. In other words, our formulas should be invariant under the exchange $r \leftrightarrow s$; in particular, we ought to have
\be
\label{qq-GL-A-prelim}
{{\cal W}}^{r,s}_{{\rm aff}, k}(\frak {su}(N)) = {{\cal W}}^{s,r}_{{\rm aff}, k'}(\frak {su}(N)),
\ee
and because $(k + h^\vee) = - \epsilon_2 / \epsilon_1$ (see (\ref{level AD})), 
 \be
 \label{qq-GL-A-level}
 (k + h^\vee) = (k' + {h}^\vee)^{-1}.
 \ee

As mentioned above, the $q$-deformation actually depends on the two parameters $s$ and $r$ given in (\ref{variables-pure-SU(N)}). But how so? To answer this, first note that the definition of $q$-deformation is that it is undone as $q \to 1$. Second, notice that in our context, even if $\epsilon_1$ or $\epsilon_2$ vanishes, $q$-deformation is not undone unless we go from 5d to 4d, i.e. as $\beta \to 0$. Altogether, this means from (\ref{variables-pure-SU(N)}) that $r$ actually corresponds to the deformation parameter $q$, whence $q$ would depend on $s$ through $q = ps$. Hence, if we relabel $s$ as $t$, from (\ref{qq-GL-A-prelim}) and (\ref{qq-GL-A-level}), we have
\be
\label{qq-GL-A}
\boxed{{\cal W}^{q,t}_{{\rm aff}, k}({\frak g}) = {{\cal W}}^{t,q}_{{\rm aff}, ^Lk}({^L\frak g}), \quad {\rm where} \quad  r^\vee(k + h^\vee) = (^Lk + {^Lh}^\vee)^{-1}}
\ee
$r^\vee$ is the lacing number of $\frak g$; $h^\vee = h^\vee (\frak g)$ and $^Lh^\vee = h^\vee (^L \frak g)$ are dual Coxeter numbers; and $\frak g = \frak {su}(N)$.\footnote{To arrive at this statement, we have made use of the fact that for simply-laced $\frak g$ and ${\frak g}_{\rm aff}$, we have ${\frak g}_{\rm aff} \cong {^L\frak g}_{\rm aff}$, $r^\vee =1$ and $h^\vee(^L\frak g) = h^\vee(\frak g)$.}   

In arriving at the identity (\ref{qq-GL-A}), we have just derived Frenkel-Reshetikhin's result in~\cite[$\S$4.1]{FR}, which defines a quantum $q$-geometric Langlands duality for the $A$ groups! 

\bigskip\noindent{\it Reduction to 4d and a Quantum Geometric Langlands Duality for the $A$ Groups}

Now, let $q \to 1$ with $t = q^\alpha$, i.e. let $t \to 1$ with $\alpha \neq 0$, where $\alpha = \sqrt{\epsilon_1 \epsilon_2} \,   (\epsilon_1 + \epsilon_2 + \sqrt{\epsilon_1 \epsilon_2})^{-1}$. In other words, with $\epsilon_{1,2} \neq 0$, let $\beta \to 0$ (c.f.~(\ref{variables-pure-SU(N)})), whence we simply have a reduction from the 5d to 4d case. Then, (\ref{qq-GL-A}) would become
\be
\label{q-GL-A}
\boxed{{\cal W}^{1,1}_{{\rm aff}, k}({\frak g}) = {{\cal W}}^{1,1}_{{\rm aff}, ^Lk}({^L\frak g}), \quad {\rm where} \quad  r^\vee(k + h^\vee) = (^Lk + {^Lh}^\vee)^{-1}}
\ee
and
\be
\label{W11}
\boxed{{\cal W}^{1,1}_{{\rm aff}, k}({\frak g}) = {\cal W}_{{\rm aff}, k}({\frak g})}
\ee
is the affine $\cal W$-algebra of (\ref{GL-AD}).

Note that the identification (\ref{W11}) coincides with~\cite[Theorem 2]{FR},  while (\ref{q-GL-A})--(\ref{W11}) define a quantum geometric Langlands duality for the $A$ groups, as expected.

\bigskip\noindent{\it A Classical $q$-Geometric Langlands Duality for the $A$ Groups}

Next, let $t \to 1$ with $q$ fixed, where again, $t = q^\alpha$, i.e. let $\alpha \to 0$. In other words, consider the limit $\epsilon_2 \to 0$ whence $k \to - h^\vee$ and $q \to e^{-i \beta \epsilon_1}$. Then, from (\ref{qq-GL-A}), we would have 
\be
\label{cq-GL-A}
{{\cal W}^{q,1}_{{\rm aff}, -h^\vee}({\frak g}) = {{\cal W}}^{1,q}_{{\rm aff}, \infty}({^L\frak g})}.
\ee

The fact that ${\cal W}^{1,1}_{{\rm aff}, -h^\vee}({\frak g}) = {\cal W}_{{\rm aff}, -h^\vee}({\frak g})$ can be identified~\cite{FF} with the center $Z(U(\hat{\frak g})_{\rm crit})$ of the completed enveloping algebra  $U(\hat{\frak g})_{\rm crit}$ of $\hat {\frak g} = {\frak g}_{\mathbb C \,  {\rm aff}}$ at critical level, and that going from 4d to 5d will effect a $q$-deformation of the underlying algebra,\footnote{According to diagrams (\ref{dia 1}) and (\ref{dia 2}), this is true as long as $\epsilon_1 + \epsilon_2 \neq 0$, which is indeed the case here since $\epsilon_2 \to 0$ but $\epsilon_1 \neq 0$.} means that we have the diagram 
 \tikzset{node distance=4.0cm, auto}
 \begin{center}
 \be
   \label{dia cq-GL-A-1}
 \begin{tikzpicture}
  \node (P) {${\cal W}^{q,1}_{{\rm aff}, -h^\vee}({\frak g}) $};
  \node (B) [right of=P] {${\cal W}^{1,1}_{{\rm aff}, -h^\vee}({\frak g}) $};
  \node (A) [below of=P, yshift = 1.6cm] {$Z(U_q(\hat{\frak g})_{\rm crit})$};
  \node (C) [below of=B, yshift = 1.6cm] {$Z(U(\hat{\frak g})_{\rm crit})$};
   \draw[transform canvas={yshift=0.5ex},->] (P) --(B) node[midway] {\footnotesize{$\beta \rightarrow 0$}};
\draw[transform canvas={yshift=-0.5ex},->](B) -- (P) node[midway] {\footnotesize{$\beta \nrightarrow 0$}}; 
 \draw[transform canvas={xshift=0.0ex},->] (P) --(A) node[midway] {};
\draw[transform canvas={xshift=-0.0ex},->](A) -- (P) node[midway] {}; 
 \draw[transform canvas={yshift=0.5ex},->] (A) --(C) node[midway] {\footnotesize{$\beta \rightarrow 0$}};
\draw[transform canvas={yshift=-0.5ex},->](C) -- (A) node[midway] {\footnotesize{$\beta \nrightarrow 0$}}; 
\draw[transform canvas={xshift=0.0ex},->] (B) --(C) node[midway] {};
\draw[transform canvas={xshift=-0.0ex},->](C) -- (B) node[midway] {};  
 \end{tikzpicture}
\ee
 \end{center}
where the quantum affine algebra $U_q(\hat{\frak g})_{\rm crit}$ is a $q$-deformed version of $U(\hat{\frak g})_{\rm crit}$. 

Notably, the left vertical relation means that we have the identification 
\be
\label{cq-GL-A-1}
\boxed{{\cal W}^{q,1}_{{\rm aff}, -h^\vee}({\frak g})  = Z(U_q(\hat{\frak g})_{\rm crit})}
\ee
Furthermore, in the limit $t \to 1$ with $q$ fixed, the algebra underlying the LHS of (\ref{cq-GL-A-1}) becomes commutative (c.f.~(\ref{he-pure-SU(N)})). In turn, this means that the LHS of (\ref{cq-GL-A-1}) also becomes commutative whence it is a Poisson algebra. 

Thus, in arriving at (\ref{cq-GL-A-1}) and the statement that followed, we have just derived~\cite[Conjecture 2]{FR}!

The fact that ${\cal W}^{1,1}_{{\rm aff}, \infty}({^L\frak g}) = {\cal W}_{{\rm aff}, \infty}({^L\frak g})$ can be identified~\cite{FF} with the classical $\cal W$-algebra $\cal W_{\rm cl}(^L\frak g)$ obtained by a Drinfeld-Sokolov reduction of the dual space to $(^L \frak g)_{\rm aff}$, and that going from 4d to 5d simply effects a $q$-deformation of the underlying algebra, means that we have the diagram 
 \tikzset{node distance=4.0cm, auto}
 \begin{center}
 \be
   \label{dia cq-GL-A-2}
 \begin{tikzpicture}
  \node (P) {${{\cal W}}^{1,q}_{{\rm aff}, \infty}({^L\frak g})$};
  \node (B) [right of=P] {${{\cal W}}^{1,1}_{{\rm aff}, \infty}({^L\frak g})$};
  \node (A) [below of=P, yshift = 1.6cm] {${\cal W}^q_{\rm cl}(^L\frak g)$};
  \node (C) [below of=B, yshift = 1.6cm] {$\cal W_{\rm cl}(^L\frak g)$};
   \draw[transform canvas={yshift=0.5ex},->] (P) --(B) node[midway] {\footnotesize{$\beta \rightarrow 0$}};
\draw[transform canvas={yshift=-0.5ex},->](B) -- (P) node[midway] {\footnotesize{$\beta \nrightarrow 0$}}; 
 \draw[transform canvas={xshift=0.0ex},->] (P) --(A) node[midway] {};
\draw[transform canvas={xshift=-0.0ex},->](A) -- (P) node[midway] {}; 
 \draw[transform canvas={yshift=0.5ex},->] (A) --(C) node[midway] {\footnotesize{$\beta \rightarrow 0$}};
\draw[transform canvas={yshift=-0.5ex},->](C) -- (A) node[midway] {\footnotesize{$\beta \nrightarrow 0$}}; 
\draw[transform canvas={xshift=0.0ex},->] (B) --(C) node[midway] {};
\draw[transform canvas={xshift=-0.0ex},->](C) -- (B) node[midway] {};  
 \end{tikzpicture}
\ee
 \end{center}
where ${\cal W}^q_{\rm cl}(^L\frak g)$ is a $q$-deformed version of ${\cal W}_{\rm cl}(^L\frak g)$. 

Notably, the left vertical relation means that we have the identification 
\be
\label{cq-GL-A-2}
\boxed{{{\cal W}}^{1,q}_{{\rm aff}, \infty}({^L\frak g})  = {\cal W}^q_{\rm cl}(^L\frak g)}
\ee
With regard to the algebra ${{\cal W}}^{q',t'}_{{\rm aff}, k'}(\frak g')$, this identification means that in the limit $q' \to 1$ with $t'$ fixed, it becomes ${\cal W}^{p'}_{\rm cl}(\frak g')$, where $p' = t'$. 

Thus, in arriving at (\ref{cq-GL-A-2}) and the statement that followed, we have just derived~\cite[Conjecture 3]{FR}!

Last but not least, from (\ref{cq-GL-A}), (\ref{cq-GL-A-1}) and (\ref{cq-GL-A-2}), we have
\be
\label{cq-GL-A-final}
\boxed{Z(U_q(\hat{\frak g})_{\rm crit}) = {\cal W}^q_{\rm cl}(^L\frak g)}
\ee
where $\frak g = \frak {su}(N)$. 

In arriving at the identification (\ref{cq-GL-A-final}) of classical algebras, we have just derived Frenkel-Reshetikhin's result in~\cite[$\S$4.1]{FR}, which defines a classical $q$-geometric Langlands duality for the $A$ groups!

\bigskip\noindent{\it A Quantum $q$-Geometric Langlands Duality and its Variants for the $D$ Groups}

What about the $D$ groups? Do the above identities of $q$-deformed affine $\cal W$-algebras also apply to them? The answer is `yes', as we shall now explain.  

Firstly, from diagram (\ref{dia 6}), it would mean that  ${\cal W}^{q}(\frak {so}(2N)_{{\rm aff}, k'})$ ought to be associated with ${\bf U}_q({\bf L}{\frak {so}}(2)_{{\rm aff}, 1})$, an $\frak {so}(2)_{{\rm aff}, 1}$-version of ${\bf U}_q({\bf L}{\frak {gl}}(1)_{{\rm aff}, 1})$, the Ding-Iohara algebra at level 1 defined by (\ref{he-pure-SU(N)}). Secondly, from diagram (\ref{dia 5}), we see that when $\epsilon_1 + \epsilon_2 = 0$, the aforementioned algebra reduces to a $\beta$-dependent loop algebra ${\bf L}{\frak {so}}(2)_{{\rm aff}, 1}$; in the $\beta \to 0$ limit whence it reduces to $\frak {so}(2)_{{\rm aff}, 1}$, it is also a Heisenberg algebra~\cite[eqn.~(15.195)]{Yellow CFT}, just like ${\frak {gl}}(1)_{{\rm aff}, 1}$. Thus, by reversing the arguments in the second point, and by noting from the explanations below (\ref{dia 2}) that when $\epsilon_1 + \epsilon_2 \neq 0$, one must, in our context, replace the loop algebra ${\bf L}\frak H$ of the Heisenberg algebra $\frak H$ with its quantum version $U_q ({\bf L}\frak H)$, we find that at the purely algebraic level, we can also regard  ${\bf U}_q({\bf L}{\frak {so}}(2)_{{\rm aff}, 1})$ as being defined by  (\ref{he-pure-SU(N)}). In turn, the first point means that ${\cal W}^{q}(\frak {so}(2N)_{{\rm aff}, k'})$ ought to be associated with (\ref{he-pure-SU(N)}), too; in particular, we can also write  ${\cal W}^{q}(\frak {so}(2N)_{{\rm aff}, k'}) =  {\cal W}^{r,s}_{{\rm aff}, k'}(\frak {so}(2N))$, and apply the expressions in (\ref{variables-pure-SU(N)}). Then, by repeating verbatim the earlier arguments for the case of $\frak g = \frak {su}(N)$, we find that we would get exactly the same identities for the case of $\frak g = \frak {so}(2N)$. 

In short, our physical arguments show that the identities (\ref{qq-GL-A}), (\ref{q-GL-A})--(\ref{W11}), (\ref{cq-GL-A-1}), (\ref{cq-GL-A-2}) and (\ref{cq-GL-A-final}), also hold for $\frak g = \frak {so}(2N)$. In arriving at this conclusion, we have just derived Frenkel-Reshetikhin's results in~\cite[$\S$4.1]{FR}, which define a quantum $q$-geometric Langlands duality and its variants for the $D$ groups!

\bigskip\noindent{\it About the case of the Nonsimply-Laced $B$, $C$ and $G_2$ Groups}

Regarding the derivation of the identities for the nonsimply-laced $B$, $C$ and $G_2$ groups, just like in the 4d case of the previous subsection, one would need to appeal to an $S$-duality which involves a modular transformation of an underlying two-torus ${\bf T}^2_\sigma = {\bf S}^1_n \times {\bf S}^1_t$ which effects the swop ${\bf S}^1_n \leftrightarrow {\bf S}^1_t$. However, in the 5d case, the $\mathbb Z_n$-twisted circle ${\bf S}^1_n$ is a \emph{preferred}  circle in the sense that the relevant states are projected onto it (as explained in~\cite[$\S$3]{5d AGT}). Consequently, it would mean that it cannot be true that (\ref{qq-GL-A}) (and therefore its variants) would hold for nonsimply-laced $\frak g$ (since $n \neq 1$). This conclusion is consistent with Frenkel-Reshetikhin's results in~\cite[$\S$4.1]{FR}.

\newsubsection{An M-Theoretic Realization of Elliptic Affine $\cal W$-algebras and a Quantum $q,v$-Geometric Langlands Duality}

Let us now analyze, within our M-theoretic framework, the properties of elliptic affine $\cal W$-algebras associated with the 6d AGT correspondence, and derive various novel identities which define a quantum $q, v$-geometric Langlands duality and its variants. For brevity, we shall only consider the case of the simply-laced $A$ and $D$ groups. The case of the nonsimply-laced $B$, $C$ and $G_2$ groups is somewhat more intricate, and we shall leave it for future discussion. 

\bigskip\noindent{\it An Elliptic Affine $\cal W$-algebra and a Quantum $q, v$-Geometric Langlands Duality for the $A$ Groups}

Consider the elliptic affine $\cal W$-algebra ${{\cal W}}^{q, v}({\frak{su}(N})_{\textrm{aff},k})$ that appears in the 6d AGT correspondence for $A_{N-1}$ groups  with $N_f = 2N$ fundamental matter in~\cite[$\S$5.1]{5d AGT}. It is characterized by an elliptic Ding-Iohara algebra~\cite{Saito} defined by the following commutation relations:
\be
{[{\tilde a}_m, {\tilde a}_n] = m (1 - v^{|m|}) {1- { s}^{|m|} \over 1- { r}^{|m|}} \delta_{m+n, 0}},
\label{he-6d-a}
\ee
and
\be
{[{\tilde b}_m, {\tilde b}_n] = {m (1 - v^{|m|}) \over (s r^{-1} v)^{|m|}} \, {1- { s}^{|m|} \over 1- { r}^{|m|}} \delta_{m+n, 0}},
\label{he-6d-b-final}
\ee
where ${[{\tilde a}_m, {\tilde b}_n] = 0}$, and
\be
\label{variables-U(1)-6d-2}
{s = e^{- i \beta \sqrt{\epsilon_1 \epsilon_2}}, \quad r = e^{- i \beta ( \epsilon_1 + \epsilon_2 + \sqrt{\epsilon_1 \epsilon_2})}, \quad v = e^{- {1 \over R_6}}}. 
\ee
From these relations, it is clear that we can regard the elliptic algebra $ {{\cal W}}^{q, v}({\frak{su}(N})_{\textrm{aff},k})$ as a three-parameter generalization of an affine $\cal W$-algebra of level $k$ associated with the Lie algebra $\frak {su}(N)$, i.e. ${{\cal W}}^{r,s,v}_{{\rm aff}, k}({\frak {su}(N)})$.

Notice that the variable $v$ is independent of $\epsilon_{1,2}$. Also, in the 6d $\to$ 5d limit effected by $R_6 \to 0$, i.e. $v \to 0$, the above relations reduce to those that underlie our discussion in the previous subsection, reflecting the fact that the elliptic algebra ${{\cal W}}^{r,s,v}_{{\rm aff}, k}({\frak {su}(N)})$ ought to reduce to the $q$-deformed algebra ${{\cal W}}^{r,s}_{{\rm aff}, k}({\frak {su}(N)})$, whereby the $q$-deformation, characterized by the first term  on the RHS of (\ref{he-6d-a}) (involving $s$ and $r$ but not $v$), is therefore independent of the value of $R_6$. As such, by repeating verbatim the arguments which led us to (\ref{qq-GL-A}), we get   
\be
\label{qqv-GL-A}
\boxed{{\cal W}^{q,t,v}_{{\rm aff}, k}({\frak g}) = {{\cal W}}^{t,q,v}_{{\rm aff}, ^Lk}({^L\frak g}), \quad {\rm where} \quad  r^\vee(k + h^\vee) = (^Lk + {^Lh}^\vee)^{-1}}
\ee
 $r^\vee$ is the lacing number of $\frak g$; $h^\vee = h^\vee (\frak g)$ and $^Lh^\vee = h^\vee (^L \frak g)$ are dual Coxeter numbers; and $\frak g = \frak {su}(N)$.

Notice that in arriving at (\ref{qqv-GL-A}), we have just derived a novel identity which defines a quantum $q, v$-geometric Langlands duality for the $A$ groups! 

\bigskip\noindent{\it A Classical $q,v$-Geometric Langlands Duality for the $A$ Groups}

Consider the limit $\epsilon_2 \to 0$ whence $k \to - h^\vee$, $t \to 1$ and $q \to e^{-i \beta \epsilon_1}$.\footnote{Here, we recall that $(k+ h^\vee) = - \epsilon_2/ \epsilon_1$, and $(q,t) = (r,s)$. } Then, from (\ref{qqv-GL-A}), we would have 
\be
\label{cqv-GL-A}
{{\cal W}^{q,1,v}_{{\rm aff}, -h^\vee}({\frak g}) = {{\cal W}}^{1,q,v}_{{\rm aff}, \infty}({^L\frak g})}.
\ee

The fact that ${\cal W}^{q,1, 0}_{{\rm aff}, -h^\vee}({\frak g}) = {\cal W}^{q,1}_{{\rm aff}, -h^\vee}({\frak g})$ can be identified with the center $Z(U_q(\hat{\frak g})_{\rm crit})$ of the quantum affine algebra  $U_q(\hat{\frak g})_{\rm crit}$  (see (\ref{cq-GL-A-1})), and that going from 5d to 6d simply effects an elliptic-deformation of the underlying algebra, means that we have the diagram 
 \tikzset{node distance=4.0cm, auto}
 \begin{center}
 \be
   \label{dia cqv-GL-A-1}
 \begin{tikzpicture}
  \node (P) {${\cal W}^{q,1,v}_{{\rm aff}, -h^\vee}({\frak g}) $};
  \node (B) [right of=P] {${\cal W}^{q,1,0}_{{\rm aff}, -h^\vee}({\frak g}) $};
  \node (A) [below of=P, yshift = 1.6cm] {$Z(U_{q,v}(\hat{\frak g})_{\rm crit})$};
  \node (C) [below of=B, yshift = 1.6cm] {$Z(U_q(\hat{\frak g})_{\rm crit})$};
   \draw[transform canvas={yshift=0.5ex},->] (P) --(B) node[midway] {\footnotesize{$R_6 \rightarrow 0$}};
\draw[transform canvas={yshift=-0.5ex},->](B) -- (P) node[midway] {\footnotesize{$R_6 \nrightarrow 0$}}; 
 \draw[transform canvas={xshift=0.0ex},->] (P) --(A) node[midway] {};
\draw[transform canvas={xshift=-0.0ex},->](A) -- (P) node[midway] {}; 
 \draw[transform canvas={yshift=0.5ex},->] (A) --(C) node[midway] {\footnotesize{$R_6 \rightarrow 0$}};
\draw[transform canvas={yshift=-0.5ex},->](C) -- (A) node[midway] {\footnotesize{$R_6 \nrightarrow 0$}}; 
\draw[transform canvas={xshift=0.0ex},->] (B) --(C) node[midway] {};
\draw[transform canvas={xshift=-0.0ex},->](C) -- (B) node[midway] {};  
 \end{tikzpicture}
\ee
 \end{center}
where the elliptic affine algebra $U_{q,v}(\hat{\frak g})_{\rm crit}$ is an elliptic-deformed version of $U_q(\hat{\frak g})_{\rm crit}$. 

Notably, the left vertical relation means that we have the identification 
\be
\label{cqv-GL-A-1}
{{\cal W}^{q,1,v}_{{\rm aff}, -h^\vee}({\frak g})  = Z(U_{q,v}(\hat{\frak g})_{\rm crit})}.
\ee
Furthermore, in this limit of $t \to 1$ with $q \neq 1$, the algebra underlying the LHS of (\ref{cqv-GL-A-1}) becomes commutative (c.f.~(\ref{he-6d-a})--(\ref{he-6d-b-final}), where $(r, s) = (q,t)$). In turn, this means that the LHS of (\ref{cqv-GL-A-1}) is also a classical algebra.

The fact that ${\cal W}^{1,q, 0}_{{\rm aff}, \infty}({^L\frak g}) = {\cal W}^{1,q}_{{\rm aff}, \infty}({^L\frak g})$ can be identified with the $q$-deformed classical algebra ${\cal W}^q_{\rm cl}(^L\frak g)$ (see (\ref{cq-GL-A-2})), and that going from 5d to 6d simply effects an elliptic-deformation of the underlying algebra, means that we have the diagram 
 \tikzset{node distance=4.0cm, auto}
 \begin{center}
 \be
   \label{dia cqv-GL-A-2}
 \begin{tikzpicture}
  \node (P) {${{\cal W}}^{1,q, v}_{{\rm aff}, \infty}({^L\frak g})$};
  \node (B) [right of=P] {${{\cal W}}^{1,q,0}_{{\rm aff}, \infty}({^L\frak g})$};
  \node (A) [below of=P, yshift = 1.6cm] {${\cal W}^{q,v}_{\rm cl}(^L\frak g)$};
  \node (C) [below of=B, yshift = 1.6cm] {${\cal W}^q_{\rm cl}(^L\frak g)$};
   \draw[transform canvas={yshift=0.5ex},->] (P) --(B) node[midway] {\footnotesize{$R_6 \rightarrow 0$}};
\draw[transform canvas={yshift=-0.5ex},->](B) -- (P) node[midway] {\footnotesize{$R_6 \nrightarrow 0$}}; 
 \draw[transform canvas={xshift=0.0ex},->] (P) --(A) node[midway] {};
\draw[transform canvas={xshift=-0.0ex},->](A) -- (P) node[midway] {}; 
 \draw[transform canvas={yshift=0.5ex},->] (A) --(C) node[midway] {\footnotesize{$R_6 \rightarrow 0$}};
\draw[transform canvas={yshift=-0.5ex},->](C) -- (A) node[midway] {\footnotesize{$R_6 \nrightarrow 0$}}; 
\draw[transform canvas={xshift=0.0ex},->] (B) --(C) node[midway] {};
\draw[transform canvas={xshift=-0.0ex},->](C) -- (B) node[midway] {};  
 \end{tikzpicture}
\ee
 \end{center}
where ${\cal W}^{q,v}_{\rm cl}(^L\frak g)$ is an elliptic-deformed version of  ${\cal W}^q_{\rm cl}(^L\frak g)$. 

Notably, the left vertical relation means that we have the identification 
\be
\label{cqv-GL-A-2}
{{{\cal W}}^{1,q, v}_{{\rm aff}, \infty}({^L\frak g})  = {\cal W}^{q,v}_{\rm cl}(^L\frak g)}.
\ee

Finally, from (\ref{cqv-GL-A}), (\ref{cqv-GL-A-1}) and (\ref{cqv-GL-A-2}), we have
\be
\label{cqv-GL-A-final}
\boxed{Z(U_{q,v}(\hat{\frak g})_{\rm crit}) = {\cal W}^{q,v}_{\rm cl}(^L\frak g)}
\ee
where $\frak g = \frak {su}(N)$. 

In arriving at the classical algebra relation (\ref{cqv-GL-A-final}), we have just derived an identity which defines a classical $q,v$-geometric Langlands duality for the $A$ groups! 

\bigskip\noindent{\it A Quantum $q,v$-Geometric Langlands Duality and its Variants for the $D$ Groups}

What about the $D$ groups? Do the above identities of elliptic affine $\cal W$-algebras also apply to them? The answer is `yes', as we shall now explain.  

Firstly, note that according to $\S$3.3\ref{s3.3}, ${{\cal W}}^{q,v}({\frak {su}(N)_{{\rm aff}, k}})$ ought to be associated with the elliptically-deformed quantum toroidal algebra ${\bf U}_{q,v}({\bf L}{\frak {gl}}(1)_{{\rm aff}, 1})$, whence ${{\cal W}}^{q,0}({\frak {su}(N)_{{\rm aff}, k}})$ ought to be associated with the quantum toroidal algebra ${\bf U}_{q,0}({\bf L}{\frak {gl}}(1)_{{\rm aff}, 1})$ (as noted in $\S$2.1\ref{s2.1}); likewise, ${{\cal W}}^{q,0}({\frak {so}(2N)_{{\rm aff}, k}})$  ought to be associated with ${\bf U}_{q,0}({\bf L}{\frak {so}}(2)_{{\rm aff}, 1})$ (as noted in $\S$2.2\ref{s2.2}). Secondly, as explained in the previous subsection, at the purely algebraic level, one can also regard ${\bf U}_{q,0}({\bf L}{\frak {so}}(2)_{{\rm aff}, 1})$ as being defined by  (\ref{he-6d-a})--(\ref{variables-U(1)-6d-2}) but with $v= 0$, just like ${\bf U}_{q,0}({\bf L}{\frak {gl}}(1)_{{\rm aff}, 1})$; this means that just like ${{\cal W}}^{q,v}({\frak {su}(N)_{{\rm aff}, k}})$, one can express ${\cal W}^{q,v}(\frak {so}(2N)_{{\rm aff}, k})$ in terms of the three parameters $r$, $s$ and $v$, too, i.e. ${\cal W}^{q,v}(\frak {so}(2N)_{{\rm aff}, k}) = {\cal W}^{r,s, v}_{{\rm aff}, k}(\frak {so}(2N))$, and apply the expressions in (\ref{variables-U(1)-6d-2}). Then, by repeating verbatim the earlier arguments for the case of $\frak g = \frak {su}(N)$, we find that we would get exactly the same identities for the case of $\frak g = \frak {so}(2N)$. 

In short, our physical arguments show that the quantum and classical relations (\ref{qqv-GL-A}) and (\ref{cqv-GL-A-final}), respectively, also hold for $\frak g = \frak {so}(2N)$. In arriving at this conclusion, we have just derived identities which define a quantum $q, v$-geometric Langlands duality and its classical variant for the $D$ groups!

\bigskip\noindent{\it About the case of the Nonsimply-Laced $B$, $C$ and $G_2$ Groups}

Regarding the derivation of the identities for the nonsimply-laced $B$, $C$ and $G_2$ groups, since the $q$-deformed affine $\cal W$-algebras are just $v = 0$ versions of the elliptic affine $\cal W$-algebras, and since the former does not even obey these aforementioned identities in the nonsimply-laced case, it will mean that it cannot be true that (\ref{qqv-GL-A}) (and therefore its variants) would hold for nonsimply-laced $\frak g$.

\newsection{Supersymmetric Gauge Theory, $\cal W$-algebras and a Quantum Geometric Langlands Correspondence}
\label{s5}

\newsubsection{A Quantum Geometric Langlands Correspondence as an $S$-duality and a Quantum $\cal W$-algebra Duality}

We shall now demonstrate, via our M-theoretic framework, how a quantum geometric Langlands correspondence can be understood simultaneously as a 4d gauge-theoretic $S$-duality and a 2d conformal field-theoretic quantum $\cal W$-algebra duality. In doing so, we would be able to affirm the conjecture in~\cite[$\S$8.6]{Frenkel} that quantum $\cal W$-algebra duality ought to play a prominent role in defining a Fourier-Mukai transform.

\bigskip\noindent{\it A GL-twisted ${\cal N} = 4$ Gauge Theory}

Consider the low energy worldvolume theory of the five-branes/planes on the LHS of (\ref{4d dual AB}) and (\ref{4d dual CDG}), i.e. 
\be
\label{LHS-4d dual AB}
\underbrace{{\bf R}^4\vert_{\epsilon_1, \epsilon_2}  \times {{\bf S}^1_t} \times {\bf S}^1_n}_{\textrm{$N$ M5-branes}},
\ee
where we set $n=1$ or $2$ if we want a $G = SU(N)$ or $SO(N+1)$ (with $N$ even) gauge theory along ${\bf R}^4\vert_{\epsilon_1, \epsilon_2} \times  {\bf S}^1_t$, and
\be
\label{LHS-4d dual CDG}
\underbrace{{\bf R}^4\vert_{\epsilon_1, \epsilon_2}  \times {{\bf S}^1_t} \times {\bf S}^1_n}_{\textrm{$N$ M5-branes + OM5-plane}},
\ee
where we set $n=1$, $2$ or $3$ if we want a $G = SO(2N)$, $USp(2N-2)$ or $G_2$ (with $N=4$) gauge theory along ${\bf R}^4\vert_{\epsilon_1, \epsilon_2} \times  {\bf S}^1_t$.

As noted in $\S$3.1\ref{s3.1}, the theory is topological along the directions spanned by ${\bf R}^4\vert_{\epsilon_1, \epsilon_2}$. This means that the precise metric on each of the ${\bf R}^2\vert_{\epsilon_i}$ planes in ${\bf R}^4\vert_{\epsilon_1, \epsilon_2} = {\bf R}^2\vert_{\epsilon_1} \times {\bf R}^2\vert_{\epsilon_2}$, is not essential in our forthcoming analysis. In particular, one can place on each of the ${\bf R}^2\vert_{\epsilon_i}$ planes,  a ``cigar-like'' metric 
\be
ds^2 = dr^2 + f(r) d\theta, \quad 0 \leq r < \infty, \quad 0 \leq \theta \leq 2 \pi, 
\ee
where $f(r) \sim r^2$ for $r \to 0$; $f(r) \to \rho^2$ for $r \to \infty$; $f(r) = \rho$ for sufficiently large $r$, say $r \geq r_0$; and $\rho$ is the asymptotic radius of the circle ${{\bf S}}^1$ parameterized by $\theta$. Then, the rotation of the plane associated with a nonzero $\epsilon_i$ would therefore correspond to a rotation of the  circle ${ {\bf S}}^1$ of the cigar. Let us denote this rotated circle as  ${ {\bf S}}^1_{\epsilon_i}$. Moreover, since the rotation of the plane confines the physical excitations close to the origin, i.e. close to the tip of the cigar, we can conveniently consider the truncated cigar with length $r \leq R$, where $R >> \rho, r_0$~\cite{NW}. 

So, if we denote the planes ${\bf R}^2\vert_{\epsilon_1}$ and  ${\bf R}^2\vert_{\epsilon_2}$ endowed with the above ``cigar-like'' metrics as $D_{R, \epsilon_1}$ and $D_{R, \epsilon_2}$,  we can also express (\ref{LHS-4d dual AB}) and (\ref{LHS-4d dual CDG}) as 
\be
\label{D-AB}
\underbrace{D_{R, \epsilon_1} \times D_{R, \epsilon_2} \times {{\Sigma}^{1,0}_{t, n}}}_{\textrm{$N$ M5-branes}},
\ee
and
\be
\label{D-CDG}
\underbrace{D_{R, \epsilon_1} \times D_{R, \epsilon_2} \times {{\Sigma}^{1,0}_{t, n}}}_{\textrm{$N$ M5-branes + OM5-plane}},
\ee
where ${{\Sigma}^{1,0}_{t, n}} =  {{\bf S}^1_t} \times  {\bf S}^1_n$ is a Riemann surface of genus one with zero punctures. 

Now, notice that we can regard $D_{R, \epsilon_1} \times D_{R, \epsilon_2}$ as a nontrivial ${ {\bf S}}^1_{\epsilon_1} \times { {\bf S}}^1_{\epsilon_2}$ fibration of ${\bf I}_1 \times {\bf I}_2$, where ${\bf I}_i$ is an interval associated with the $i^{\rm th}$ plane. However, macroscopically at low energies whence the curvature of the cigar tips is not observable, $D_{R, \epsilon_1} \times D_{R, \epsilon_2}$ is effectively a \emph{trivial}  ${ {\bf S}}^1_{\epsilon_1} \times { {\bf S}}^1_{\epsilon_2}$ fibration of ${\bf I}_1 \times {\bf I}_2$. Therefore, with regard to the minimal energy limit of the M5-brane worldvolume theory relevant to us, we can simply take (\ref{D-AB}) and (\ref{D-CDG}) to be
\be
\label{SxI-AB}
\underbrace{ {\bf T}^2_{\epsilon_1, \epsilon_2}\times {\bf I}_1 \times   {\bf I}_2 \times {{\Sigma}^{1,0}_{t, n}}}_{\textrm{$N$ M5-branes}},
\ee
and
\be
\label{SxI-CDG}
\underbrace{{\bf T}^2_{\epsilon_1, \epsilon_2} \times {\bf I}_1  \times {\bf I}_2 \times {{\Sigma}^{1,0}_{t, n}}}_{\textrm{$N$ M5-branes + OM5-plane}},
\ee 
where the torus $ {\bf T}^2_{\epsilon_1, \epsilon_2} = {\bf S}^1_{\epsilon_1} \times  {\bf S}^1_{\epsilon_2}$. 

The rotation of the circles in $ {\bf T}^2_{\epsilon_1, \epsilon_2}$ does not break any supersymmetries in the above flat worldvolumes. Hence,  from the perspective of a worldvolume compactification on $ {\bf T}^2_{\epsilon_1, \epsilon_2}$, we have an ${\cal N} = 4$ theory along  ${\bf I}_1 \times   {\bf I}_2 \times {{\Sigma}^{1,0}_{t, n}}$ with gauge group $G$.\footnote{As mentioned earlier, the compactification of the worldvolume on ${\bf S}^1_n$ will result in a 5d Yang-Mills theory with gauge group $G$. According to~\cite{MD}, this 5d Yang-Mills theory will capture all the degrees of freedom of the 6d ${\cal N} =(2,0)$ from which it descended from, at least where the BPS spectrum is concerned. Together, this means that for all our purposes, we can regard the worldvolume theory to be a  6d ${\cal N} =(2,0)$ theory of type $G$, whence our claim follows.} Its complexified gauge coupling is given by $\tau = i R_2 / R_1$, the complex modulus of $ {\bf T}^2_{\epsilon_1, \epsilon_2}$, where $R_i$ the radius of ${\bf S}^1_{\epsilon_i}$. 

Notice that the low energy worldvolume theories in (\ref{SxI-AB}) and (\ref{SxI-CDG}) are just a specialization to $(g, p) =  (1,0)$ of 6d ${\cal N} = (2,0)$ compactifications on  ${{\Sigma}^{g, p}_{t, n}}$ which result in 4d ${\cal N} =2$ theories of class $S$~\cite{Gaiotto} along ${\bf T}^2_{\epsilon_1, \epsilon_2} \times {\bf I}_1  \times {\bf I}_2$, where ${{\Sigma}^{g, p}_{t, n}}$  is a genus $g \geq 0$ Riemann surface with $p \geq 0$ punctures. According to~\cite{XN-regular} and~\cite[$\S$3]{NW}, based on the fact that the integrable Hitchin systems associated with theories of class $S$ ought to be defined over ${{\Sigma}^{g, p}_{t, n}}$, the ${\cal N} = 4$ theory along $ {\bf I}_1  \times {\bf I}_2 \times {{\Sigma}^{g,p}_{t, n}}$ must be topologically-twisted in the GL sense of~\cite{KW}. 

In other words, from the perspective of a compactification on $ {\bf T}^2_{\epsilon_1, \epsilon_2}$, we can effectively regard our low energy worldvolume theory in (\ref{SxI-AB}) or (\ref{SxI-CDG}) as a GL-twisted ${\cal N} = 4$ theory on $ {\bf I}_1  \times {\bf I}_2 \times {{\Sigma}^{1,0}_{t, n}}$ with gauge group $G$. This twisted theory also enjoys an $S$-duality symmetry that maps $G$ (with lacing number $r^\vee$) to its Langlands dual $^LG$ via  $\tau \to -1 / r^\vee \tau$, i.e. a modular transformation of $ {\bf T}^2_{\epsilon_1, \epsilon_2}$. The minimal energy (maximal) BPS states of the worldvolume theory -- these underlie the spectrum of spacetime BPS states on the LHS of (\ref{4d dual AB}) or (\ref{4d dual CDG}) that ought to be invariant under string dualities -- would then be furnished by the Hilbert space ${\cal H}^\sigma_{{\bf I}_1 \times {\bf I}_2}(X^{\Sigma_1}_G)_{{\cal B}}$ of a topological sigma-model on ${\bf I}_1 \times {\bf I}_2$ with boundary ${\cal B}$ and target $X^{\Sigma_1}_G$ determined by $G$ and $\Sigma_1 = {{\Sigma}^{1,0}_{t, n}}$. 

\bigskip\noindent{\it $S$-Duality as a Quantum $\cal W$-algebra Duality for genus $g =1$}

On the other hand, from the \emph{equivalent} perspective of a compactification on $ {{\Sigma}^{1,0}_{t, n}}$, we can regard the low energy worldvolume theory in (\ref{SxI-AB}) or (\ref{SxI-CDG}) as a massless ${\cal N} = 2^\ast$ theory  on ${\bf R}^4\vert_{\epsilon_1, \epsilon_2}$ with gauge group $G$. Therefore, according to our explanations leading up to (\ref{AGT-duality}), the minimal energy (maximal) BPS states of the worldvolume theory ought to also be given  by the module $\widehat{{\cal W}_{{\rm aff}, ^Lk}}(^L\frak g)$,\footnote{Here, ${\cal W}_{{\rm aff}, ^Lk}(^L\frak g) = {\cal W}_{^L k}(^L{\frak g}_{\rm aff}) = {\cal W}_{^L k}(^L[({\frak g})_{\rm aff}]) =  {^L{\cal W}}_{k}(({\frak g})_{\rm aff}) = {^L{\cal W}_{{\rm aff}, k}}(\frak g)$, where ${^L{\cal W}_{{\rm aff}, k}}(\frak g)$ can be interpreted as the ``Langlands dual'' of ${\cal W}_{{\rm aff}, k}(\frak g)$.  \label{LW(g)}} defined on $ \Sigma_1 = {{\Sigma}^{1,0}_{t, n}}$. In short, we have
\be
\label{H=W, g=1}
{\cal H}^\sigma_{{\bf I}_1 \times {\bf I}_2}(X^{\Sigma_1}_G)_{{\cal B}} =  {\widehat{{\cal W}_{{\rm aff}, ^Lk}}(^L\frak g)}_{\Sigma_1},
\ee
where $^L\frak g$ is the Lie algebra of $^LG$, and the subscript ``$\Sigma_1$'' just indicates that the module is defined on it.\footnote{The equivalence of the compactifications on $ {\bf T}^2_{\epsilon_1, \epsilon_2}$ and  $ {{\Sigma}^{1,0}_{t, n}}$ of the low energy worldvolume theories  in (\ref{SxI-AB})--(\ref{SxI-CDG}) which underlies (\ref{H=W, g=1}), has also been fruitfully exploited in~\cite[$\S$7.4]{4d AGT} to prove a conjecture by Alday-Tachikawa in ref.~[19] of \emph{loc.~cit.}} 

The level on the RHS of (\ref{H=W, g=1}) can (according to (\ref{level AD}) and (\ref{level BC}), which also holds for the $G_2$ case) be expressed as
\be
\label{level LG}
^Lk + {^Lh} = - {\epsilon_2 \over \epsilon_1},  
\ee
where $^Lh = h(^L \frak g)$ is the Coxeter number.

Let us now effect an $S$-duality transformation of the ${\cal N} = 4$ theory via $\tau \to -1 / r^\vee \tau$, i.e. a modular transformation of $ {\bf T}^2_{\epsilon_1, \epsilon_2}$ (which, as we will see below, amounts to a swop $\epsilon_1 \leftrightarrow \epsilon_2$ that is a symmetry of the worldvolume theory). The equivalence of the compactifications on $ {\bf T}^2_{\epsilon_1, \epsilon_2}$ and  $ {{\Sigma}^{1,0}_{t, n}}$ of the low energy worldvolume theories would then mean that this step ought to be accompanied by a modular transformation of $ {{\Sigma}^{1,0}_{t, n}}$ (so that the $S$-dual gauge symmetry can be consistently defined  over ${\bf I}_1 \times {\bf I}_2$, where this transformation is also a symmetry of the worldvolume theory). Thus, in place of (\ref{SxI-AB}) and (\ref{SxI-CDG}), we have
\be
\label{SxI-AB-S dual}
\underbrace{ {\tilde {\bf T}}^2_{\epsilon_1, \epsilon_2}\times {\bf I}_1 \times   {\bf I}_2 \times {{\tilde \Sigma}^{1,0}_{t, n}}}_{\textrm{$N$ M5-branes}},
\ee
and
\be
\label{SxI-CDG-S dual}
\underbrace{{\tilde {\bf T}}^2_{\epsilon_1, \epsilon_2} \times {\bf I}_1  \times {\bf I}_2 \times {{\tilde \Sigma}^{1,0}_{t, n}}}_{\textrm{$N$ M5-branes + OM5-plane}},
\ee 
where ${\tilde {\bf T}}^2_{\epsilon_1, \epsilon_2}$ and ${{\tilde \Sigma}^{1,0}_{t, n}}$ are the modular transformations of the original tori.   

From the perspective of a worldvolume compactification on ${\tilde {\bf T}}^2_{\epsilon_1, \epsilon_2}$, the minimal energy (maximal) BPS states would be furnished by ${\cal H}^{^L\sigma}_{{\bf I}_1 \times {\bf I}_2}(X^{{\Sigma}_1}_{^LG})_{^L{\cal B}}$,\footnote{The topological sigma-model which underlies these states is defined by equations over ${\tilde \Sigma}^{1,0}_{t, n}$ that depend on $^LG$ and its genus only. As such, we have, for later convenience, replaced ${\tilde \Sigma}^{1,0}_{t, n}$ with ${\Sigma}_1$ in the expression for $\cal H$.} the Hilbert space of an $S$-dual topological sigma-model on ${\bf I}_1 \times {\bf I}_2$ with boundary ${^L{\cal B}}$ and target $X^{\Sigma_1}_{^LG}$. From the perspective of a worldvolume compactification on ${{\tilde \Sigma}^{1,0}_{t, n}}$,  the minimal energy (maximal) BPS states would be furnished by ${\widehat{{\cal W}_{{\rm aff}, \tilde k}}(\frak g)}_{{\tilde \Sigma}_1}$, where $ {\tilde \Sigma}_1 = {\tilde \Sigma}^{1,0}_{t, n}$. The equivalence of these perspectives would then mean that

\be
\label{H=W, g=1, S-dual}
{\cal H}^{^L\sigma}_{{\bf I}_1 \times {\bf I}_2}(X^{{\Sigma}_1}_{^LG})_{^L{\cal B}} =  {\widehat{{\cal W}_{{\rm aff}, \tilde k}}(\frak g)}_{{\tilde \Sigma}_1}.
\ee

Notice that the transformation $\tau \to - 1 / r^\vee \tau$ effected by the replacements $R_2 \to R_1$ and $R_1 \to r^\vee R_2$, can be regarded as an exchange ${\bf S}^1_{\epsilon_1} \leftrightarrow  {\bf S}^1_{\epsilon_2}$ of the circles (up to some scaling of one of the radii), i.e. where the rotations of the planes are concerned, we have a swop $\epsilon_1 \leftrightarrow \epsilon_2$. In other words, on the RHS of (\ref{H=W, g=1, S-dual}), we have 
\be
\label{tilde k}
\tilde k + h = - {\epsilon_1 \over \epsilon_2},
\ee 
where $h = h(\frak g)$ is the Coxeter number. 

The RHS of (\ref{H=W, g=1, S-dual}) is defined in terms of ${\tilde \Sigma}_1$, which is a modular transformation of ${\Sigma}_1$. In order to define it in  terms of ${\Sigma}_1$ so as to be consistent with the LHS, recall from our explanations below (\ref{level BC}) that we can express ${\widehat{{\cal W}_{{\rm aff}, \tilde k}}(\frak g)}_{{\tilde \Sigma}_1}$ as ${\widehat{{\cal W}_{{\rm aff}, k}}(\frak g)}_{{ \Sigma}_1}$,\footnote{Here, ${\cal W}_{{\rm aff}, k}(\frak g) = {\cal W}_{^L(^Lk)}(^L[({^L\frak g})_{\rm aff}]) =  {^L{\cal W}}_{^Lk}((^L{\frak g})_{\rm aff}) = {^L{\cal W}}_{{\rm aff}, ^Lk}(^L{\frak g})$, where ${^L{\cal W}}_{{\rm aff}, ^Lk}(^L{\frak g})$ can be interpreted as the ``{Langlands dual}'' of ${{\cal W}}_{{\rm aff}, ^Lk}(^L{\frak g})$. \label{LW(Lg)}} whence we can also rewrite (\ref{H=W, g=1, S-dual}) as    
\be
\label{H=W, g=1, S-dual n-corrected}
{\cal H}^{^L\sigma}_{{\bf I}_1 \times {\bf I}_2}(X^{{\Sigma}_1}_{^LG})_{^L{\cal B}} =  {\widehat{{\cal W}_{{\rm aff},  k}}(\frak g)}_{{ \Sigma}_1},
\ee
where
\be
\label{level G n-corrected}
r^\vee(k + h) = - {\epsilon_1 \over \epsilon_2}.
\ee 

Altogether from our discussions which led us to (\ref{H=W, g=1}), (\ref{level LG}), (\ref{H=W, g=1, S-dual n-corrected}) and (\ref{level G n-corrected}), and footnotes~\ref{LW(g)} and~\ref{LW(Lg)}, we have the following diagram
\tikzset{node distance=7.0cm, auto}
\begin{center}
\be
\label{dia S=W}
 \begin{tikzpicture}
  \node (P) {${\cal H}^\sigma_{{\bf I}_1 \times {\bf I}_2}(X^{\Sigma_1}_G)_{{\cal B}}$};
  \node (B) [right of=P] {$  {\widehat{^L{\cal W}_{{\rm aff}, \kappa}}(\frak g)}_{\Sigma_1}$};
  \node (A) [below of=P, yshift = 3.2cm] {${\cal H}^{^L\sigma}_{{\bf I}_1 \times {\bf I}_2}(X^{{\Sigma}_1}_{^LG})_{^L{\cal B}}$};
  \node (C) [below of=B, yshift = 3.2cm] {$\widehat{{^L{\cal W}}_{{\rm aff}, ^L\kappa}}(^L{\frak g})_{{ \Sigma}_1}$};
   \draw[transform canvas={yshift=0.0ex},->] (P) --(B) node[midway] {{\scriptsize String Duality}};
\draw[transform canvas={yshift=-0.0ex},->](B) -- (P) node[midway] {}; 
 \draw[transform canvas={xshift=0.0ex},->] (P) --(A) node[midway] {{\scriptsize $\tau \to -{1 \over r^\vee \tau}$}};
\draw[transform canvas={xshift=-0.0ex},->](A) -- (P) node[midway] {{\scriptsize $S$-duality}}; 
 \draw[transform canvas={yshift=0.0ex},->] (A) --(C) node[midway] {{\scriptsize String Duality}};
\draw[transform canvas={yshift=-0.0ex},->](C) -- (A) node[] {}; 
\draw[transform canvas={xshift=0.0ex},->] (B) --(C) node[midway] {{\scriptsize $\cal W$-duality}};
\draw[transform canvas={xshift=-0.0ex},->](C) -- (B) node[midway] {{\scriptsize $r^\vee (\kappa + {h})  = (^L\kappa + {^Lh})^{-1}$}};   
\end{tikzpicture}
\ee
 \end{center}
where ${{^L{\cal W}}_{{\rm aff}, \kappa}}({\frak g})$ is the ``Langlands dual'' of ${\cal W}_{{\rm aff}, \kappa}(\frak g)$, an affine $\cal W$-algebra of level $\kappa$ labeled by the Lie algebra $\frak g$, and $\kappa + h = - \epsilon_2 / \epsilon_1$. 

From the diagram, it is clear that there is, in our context, a one-to-one correspondence between an $S$-duality of the GL-twisted ${\cal N} = 4$ theory and a $\cal W$-duality that is in fact a quantum $\cal W$-algebra duality of 2d CFT! This is an important fact that we will exploit to elucidate the connection between the gauge-theoretic realization and algebraic CFT formulation of the quantum geometric Langlands correspondence for $G$.  To do that however, we will first need to show that the above diagram also holds for $\Sigma_g = {{\Sigma}^{g,0}_{t, n}}$, where $g \geq 1$. 

\begin{figure}
\centering
\includegraphics[width=1.1\textwidth]{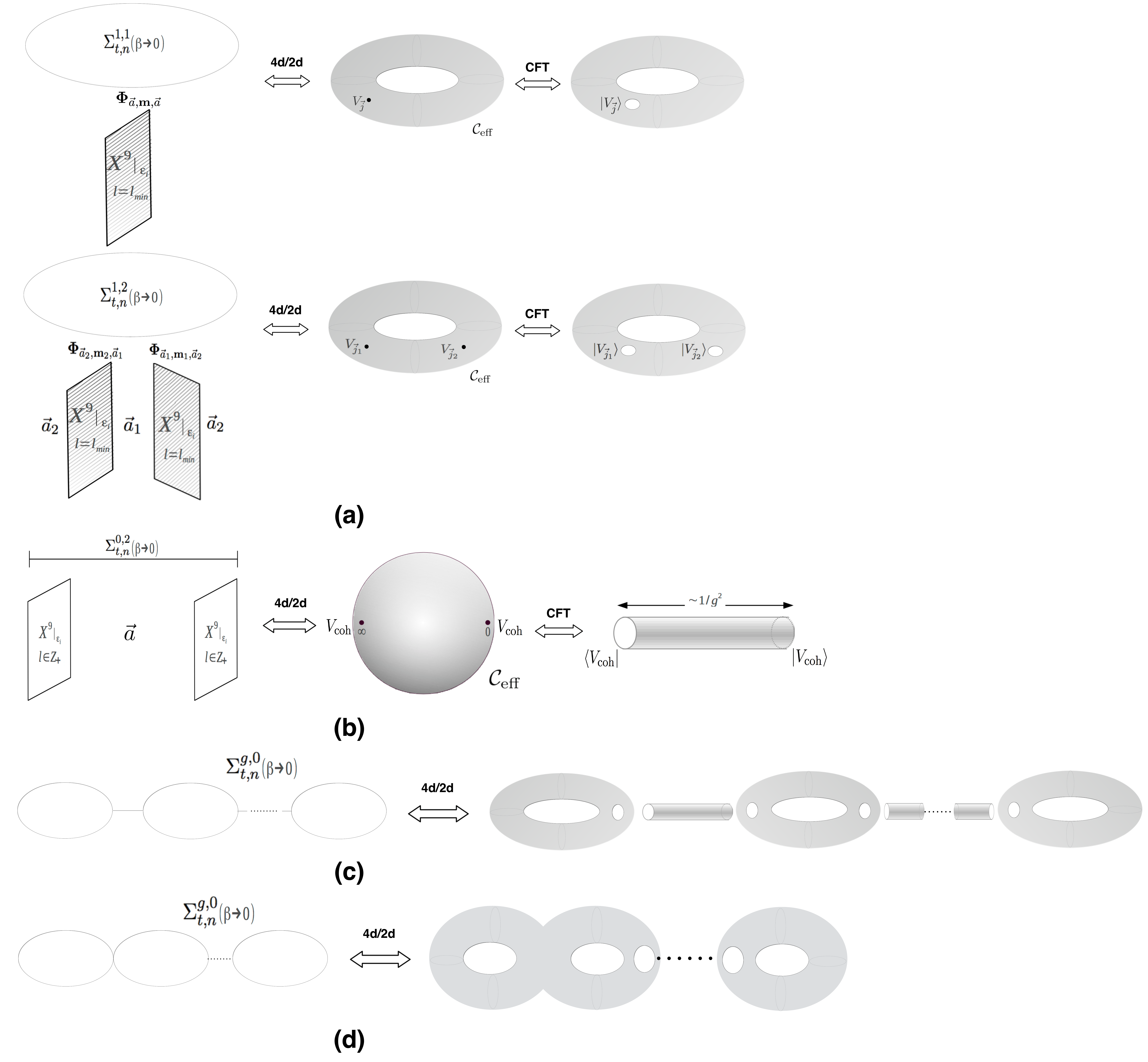}
\newpage\caption{Building blocks relevant to our discussion on the connection between the gauge-theoretic realization and algebraic CFT formulation of the geometric Langlands correspondence. (a). Dual M-theory compactifications which realize the AGT correspondence for ${\cal N} = 2^\ast$ $SU(N)$ theory and ${\cal N}= 2$ necklace quiver theory with two $SU(N)$ gauge groups, where $\beta$ denotes the size of $\Sigma$.  (b). Dual M-theory compactifications which realize the AGT correspondence for ${\cal N}= 2$ pure $SU(N)$ theory. (c).  Gluing together $g$ copies of the (a)-compactifications using the (b)-compactifications. (d).  Finally, a single compactification on $\Sigma^{g,0}_{t,n}$ without M9-planes on the 4d side, corresponding to a genus $g$ surface with no punctures on the 2d side.}
\label{fig 1}
\end{figure}

\bigskip\noindent{\it $S$-Duality as a Quantum $\cal W$-algebra Duality for genus $g \geq 1$}


To this end, consider the 4d AGT correspondence for a necklace $SU(N)$ quiver theory with one and two gauge groups; according to~\cite[Fig.~9]{4d AGT}, the correspondence can, in our framework, be understood as Fig.~\ref{fig 1}(a), where $n=1$. Consider also the 4d AGT correspondence for a pure $SU(N)$ theory; according to~\cite[Fig.~3]{4d AGT}, the correspondence can, in our framework, be understood as Fig.~\ref{fig 1}(b), where $n=1$.   

In the same way that we can construct in~\cite[$\S$6.3]{4d AGT}, a generalization from pure to linear/necklace quiver theories of the 4d AGT correspondence via~\cite[Figs.~4--6 and 8--9]{4d AGT}, we can also construct a generalization from  $\Sigma_1$ to  $\Sigma_g$ of the above correspondences in diagram (\ref{dia S=W}). Essentially, we would like to construct a  generalization of  (\ref{D-AB}) and (\ref{D-CDG}) in which ${{\Sigma}^{1,0}_{t, n}}$ is replaced by ${{\Sigma}^{g,0}_{t, n}}$. 

To do so, firstly, consider the massless limit of the 4d theories in Fig.~\ref{fig 1}(a). This corresponds to setting $\vec m_{s-1} = 0$ in~\cite[eqn.~(6.98)]{4d AGT} (where $n=1$ and $2$ therein as we are considering only one and two gauge groups). In turn, this means from~\cite[eqn.~(6.102)]{4d AGT} that the vertex operators ${V}_{\vec j}$ and ${V}_{\vec j_{1,2}}$ in Fig.~\ref{fig 1}(a) have conformal dimension zero and are thus trivial. The vertex operators $V_{\rm coh}$ in Fig.~\ref{fig 1}(b) are also trivial, as it is a coalescence of the aforementioned vertex operators. Consequently, the corresponding M9-planes on the left of Fig.~\ref{fig 1}(a)--(b) must also be trivial. Secondly, since all the M9-planes and vertex operators are now \emph{similar} (and trivial), it would mean that we can glue along the M9-planes and holes, $g$ copies of the (a)-compactifications using the (b)-compactifications. This is shown in Fig.~\ref{fig 1}(c). Finally, as the correspondence  in Fig.~\ref{fig 1}(b) is conformally invariant along $\Sigma^{0,2}_{t, n}$ and  ${\cal C}_{\rm eff}$, it would mean that we can scale the lines and cylinders in Fig.~\ref{fig 1}(c) down to zero length, whence we have a single compactification on $\Sigma^{g,0}_{t,n}$ without M9-planes on the gauge theory side, corresponding to a genus $g$ surface with no punctures on the CFT side, as shown in Fig.~\ref{fig 1}(d). The worldvolume theory of the M5-branes in this single compactification is a generalization of  (\ref{D-AB}) in which ${{\Sigma}^{1,0}_{t, 1}}$ is replaced by ${{\Sigma}^{g,0}_{t, 1}}$. Since the construction leading up to this final correspondence is actually independent of $n$, we actually have a generalization of  (\ref{D-AB}) in which ${{\Sigma}^{1,0}_{t, n}}$ is replaced by ${{\Sigma}^{g,0}_{t, n}}$ for $n = 1$ or 2.  

One could also add an OM5-plane to the stack of $N$ M5-branes, and (\ref{4d dual CDG}) would mean that we would have the same configurations as those in Fig.~\ref{fig 1} after we add the appropriate M9-planes. Hence, by repeating the construction, we would arrive at a generalization of  (\ref{D-CDG}) in which ${{\Sigma}^{1,0}_{t, n}}$ is replaced by ${{\Sigma}^{g,0}_{t, n}}$ for $n =1, 2$ or 3. 

As such, in place of (\ref{SxI-AB}) and (\ref{SxI-CDG}), we now have  
\be
\label{SxI-AB-all g}
\underbrace{ {\bf T}^2_{\epsilon_1, \epsilon_2}\times {\bf I}_1 \times   {\bf I}_2 \times {{\Sigma}^{g,0}_{t, n}}}_{\textrm{$N$ M5-branes}},
\ee
for $g \geq 1$ and $n =1$ or 2, and
\be
\label{SxI-CDG-all g}
\underbrace{{\bf T}^2_{\epsilon_1, \epsilon_2} \times {\bf I}_1  \times {\bf I}_2 \times {{\Sigma}^{g,0}_{t, n}}}_{\textrm{$N$ M5-branes + OM5-plane}},
\ee 
for $g \geq 1$ and $n =1, 2$ or 3. 

In turn, in place of (\ref{H=W, g=1})--(\ref{level LG}), we now have
\be
\label{H=W, g}
{\cal H}^\sigma_{{\bf I}_1 \times {\bf I}_2}(X^{\Sigma_g}_G)_{{\cal B}} =  {\widehat{{\cal W}_{{\rm aff}, ^Lk}}(^L\frak g)}_{\Sigma_g}, \qquad ^Lk + {^Lh} = - {\epsilon_2 \over \epsilon_1}.
\ee

Notice from the LHS of Fig.~\ref{fig 1}(d) that the single compactification on $\Sigma^{g,0}_{t,n}$ without M9-planes, can be regarded as a composition of $g$ compactifications  on $\Sigma^{1,0}_{t,n}$ without M9-planes. If we were to effect an $S$-duality transformation which brought us from (\ref{SxI-AB})/(\ref{SxI-CDG}) to (\ref{SxI-AB-S dual})/(\ref{SxI-CDG-S dual}) on each of the $g$ compactifications, we would get, in place of (\ref{SxI-AB-S dual}) and (\ref{SxI-CDG-S dual}), 
\be
\label{SxI-AB-S dual- all g}
\underbrace{ {\tilde {\bf T}}^2_{\epsilon_1, \epsilon_2}\times {\bf I}_1 \times   {\bf I}_2 \times {{\tilde \Sigma}^{g,0}_{t, n}}}_{\textrm{$N$ M5-branes}},
\ee
for $g \geq 1$ and $n =1$ or 2, and
\be
\label{SxI-CDG-S dual-all g}
\underbrace{{\tilde {\bf T}}^2_{\epsilon_1, \epsilon_2} \times {\bf I}_1  \times {\bf I}_2 \times {{\tilde \Sigma}^{g,0}_{t, n}}}_{\textrm{$N$ M5-branes + OM5-plane}},
\ee 
for $g \geq 1$ and $n =1, 2$ or 3. Here, ${{\tilde \Sigma}^{g,0}_{t, n}}$ is a genus $g$ surface with no punctures that is a composition of $g$ copies of ${{\tilde \Sigma}^{1,0}_{t, n}}$. 

In turn, in place of (\ref{H=W, g=1, S-dual n-corrected})--(\ref{level G n-corrected}), we now have
\be
\label{H=W, g, S-dual n-corrected}
{\cal H}^{^L\sigma}_{{\bf I}_1 \times {\bf I}_2}(X^{\Sigma_g}_{^LG})_{^L{\cal B}} =  {\widehat{{\cal W}_{{\rm aff}, k}}(\frak g)}_{\Sigma_g}, \qquad r^\vee(k + h) = - {\epsilon_1 \over \epsilon_2}.
\ee


 Altogether from our discussions which led us to (\ref{H=W, g}) and (\ref{H=W, g, S-dual n-corrected}), and footnotes~\ref{LW(g)} and~\ref{LW(Lg)}, we have the following diagram
\tikzset{node distance=7.0cm, auto}
\begin{center}
\be
\label{dia S=W for g}
 \begin{tikzpicture}
  \node (P) {${\cal H}^\sigma_{{\bf I}_1 \times {\bf I}_2}(X^{\Sigma_g}_G)_{{\cal B}}$};
  \node (B) [right of=P] {$  {\widehat{^L{\cal W}_{{\rm aff}, \kappa}}(\frak g)}_{\Sigma_g}$};
  \node (A) [below of=P, yshift = 3.2cm] {${\cal H}^{^L\sigma}_{{\bf I}_1 \times {\bf I}_2}(X^{\Sigma_g}_{^LG})_{^L{\cal B}}$};
  \node (C) [below of=B, yshift = 3.2cm] {$\widehat{{^L{\cal W}}_{{\rm aff}, ^L\kappa}}(^L{\frak g})_{{ \Sigma}_g}$};
   \draw[transform canvas={yshift=0.0ex},->] (P) --(B) node[midway] {{\scriptsize String Duality}};
\draw[transform canvas={yshift=-0.0ex},->](B) -- (P) node[midway] {}; 
 \draw[transform canvas={xshift=0.0ex},->] (P) --(A) node[midway] {{\scriptsize $\tau \to -{1 \over r^\vee \tau}$}};
\draw[transform canvas={xshift=-0.0ex},->](A) -- (P) node[midway] {{\scriptsize $S$-duality}}; 
 \draw[transform canvas={yshift=0.0ex},->] (A) --(C) node[midway] {{\scriptsize String Duality}};
\draw[transform canvas={yshift=-0.0ex},->](C) -- (A) node[] {}; 
\draw[transform canvas={xshift=0.0ex},->] (B) --(C) node[midway] {{\scriptsize $\cal W$-duality}};
\draw[transform canvas={xshift=-0.0ex},->](C) -- (B) node[midway] {{\scriptsize $r^\vee (\kappa + {h}) = (^L\kappa + {^Lh})^{-1}$}};   
\end{tikzpicture}
\ee
 \end{center}
Thus, we have a generalization of diagram (\ref{dia S=W}) to $\Sigma_g$, where $g \geq 1$, as desired.

\bigskip\noindent{\it  A Quantum Geometric Langlands Correspondence as an $S$-Duality and a Quantum $\cal W$-algebra Duality}

In the context of a  quantum geometric Langlands correspondence as realized by an $S$-duality in a GL-twisted ${\cal N} = 4$ theory in~\cite{KW, K}, a twisted D-module on the moduli space ${\rm Bun}_G(\Sigma_g)$ of principal $G_{\mathbb C}$-bundles over $\Sigma_g$ (where $g > 1$), is a module over the sheaf of (holomorphic) differential operators on ${\cal L}^{\Psi - h^\vee}$, where $\cal L$ is a certain holomorphic line bundle over ${\rm Bun}_G(\Sigma_g)$ determined by the property that its first Chern class generates the second cohomology of ${\rm Bun}_G(\Sigma_g)$; $h^\vee = h^\vee (\frak g)$ is the dual Coxeter number; and the parameter $\Psi \in \mathbb C$ transforms under $S$-duality as $\Psi \to - 1 / r^\vee \Psi$. The way that $\Psi$ and $\tau$ transform under $S$-duality is similar; this is a consequence of the fact that they are related via $t^2 = - (\Psi - \bar\tau) / (\Psi - \tau)$, where the parameter $t \in \mathbb C$ transforms under $S$-duality as $t \to \pm t |\tau| / \tau$. The classical limit of the correspondence is defined at $\Psi =0$, which for purely imaginary $\tau$ (as in our case) that is nonvanishing, is at $t =\pm 1$. In this limit, $S$-duality maps $t = \pm 1 \to t = \pm i$. 

Let ${\rm D}^{\rm mod}_{{\cal L}^{\gamma}}(X)$ be a derived category of twisted $\cal D$-modules on a complex manifold $X$. Then, a quantum geometric Langlands correspondence is the following equivalence~\cite{Frenkel, KW, K} 
\be
{\rm D}^{\rm mod}_{{\cal L}^{\Psi - h^\vee}}({\rm Bun}_G(\Sigma_g)) = {\rm D}^{\rm mod}_{{\cal L}^{{^L\Psi} - {^Lh}^\vee}}({\rm Bun}_{^LG}(\Sigma_g)), \qquad {\rm where} \qquad {^L \Psi} =  - 1 / r^\vee \Psi;
\ee
 $\Psi \neq 0$ and so ${^L\Psi} \neq \infty$; and $t \neq \pm 1$ whenever $\tau \neq 0$. 

Let us, for simplicity, consider the case without gauge theory line operators, i.e. $t = 0$ whence $\Psi = \bar \tau$~\cite{K} . Then, the above equivalence of derived categories of  twisted $\cal D$-modules, can, according to~\cite[$\S$2-$\S$3]{K}, be interpreted as an equivalence
\be
\label{H = H QGL}
{\cal H}^{\rm A}_{{\bf I} \times {\bf R}}({\cal M}_H (G, \Sigma_g))_{{\cal B}_{d.c.}, \, {\cal B}_\alpha } = {\cal H}^{\rm A}_{{\bf I} \times {\bf R}}({\cal M}_H ({^LG}, \Sigma_g))_{{\cal B}_{d.c.}, \, {\cal B}_\alpha}
\ee  
 of Hilbert spaces of string states -- specifically, those of an open topological A-model and its identical $S$-dual on ${\bf I} \times \bf R$ in complex structure $I$  with targets ${\cal M}_H (G, \Sigma_g)$ and ${\cal M}_H ({^LG}, \Sigma_g)$, respectively, that end on a distinguished coisotropic brane ${\cal B}_{d.c.}$ and a Lagrangian brane ${\cal B}_{\alpha}$ -- which furnish the aforementioned minimal energy worldvolume (maximal) BPS states.\footnote{To arrive at (\ref{H = H QGL}) and the following statement, we have (i) exploited the fact that the sigma-models underlying (\ref{H=W, g}) and (\ref{H=W, g, S-dual n-corrected}) are topological, whence we are free to stretch their worldsheets in one direction; (ii) noted that according to~\cite{K}, strings of type $({\cal B}_{d.c.}, {\cal B}_{d.c.})$ correspond not to states but a noncommutative algebra of quantized holomorphic functions; (iii) noted that as pointed out in~\cite{NW}, the brane at the end of $\bf I$ which corresponds to the tip of the cigar arises not from an actual boundary condition but from geometry  whence it can only be space-filling, i.e. strings whose excitations correspond to states must be of type $({\cal B}_{d.c.}, {\cal B}_\alpha)$.} Here, ${\cal M}_H (\mathscr G, \Sigma_g)$, the moduli space of $\mathscr G$ Hitchin equations on $\Sigma_g$, can be regarded as the moduli space of $\mathscr G$ Higgs bundles on $\Sigma_g$ (since we are in complex structure $I$).

 In turn, according to the string duality relations in (\ref{dia S=W for g}), the equivalence  of Hilbert spaces of states in (\ref{H = H QGL})  would correspond to the equivalence 
\be
 \widehat{{^L{\cal W}_{{\rm aff}, \kappa}}(\frak g)}_{\Sigma_g} = \widehat{{^L{\cal W}}_{{\rm aff}, ^L\kappa}}(^L{\frak g})_{{ \Sigma}_g}, \qquad {\rm where} \qquad  r^\vee (\kappa + {h}) = (^L\kappa + {^Lh})^{-1},
  \ee  
of modules of affine $\cal W$-algebras. This equivalence is a consequence of a  Feigin-Frenkel $\cal W$-algebra duality which we physically derived in $\S$4.1\ref{s4.1}. 

In short, from the preceding three paragraphs, we have the diagram
\tikzset{node distance=7.0cm, auto}
\begin{center}
\be
\label{dia S=W for QGL}
 \begin{tikzpicture}
  \node (P) {${\rm D}^{\rm mod}_{{\cal L}^{\Psi - h^\vee}}({\rm Bun}_G(\Sigma_g))$};
  \node (B) [right of=P] {$ \widehat{{^L{\cal W}_{{\rm aff}, \kappa}}(\frak g)}_{\Sigma_g}$};
  \node (A) [below of=P, yshift = 3.2cm] {${\rm D}^{\rm mod}_{{\cal L}^{{^L\Psi} - {^Lh}^\vee}}({\rm Bun}_{^LG}(\Sigma_g))$};
  \node (C) [below of=B, yshift = 3.2cm] {$\widehat{{^L{\cal W}}_{{\rm aff}, ^L\kappa}}(^L{\frak g})_{{ \Sigma}_g}$};
   \draw[transform canvas={yshift=0.0ex},->] (P) --(B) node[midway] {{\scriptsize String Duality}};
\draw[transform canvas={yshift=-0.0ex},->](B) -- (P) node[midway] {}; 
 \draw[transform canvas={xshift=0.0ex},->] (P) --(A) node[midway] {{\scriptsize $ {^L\Psi} = -{1 \over r^\vee \Psi}$}};
\draw[transform canvas={xshift=-0.0ex},->](A) -- (P) node[midway] {{\scriptsize $S$-duality}}; 
 \draw[transform canvas={yshift=0.0ex},->] (A) --(C) node[midway] {{\scriptsize String Duality}};
\draw[transform canvas={yshift=-0.0ex},->](C) -- (A) node[] {}; 
\draw[transform canvas={xshift=0.0ex},->] (B) --(C) node[midway] {{\scriptsize FF-duality}};
\draw[transform canvas={xshift=-0.0ex},->](C) -- (B) node[midway] {{\scriptsize $(^L\kappa + {^Lh}) = {1\over r^\vee (\kappa + {h})}$}};   
\end{tikzpicture}
\ee
 \end{center}
From the diagram, it is clear that within our M-theoretic framework, quantum geometric Langlands correspondence as an $S$-duality of 4d gauge theory, can be understood, via string duality, as a Feigin-Frenkel quantum $\cal W$-algebra duality of 2d CFT, and vice versa!~\footnote{More precisely, we have a dual version of the Feigin-Frenkel duality involving ``Langlands dual'' $\cal W$-algebras and the Coxeter number $h$, which, as explained in $\S$4.1\ref{s4.1}, is equivalent to the original duality involving regular $\cal W$-algebras and dual Coxeter number $h^\vee$. }   

Indeed, quantum geometric Langlands correspondence can be mathematically described as a Fourier-Mukai transform~\cite[$\S$6.3]{Frenkel}, and it is conjectured~\cite[$\S$8.6]{Frenkel} that a Feigin-Frenkel quantum $\cal W$-algebra duality ought to play a prominent role in defining it. Since an $S$-duality of our GL-twisted ${\cal N} = 4$ gauge theory is equivalent to a $T$-duality of our open topological A-model with branes~\cite{K}, and in mathematical terminology, this $T$-duality is just a Fourier-Mukai transform,  (\ref{dia S=W for QGL}), which relates an $S$-duality to a Feigin-Frenkel quantum $\cal W$-algebra duality via a string duality, would be a purely physical M-theoretic affirmation of this conjecture.


\newsubsection{ A Geometric Langlands Correspondence as an $S$-duality and a Classical $\cal W$-algebra Duality}
\label{s5.2}


We shall now demonstrate, via our M-theoretic framework, how a geometric Langlands correspondence can be understood simultanueously as a 4d gauge-theoretic $S$-duality and a 2d conformal field-theoretic classical $\cal W$-algebra duality. In doing so, we would be able to elucidate the sought-after connection between its gauge-theoretic realization by Kapustin-Witten~\cite{KW} and its original algebraic CFT formulation by Beilinson-Drinfeld~\cite{BD}.

\bigskip\noindent{\it A Geometric Langlands Correspondence as an $S$-Duality and a Classical $\cal W$-algebra Duality} 

 From the formula $t^2 = - (\Psi - \bar\tau) / (\Psi - \tau)$, we find that in our case of purely imaginary $\tau$, we have $\Psi = \tau (t^2 -1) / (t^2 +1)$. As noted in the previous subsection, the classical limit of the quantum geometric Langlands correspondence occurs at $\Psi =0$. For any nonvanishing value of $\tau$, this happens specifically at $t = \pm 1$; alternatively, if $\tau = 0$, this happens for \emph{any} value of $t$. We shall henceforth adopt the latter viewpoint on the classical  limit, so that our following discussion would be valid for\emph{ any} value of $t$.

 Recall that in our framework, $\tau = i R_2 / R_1$. Therefore, $\tau \to 0$ when $R_1 \gg R_2$, i.e. when ${\bf S}^1_{\epsilon_1}$ decompactifies relative to ${\bf S}^1_{\epsilon_2}$ in (\ref{SxI-AB-all g})--(\ref{SxI-CDG-all g}). This is tantamount to turning off Omega-deformation in the underlying ${\bf R}^2$-plane. In other words, in our framework, setting $\Psi = 0$ corresponds to setting $\epsilon_1 =0$.

 Let us now turn our attention to the various objects in (\ref{dia S=W for QGL}). When  $\Psi = 0$,  on the top LHS of (\ref{dia S=W for QGL}), we have ${\rm D}^{\rm mod}_{{\cal L}^{ - h^\vee}}({\rm Bun}_G(\Sigma_g)) = {\rm D}^{\rm mod}_{\rm crit}({\rm Bun}_G(\Sigma_g))$, the derived category of critically-twisted $\cal D$-modules on ${\rm Bun}_G(\Sigma_g)$. On the other hand, on the bottom LHS of (\ref{dia S=W for QGL}), we have ${\rm D}^{\rm mod}_{{\cal L}^{\infty}}({\rm Bun}_{^LG}(\Sigma_g)) = {\rm D}^{\rm coh}({\cal M}^{\rm flat}_{^LG_{\mathbb C}}(\Sigma_g))$, the derived category of coherent sheaves on the moduli space ${\cal M}^{\rm flat}_{^LG_{\mathbb C}}(\Sigma_g)$ of flat $^LG_{\mathbb C}$-connections on $\Sigma_g$, whose objects therefore define flat $^LG_{\mathbb C}$-bundles on $\Sigma_g$; in other words, we have a derived category ${\rm D}^{\rm flat}_{^L G_{\mathbb C}}(\Sigma_g)$ of flat $^LG_{\mathbb C}$-bundles on $\Sigma_g$. That these derived categories ought to be equivalent, is just the statement of the usual geometric Langlands correspondence as realized by an $S$-duality of a 4d gauge theory.  
 
 Because setting $\Psi =0$ also corresponds to setting $\epsilon_1 =0$, on the top RHS of (\ref{dia S=W for QGL}), we have (after recalling footnote~\ref{LW(g)} and (\ref{H=W, g})) a module $\widehat{{{\cal W}}_{{\rm aff}, \infty}}(^L{\frak g})_{{ \Sigma}_g}$, whose states, via a CFT state-operator isomorphism, can be identified with polynomials in the $z$-derivatives of spin-$s_i$ holomorphic fields ${\cal W}_{s_i}(z)$ ($i = 1, \dots, {\rm rank}(^L\frak g_{\mathbb C})$) which define a classical $\cal W$-algebra ${{{\cal W}_{{\rm aff}, \infty}}(^L\frak g)}_{\Sigma_g}$ on $\Sigma_g$. This is also the algebra of functions in $^L \frak g_{\mathbb C}$-opers which correspond to flat $^LG_{\mathbb C}$-connections on $\Sigma_g$~\cite{Frenkel}, whence we can identify the polynomials with the functions and associate each state in the module with a flat $^LG_{\mathbb C}$-bundle on $\Sigma_g$, i.e. we have a ``module'' ${\rm M}^{\rm flat}_{^L G_{\mathbb C}}(\Sigma_g)$ of flat $^LG_{\mathbb C}$-bundles on $\Sigma_g$.  On the other hand, on the bottom RHS of (\ref{dia S=W for QGL}), we have (according to a FF-duality starting from $\widehat{{{\cal W}}_{{\rm aff}, \infty}}(^L{\frak g})_{{ \Sigma}_g}$) the module $\widehat{{{\cal W}}_{{\rm aff}, - {h^\vee}}}({\frak g})_{{ \Sigma}_g}$ of a distinguished $\cal W$-algebra ${{{\cal W}_{{\rm aff}, -h^\vee}}(\frak g)}_{\Sigma_g}$ on $\Sigma_g$. This algebra is also the center $Z(U(\hat{\frak g})_{\rm crit})$ of the completed enveloping algebra  $U(\hat{\frak g})_{\rm crit}$ of $\hat {\frak g} = {\frak g}_{\mathbb C \,  {\rm aff}}$ at critical level~\cite{FF}. From ${\frak g}_{\mathbb C \,  {\rm aff}}$ at the critical level on $\Sigma_g$, one can also show (via~\cite[$\S$7.1]{my GL}, which physically realizes the corresponding mathematical result by Beilinson-Drinfeld in~\cite{BD}) that the states in the module of $Z(U(\hat{\frak g})_{\rm crit})$ can be interpreted as correlation functions which define critically-twisted $\cal D$-modules on ${\rm Bun}_G(\Sigma_g)$, i.e. we have a ``module'' ${\rm M}^{\rm mod}_{\rm crit}({\rm Bun}_G(\Sigma_g))$ of critically-twisted $\cal D$-modules on ${\rm Bun}_G(\Sigma_g)$.
 
 Altogether therefore, in place of (\ref{dia S=W for QGL}), we now have 
 \tikzset{node distance=7.0cm, auto}
\begin{center}
\be
\label{dia S=W for GL}
 \begin{tikzpicture}
  \node (P) {${\rm D}^{\rm mod}_{\rm crit}({\rm Bun}_G(\Sigma_g))$};
  \node (B) [right of=P] {${\rm M}^{\rm flat}_{^L G_{\mathbb C}}(\Sigma_g)$};
  \node (A) [below of=P, yshift = 3.2cm] {${\rm D}^{\rm flat}_{^L G_{\mathbb C}}(\Sigma_g)$ };
  \node (C) [below of=B, yshift = 3.2cm] {${\rm M}^{\rm mod}_{\rm crit}({\rm Bun}_G(\Sigma_g))$};
   \draw[transform canvas={yshift=0.0ex},->] (P) --(B) node[midway] {{\scriptsize String Duality}};
\draw[transform canvas={yshift=-0.0ex},->](B) -- (P) node[midway] {}; 
 \draw[transform canvas={xshift=0.0ex},->] (P) --(A) node[midway] {{\scriptsize KW realization}};
\draw[transform canvas={xshift=-0.0ex},->](A) -- (P) node[midway] {{\scriptsize $S$-duality}}; 
 \draw[transform canvas={yshift=0.0ex},->] (A) --(C) node[midway] {{\scriptsize String Duality}};
\draw[transform canvas={yshift=-0.0ex},->](C) -- (A) node[] {}; 
\draw[transform canvas={xshift=0.0ex},->] (B) --(C) node[midway] {{\scriptsize FF-duality}};
\draw[transform canvas={xshift=-0.0ex},->](C) -- (B) node[midway] {{\scriptsize BD formulation}};   
\end{tikzpicture}
\ee
 \end{center}
From the diagram, it is clear that within our M-theoretic framework, the geometric Langlands correspondence realized by Kapustin-Witten in~\cite{KW} via an $S$-duality of 4d gauge theory, can be understood, through string duality, as the geometric Langlands correspondence formulated by Beilinson-Drinfeld in~\cite{BD} via a Feigin-Frenkel classical $\cal W$-algebra duality of 2d CFT, and vice versa!  
 
%
 

\bigskip\noindent{\it Wilson and 't Hooft-Hecke Line Operators in 4d Gauge Theory as Monodromy Loop Operators in 2d CFT}

Let us now set $t = 1$ so that there can be `t Hooft line operators at $\Psi = 0$ (i.e. $\epsilon_1 = 0$). Under $S$-duality, these are mapped to Wilson line operators at $\Psi = \infty$ and $t =i$~\cite{KW}. 

In the context of M-theory, such line operators can be realized as boundary M2-branes~\cite{loop-many}. Hence, let us place M2-branes on the LHS of (\ref{4d dual AB}) and (\ref{4d dual CDG}) along ${\bf R} \subset {\bf R}^4\vert_{0, \epsilon_2}$ (in the undeformed plane), ${\bf S}^1_n$, and ${\bf R} \subset {\bf R}^5$.  Then, repeating the arguments in~\cite[eqns.~(5.1)--(5.8)]{4d AGT} that took us from the LHS to RHS of (\ref{4d dual AB}) and (\ref{4d dual CDG}), we now have, in place of  (\ref{4d dual AB}) and (\ref{4d dual CDG}),  
\be
\label{4d dual AB-line}
\underbrace{{\mathring {\bf R}}^4\vert_{0, \epsilon_2}  \times {{\mathring {\bf S}}^1_n} \times {\bf S}^1_t}_{\textrm{$N$ M5 + M2 on $\circ$}}\times {\mathring {\bf R}}^{5}\vert_{\epsilon_2; \,  x_{6,7}} 
\Longleftrightarrow {{\bf R}^{5}}\vert_{\epsilon_2; \, x_{4,5}} \times \underbrace{ {\bf S}^1_t \times {{\mathring{\bf S}}^1_n}  \times TN_N^{R\to 0}\vert_{\epsilon_2; \, x_{6,7}}}_{\textrm{$1$ M5-branes + M0 on $\circ$}},
\ee
where $n=1$ or $2$, and 
\be
\label{4d dual CDG-line}
\underbrace{{\mathring{\bf R}}^4\vert_{0, \epsilon_2}  \times {{\mathring{\bf S}}^1_n} \times {\bf S}^1_t}_{\textrm{$N$ M5 + OM5 + M2 on $\circ$}}\times {\mathring{\bf R}}^{5}\vert_{\epsilon_2; \,  x_{6,7}} \Longleftrightarrow {{\bf R}^{5}}\vert_{\epsilon_2; \, x_{4,5}} \times \underbrace{{\bf S}^1_t \times {{\mathring{\bf S}}^1_n}  \times SN_N^{R\to 0}\vert_{\epsilon_2; \, x_{6,7}}}_{\textrm{$1$ M5 + M0 on $\circ$}},
\ee
where $n=1$, $2$ or $3$. Here, the M0-brane will become a D0-brane when we reduce M-theory on a circle to type IIA string theory~\cite{M0}.

Repeating our discussion that led us to (\ref{SxI-AB-all g}) and (\ref{SxI-CDG-all g}), bearing in mind the LHS of (\ref{4d dual AB-line}) and (\ref{4d dual CDG-line}), we find that in place of (\ref{SxI-AB-all g}) and (\ref{SxI-CDG-all g}), we now have
\be
\label{SxI-AB-all g classical GL - t hooft}
\underbrace{ {\bf T}^2_{0, \epsilon_2} \times {\bf I} \times   {\bf R}_+ \times {{\Sigma}^{g,0}_{t, n}}}_{\textrm{$N$ M5-branes}} 
\llap{$\overbrace{\phantom{    {\bf T}^2_{0, \epsilon_2} \times {\bf I} \times   }}^{\textrm {M2 on} \, \, {\bf R}_+ \times {\bf S}^1_{n,g}}$}
\ee
for $g \geq 1$ and $n =1$ or 2, and
\be
\label{SxI-CDG-all g classical GL - t hooft}
\underbrace{{\bf T}^2_{0, \epsilon_2} \times {\bf I} \times   {\bf R}_+ \times {{\Sigma}^{g,0}_{t, n}}}_{\textrm{$N$ M5-branes + OM5-plane}}
\llap{$\overbrace{\phantom{    {\bf T}^2_{0, \epsilon_2} \times {\bf I} \times   }}^{\textrm {M2 on} \, \, {\bf R}_+ \times {\bf S}^1_{n, g}}$}
\ee 
for $g \geq 1$ and $n =1, 2$ or 3. Here, ${\bf S}^1_{n,g}$ is a disjoint union of a $g$ number of ${\bf S}^1_{n}$ one-cycles of ${\Sigma}^{g,0}_{t, n}$.   From the perspective of the resulting sigma-models on ${\bf I} \times {\bf R}_+$ after we compactify the above worldvolumes on ${\bf T}^2_{0, \epsilon_2} \times {\Sigma}^{g,0}_{t, n}$, there is just a static `t Hooft line operator with gauge symmetry $G$ along the ``time'' direction $ {\bf R}_+$. The line operator would act on the boundaries and therefore the branes of the sigma-model, so let us denote the branes being acted upon as ${\cal B}_{\textrm {`t-Hooft}}$.

Likewise, we find that in place of (\ref{SxI-AB-S dual- all g}) and (\ref{SxI-CDG-S dual-all g}), we now have
\be
\label{SxI-AB-all g classical GL - Wilson}
\underbrace{ {\tilde {\bf T}}^2_{0, \epsilon_2} \times {\bf I} \times   {\bf R}_+ \times {{\tilde \Sigma}^{g,0}_{t, n}}}_{\textrm{$N$ M5-branes}} 
\llap{$\overbrace{\phantom{    {\tilde \bf T}^2_{0, \epsilon_2} \times {\bf I} \times   }}^{\textrm {M2 on} \, \, {\bf R}_+ \times {\tilde {\bf S}}^1_{n,g}}$}
\ee
for $g \geq 1$ and $n =1$ or 2, and
\be
\label{SxI-CDG-all g classical GL - Wilson}
\underbrace{{ \tilde {\bf T}}^2_{0, \epsilon_2} \times {\bf I} \times   {\bf R}_+ \times {{\tilde \Sigma}^{g,0}_{t, n}}}_{\textrm{$N$ M5-branes + OM5-plane}}
\llap{$\overbrace{\phantom{    { \tilde {\bf T}}^2_{0, \epsilon_2} \times {\bf I} \times   }}^{\textrm {M2 on} \, \, {\bf R}_+ \times {\tilde {\bf S}}^1_{n, g}}$}
\ee 
for $g \geq 1$ and $n =1, 2$ or 3. Here, ${\tilde {\bf S}}^1_{n,g}$ is a disjoint union of a $g$ number of ${\bf S}^1_{n}$ one-cycles of ${\tilde \Sigma}^{g,0}_{t, n}$. From the perspective of the resulting $S$-dual sigma-models on ${\bf I} \times {\bf R}_+$ after we compactify the above worldvolumes on ${\tilde { \bf T}}^2_{0, \epsilon_2} \times {\tilde \Sigma}^{g,0}_{t, n}$, there is just a static Wilson line operator with gauge symmetry $^LG$ along the ``time'' direction $ {\bf R}_+$. The line operator would act on the boundaries and therefore the branes of the sigma-model, so let us denote the branes being acted upon as ${^L\cal B}_{\textrm {Wilson}}$.

Repeating our discussion that led us to (\ref{H=W, g}) and (\ref{H=W, g, S-dual n-corrected}), bearing in mind the RHS of (\ref{4d dual AB-line}) and (\ref{4d dual CDG-line}), we find that in place of $\Sigma_g$ on the RHS of (\ref{H=W, g}) and (\ref{H=W, g, S-dual n-corrected}), we now have $\Sigma^{\rm loop}_g$, which is $\Sigma_g$ with a loop operator that is a disjoint union of $g$ number of loop operators around its $g$ number of ${\bf S}^1_n$ one-cycles, each corresponding to a worldloop of a D0-brane. The loop operators that now appear on the RHS of (\ref{H=W, g}) and (\ref{H=W, g, S-dual n-corrected}) are \emph{dual} to each other in the sense that they are defined on dual sets of $g$ nonintersecting one-cycles of a genus $g$ surface (recall that it is actually $\tilde \Sigma_g$ which underlies (\ref{H=W, g, S-dual n-corrected}), as reflected in (\ref{SxI-AB-all g classical GL - Wilson})--(\ref{SxI-CDG-all g classical GL - Wilson})), and in the context of our 2d CFT on this surface, we can regard the former and latter as ``Wilson'' and ``'t Hooft''  loop operators, since they are associated with $^L \frak g$ and $\frak g$, respectively. In fact, one can see from (\ref{dia S=W for GL}) that string duality, apart from being a 4d-2d duality, is also an $S$-duality in the sense that it maps objects associated with $G$ to objects associated with $^LG$; the ``Wilson'' and ``'t Hooft''  loop operators therefore ought to be associated with the $S$-dual of the `t Hooft and Wilson line operators described in the last two paragraphs, i.e. the ``Wilson'' and ``'t Hooft''  CFT operators ought to be associated with Wilson and 't Hooft gauge theory operators, which thus justifies their nomenclature. At any rate, it is clear that these loop operators act on the underlying modules on the RHS of (\ref{H=W, g}) and (\ref{H=W, g, S-dual n-corrected}). Let us denote the modules being acted upon as ${\widehat{\cal W} (^L\frak g)}_{\textrm {``Wilson''}}$ and ${\widehat{\cal W} (\frak g)}_{\textrm {```t-Hooft''}}$.

In sum, the three preceding paragraphs mean that under the action of these one-dimensional operators, we would, in place of (\ref{dia S=W for GL}), have 
 \tikzset{node distance=7.0cm, auto}
\begin{center}
\be
\label{dia S=W for GL with line operators}
 \begin{tikzpicture}
  \node (P) {${\rm D}^{\rm mod}_{\rm crit}({\rm Bun}_G(\Sigma_g))_{{\cal B}_{\textrm {`t-Hooft}}}$};
  \node (B) [right of=P] {${\rm M}^{\rm flat}_{^L G_{\mathbb C}}(\Sigma_g)_{{\widehat{\cal W} (^L\frak g)}_{\textrm {``Wilson''}}}$};
  \node (A) [below of=P, yshift = 3.2cm] {${\rm D}^{\rm flat}_{^L G_{\mathbb C}}(\Sigma_g)_{{^L\cal B}_{\textrm {Wilson}}}$ };
  \node (C) [below of=B, yshift = 3.2cm] {${\rm M}^{\rm mod}_{\rm crit}({\rm Bun}_G(\Sigma_g))_{{\widehat{\cal W} (\frak g)}_{\textrm {``'t-Hooft''}}}$};
   \draw[transform canvas={yshift=0.0ex},->] (P) --(B) node[midway] {{\scriptsize String Duality}};
\draw[transform canvas={yshift=-0.0ex},->](B) -- (P) node[midway] {}; 
 \draw[transform canvas={xshift=0.0ex},->] (P) --(A) node[midway] {{\scriptsize KW realization}};
\draw[transform canvas={xshift=-0.0ex},->](A) -- (P) node[midway] {{\scriptsize $S$-duality}}; 
 \draw[transform canvas={yshift=0.0ex},->] (A) --(C) node[midway] {{\scriptsize String Duality}};
\draw[transform canvas={yshift=-0.0ex},->](C) -- (A) node[] {}; 
\draw[transform canvas={xshift=0.0ex},->] (B) --(C) node[midway] {{\scriptsize FF-duality}};
\draw[transform canvas={xshift=-0.0ex},->](C) -- (B) node[midway] {{\scriptsize BD formulation}};   
\end{tikzpicture}
\ee
 \end{center}

Indeed, as indicated in the above diagram, one can understand the action of the 't Hooft and Wilson line operators on the derived categories (of sigma-model branes) on the LHS of (\ref{dia S=W for GL}) in terms of the  action of the ``'t Hooft'' and ``Wilson'' loop operators on the modules (of affine $\cal W$-algebras on $\Sigma_g$) on the RHS of  (\ref{dia S=W for GL}), as follows. 

First, let us describe the action of the 't Hooft line operator on ${\rm D}^{\rm mod}_{\rm crit}({\rm Bun}_G(\Sigma_g))$. According to~\cite[eqn.~(8.9)]{KW}, the  't Hooft line operator maps the gauge theory magnetic flux ${\bf m}_0 \to {\bf m}_0 + \xi({^LR})$, where ${\bf m}_0$ and $\xi({^LR})$ are characteristic classes that classify the topology of $G$-bundles over $\Sigma_g$ and ${\bf S}^2$, respectively. The latter bundle characterizes the 't Hooft line operator, and $^LR$ is a representation of $^LG$. As the relevant branes which correspond to the critically-twisted $\cal D$-modules wrap different components of the sigma-model target labeled by all possible values of ${\bf m}_0$, the 't Hooft line operator acts by mapping each object in ${\rm D}^{\rm mod}_{\rm crit}({\rm Bun}_G(\Sigma_g))$ labeled by ${\bf m}_0$, to another labeled by ${\bf m}_0 + \xi({^LR})$. The derived category after this action is ${\rm D}^{\rm mod}_{\rm crit}({\rm Bun}_G(\Sigma_g))_{{\cal B}_{\textrm {`t-Hooft}}}$, as shown in (\ref{dia S=W for GL with line operators}).     

On the other hand, the ``'t Hooft'' loop operator is (via~\cite[$\S$3.2]{gomis} in the  massless ${\cal N} = 2^\ast$ case) a monodromy operator which acts on 
\be
Z_{\frak g}({\bf a}) = {\rm Tr}_{[V_{\bf a}]} \, {\rm q}^{L_0},
\ee 
the chiral partition functions (i.e. $\Sigma_1$ conformal blocks) of the module ${\rm M}^{\rm mod}_{\rm crit}({\rm Bun}_G(\Sigma_1))$, as 
\be
Z_{\frak g}({\bf a}) \to \sum_{{\bf p}_k}  \lambda_{{\bf a}, {\bf p}} \, Z_{\frak g}({\bf p}_k).
\ee
Here, $[V_{\bf a}]$ is the submodule labeled by the highest coweight vector ${\bf a}$ of ${\frak g}$;\footnote{Here, $\bf a$ is a coweight and not weight vector because on the 2d CFT side, we are actually in a Langlands dual frame where roots and coroots are exchanged, just as it is $h$ and not $h^\vee$ that appears in (\ref{dia S=W for QGL}).} ${\rm q} = e^{2\pi i \tau_{\Sigma_1}}$, where $\tau_{\Sigma_1}$ is the complex structure of $\Sigma_1$; $L_0$ is a translation operator along $\Sigma_1$; ${\bf p}_k = {\bf a} + {\rm b} {\bf h}_k$, where ${\rm b}$ is a constant and ${\bf h}_k$ are coweights of a representation $R$ of $G$; and the $\lambda_{{\bf a}, {\bf p}}$'s are constants. On \emph{all} the $g$ identical $\Sigma_1$ surfaces -- each with a ``'t Hooft'' loop operator around the same one-cycle -- that compose $\Sigma_g$ (shown on the RHS of Fig.~\ref{fig 1}(d)), what we have just said holds. Therefore, it would mean that the overall ``'t Hooft'' loop operator acts by mapping each state in ${\rm M}^{\rm mod}_{\rm crit}({\rm Bun}_G(\Sigma_g))$ labeled by $\bf a$, to another labeled by ${\bf a} + {\bf h}$, where $\bf h$ is a weight of a representation $^LR$ of $^LG$. The module after this action is ${\rm M}^{\rm mod}_{\rm crit}({\rm Bun}_G(\Sigma_g))_{{\widehat{\cal W} (\frak g)}_{\textrm {``'t-Hooft''}}}$, as shown in (\ref{dia S=W for GL with line operators}). This is just the 2d CFT version of the 4d gauge theory statement about the action of a 't Hooft line operator on  ${\rm D}^{\rm mod}_{\rm crit}({\rm Bun}_G(\Sigma_g))$.


Next, let us describe the action of the Wilson line operator on ${\rm D}^{\rm flat}_{^L G_{\mathbb C}}(\Sigma_g)$. According to~\cite[eqn.~(8.8)]{KW}, the  Wilson line operator maps the gauge theory electric flux ${\bf e}_0 \to {\bf e}_0 + {\theta}_{^LR}$, where ${\bf e}_0$ and ${\theta}_{^LR}$ are characters  of the center of (the universal cover of) $^LG$, where the latter, determined by a representation ${^LR}$ of $^LG$, characterizes the Wilson line operator. The relevant branes which correspond to the flat $^LG_{\mathbb C}$-bundles are zerobranes labeled by ${\bf e}_0$. Because they are supported on points, the effect of the shift ${\bf e}_0 \to {\bf e}_0 + {\theta}_{^LR}$ which twists them by ${\theta}_{^LR}(\zeta)$ ($\zeta$ being the obstruction to the existence of a universal $\overline{^LG}$ Higgs bundle), is trivial, i.e. the zerobranes remain as they are under the shift. Thus, the Wilson line operator acts by mapping each object in ${\rm D}^{\rm flat}_{^L G_{\mathbb C}}(\Sigma_g)$ to itself. Therefore, the derived category  ${\rm D}^{\rm flat}_{^L G_{\mathbb C}}(\Sigma_g)_{{^L\cal B}_{\textrm {Wilson}}}$ after this action in (\ref{dia S=W for GL with line operators}), is simply ${\rm D}^{\rm flat}_{^L G_{\mathbb C}}(\Sigma_g)$. 

On the other hand, the ``Wilson'' loop operator is (via~\cite[Appendix D]{gomis} in the massless ${\cal N} = 2^\ast$ case) a monodromy operator which acts on the chiral partition functions $Z_{^L\frak g}({\bf a}^\vee)$ (i.e. $\Sigma_1$ conformal blocks) of the module ${\rm M}^{\rm flat}_{^L G_{\mathbb C}}(\Sigma_1)$, as 
\be
Z_{^L\frak g}({\bf a}^\vee) \to \lambda_{{\bf a}^\vee} \, Z_{^L\frak g} ({\bf a}^\vee),
\ee
where the highest coweight vector ${\bf a}^\vee$ of $^L {\frak g}$ labels a submodule, and $\lambda_{{\bf a}^\vee}$ is a constant. On \emph{all }the $g$ identical $\Sigma_1$ surfaces -- each with a ``Wilson'' loop operator around the same one-cycle -- that compose $\Sigma_g$ (shown on the RHS of Fig.~\ref{fig 1}(d)), what we have just said holds. Thus, it would mean that the overall ``Wilson'' loop operator acts by mapping each state in ${\rm M}^{\rm flat}_{^L G_{\mathbb C}}(\Sigma_g)$ to itself. Therefore, the module ${\rm M}^{\rm flat}_{^L G_{\mathbb C}}(\Sigma_g)_{{\widehat{\cal W} (^L\frak g)}_{\textrm {``Wilson''}}}$  after this action, shown in (\ref{dia S=W for GL with line operators}), is simply ${\rm M}^{\rm flat}_{^L G_{\mathbb C}}(\Sigma_g)$. This is just the 2d CFT version of the 4d gauge theory statement about the action of a Wilson line operator on ${\rm D}^{\rm flat}_{^L G_{\mathbb C}}(\Sigma_g)$. 

\bigskip\noindent{\it About Hecke Modifications and Correspondence}

One could certainly proceed to discuss the topic of Hecke modifications and correspondence in the context of 2d CFT via the monodromy ``'t Hooft'' loop operator. The discussion would parallel that in~\cite[$\S$7.3]{my GL}. However, for brevity, we shall defer it to another occasion and now turn to an even more interesting offshoot of our discussion hitherto -- higher geometric Langlands correspondences.

\newsection{Higher Geometric Langlands Correspondences from M-Theory}
\label{s6}


\newsubsection{A $q$-Geometric Langlands Correspondence for Simply-Laced Lie Groups}

We would now like to ascertain, for simply-laced Lie groups $G$, what a $q$-geometric Langlands correspondence that is related to the classical $q$-geometric Langlands duality in (\ref{cq-GL-A-final}), means. 

To this end, note that in the 5d AGT correspondence behind our derivation of (\ref{cq-GL-A-final}), the 2d CFT is defined over the \emph{same} (punctured) Riemann surface as that in the 4d AGT correspondence~\cite{5d AGT}. This means that the construction in Fig.~\ref{fig 1} is also valid in the 5d case, whence we again have the effective worldvolume configurations (\ref{SxI-AB-all g}) and (\ref{SxI-CDG-all g}). 

That said, note that according to our explanations leading up to~\cite[eqns.~(3.57)]{5d AGT}, we must, in the 5d case where $\beta \nrightarrow 0$ whence the fifth circle decompactifies, associate to each and every point on the compactified Riemann surface ${\Sigma}^{g,0}_{t, 1} = \Sigma_g$, a loop which effectively represents the decompactified fifth circle. As such, this means that in place of (\ref{SxI-AB-all g}) and (\ref{SxI-CDG-all g}), we ought to have 
\be
\label{qGL-AB-all g}
\underbrace{ {\bf T}^2_{\epsilon_1, \epsilon_2}\times {\bf I}_1 \times   {\bf I}_2 \times {{\Sigma}^{{\rm S}^1}_{g}}}_{\textrm{$N$ M5-branes}}
\ee
for the simply-laced $A$ groups, and
\be
\label{qGL-CDG-all g}
\underbrace{{\bf T}^2_{\epsilon_1, \epsilon_2} \times {\bf I}_1  \times {\bf I}_2 \times {{\Sigma}^{{\rm S}^1}_{g}}}_{\textrm{$N$ M5-branes + OM5-plane}}
\ee 
for  the simply-laced $D$ groups, where ${{\Sigma}^{{\rm S}^1}_{g}}$ is the compactified Riemann surface $\Sigma_g$ with an ${\rm S}^1$ loop of radius $\beta$ over every point.   


\newpage

\bigskip\noindent{\it Quantization of Circle-Valued $G$ Hitchin Systems and  Circle-Valued $^LG$-bundles on a Complex Curve}

Since this $q$-geometric Langlands correspondence of interest is supposed to be a higher 5d analog of the standard geometric Langlands correspondence discussed in $\S$5.2\ref{s5.2}, we shall, as was done in $\S$5.2\ref{s5.2}, set $\epsilon_1 = 0$, $t =1$ (although we will not consider gauge theory line operators for simplicity) and $g > 1$. This means that in place of (\ref{qGL-AB-all g}) and (\ref{qGL-CDG-all g}), we have 
\be
\label{cqGL-AB-all g}
\underbrace{ {\bf T}^2_{0, \epsilon_2}\times {\bf R}_+ \times   {\bf I} \times {{\Sigma}^{{\rm S}^1}_{g}}}_{\textrm{$N$ M5-branes}}
\ee
for the simply-laced $A$ groups, and
\be
\label{cqGL-CDG-all g}
\underbrace{{\bf T}^2_{0, \epsilon_2} \times {\bf R}_+  \times {\bf I} \times {{\Sigma}^{{\rm S}^1}_{g}}}_{\textrm{$N$ M5-branes + OM5-plane}}
\ee
for  the simply-laced $D$ groups, whence in place of (\ref{H=W, g}), we now have, from the RHS of (\ref{GSU(N)}) and (\ref{GSO(2N)}) and the fact that they are given by the LHS of (\ref{cq-GL-A-2}) when $\epsilon_1 = 0$ with $q = e^{-i \beta \epsilon_2}$,
\be
\label{H=W, g, cqGL}
{\cal H}^{\rm A}_{{\bf I} \times {\bf R}_+}({\cal M}^{{\rm S}^1}_H (G, \Sigma_g))_{{\cal B}^\beta_{c.c.}, \, {\cal B}^\beta_\alpha }  =  { \widehat{{\cal W}^q_{\rm cl}}(^L\frak g)_{\Sigma_g}}.
\ee
The LHS is the Hilbert space of a topological A-model on ${\bf I} \times {\bf R}_+$ with target ${\cal M}^{{\rm S}^1}_H (G, \Sigma_g)$, the moduli space of circle-valued $G$ Hitchin equations on $\Sigma_g$, which ends on a space-filling and middle-dimensional brane ${\cal B}^{\beta}_{c.c.}$ and ${\cal B}^{\beta}_\alpha$, respectively, where  ${\cal B}^{0}_{c.c.} = {\cal B}_{c.c.}$ and ${\cal B}^{0}_\alpha = {\cal B}_\alpha$ are a canonical coisotropic and Lagrangian brane on ${\cal M}_H (G, \Sigma_g)$. The RHS is a module of a $q$-deformed version of the classical $\cal W$-algebra ${{\cal W}_{\rm cl}}(^L\frak g)$ on $\Sigma_g$ obtained by a Drinfeld-Sokolov reduction of the dual space to $(^L \frak g)_{\rm aff}$. Here, $G = SU(N)$ or $SO(2N)$, with Lie algebra $\frak g$ and Langlands dual $^L \frak g$. 

Note that according to the axioms of open-closed topological field theory, a $({\cal B}^{\beta}_{c.c.}, {\cal B}^{\beta}_\alpha)$ string would be a module for a $({\cal B}^{\beta}_{c.c.}, {\cal B}^{\beta}_{c.c})$ string (see~\cite[Fig.~6e]{KW} for illustration). Also, generalizing to our case, the arguments in~\cite[$\S$11.1]{K} (which show that  $({\cal B}^{0}_{c.c.}, {\cal B}^{0}_{c.c})$ strings furnish a noncommutative algebra of holomorphic functions on ${\cal M}_H (G, \Sigma_g)$ which captures a classical $G$ Hitchin integrable system on $\Sigma_g$), it is clear that  the $({\cal B}^{\beta}_{c.c.}, {\cal B}^{\beta}_{c.c})$ strings would furnish a noncommutative algebra ${\cal O}_{\hbar}({\cal M}^{{\rm S}^1}_{{\rm H.S.}} (G, \Sigma_g))$ of holomorphic functions on ${\cal M}^{{\rm S}^1}_{{\rm H.S.}} (G, \Sigma_g)$ which captures a classical circle-valued $G$ Hitchin integrable system on $\Sigma_g$, where $\hbar = 1 / {\rm Im}\,\tau = R_1/R_2 \sim \epsilon_2$ is the nonvanishing Planck constant which effects the noncommutativity.\footnote{That $\hbar \sim \epsilon_2$, can be understood as follows. Firstly, $R_{1,2}$ are the radii of the circles in ${\bf T}^2_{\epsilon_1, \epsilon_2}$ which are rotated by an angle $\epsilon_{1,2}$. Secondly, in the present case where we have ${\bf T}^2_{0, \epsilon_2}$, $R_1$ can be fixed while $R_2$ would depend on $\epsilon_2$: the larger $\epsilon_2$ is, the greater the Omega-deformation, and the closer to the origin of the ${\bf R}^2$-plane the excitations would be localized, whence $R_2$ would be smaller. In sum, this means that $\hbar \sim R^{-1}_2 \sim \epsilon_2$ (up to a dimensionful constant).} Last but not least, according to our earlier explanation (leading up to (\ref{dia S=W for GL})) that one can associate each state in the module $\widehat{{\cal W}_{\rm cl}}(^L\frak g)_{\Sigma_g}$ with a flat $^LG$-bundle on $\Sigma_g$, and our explanation leading up to~\cite[eqn.~(2.39)]{5d AGT},  one can associate each state in the module $\widehat{{\cal W}^q_{\rm cl}}(^L\frak g)_{\Sigma_g}$ with a circle-valued flat $^LG$-bundle on $\Sigma_g$, where the radius of the circle is given by $\beta$. Altogether, this means that we can also express (\ref{H=W, g, cqGL}) as
\be
\label{cat cqGL}
{{\rm C}^{\rm mod}_{{\cal O}_{\hbar}}({\cal M}^{{\rm S}^1}_{{\rm H.S.}} (G, \Sigma_g)) = {\rm M}^{{{\rm S}^1}}_{^L G}(\Sigma_g)_{\rm flat} },
\ee
where the LHS is a category of modules of ${\cal O}_{\hbar}({\cal M}^{{\rm S}^1}_{{\rm H.S.}} (G, \Sigma_g))$, and the RHS is a ``module'' of circle-valued flat $^LG$-bundles on $\Sigma_g$. 

In turn, (\ref{cat cqGL}) means that we have the following correspondence
\be
\label{cqGL-corr}
 \boxed{{{\cal O}_{\hbar}({\cal M}^{{\rm S}^1}_{{\rm H.S.}} (G, \Sigma_g)){\textnormal{-}}{\rm{module}}}} \, \,
  {\Longleftrightarrow} \,  \, \boxed{ {\textnormal {circle-valued flat}}~{^LG}{\textnormal{-bundle on}}~{\Sigma_g} }
  \ee 
 where $g > 1$. Clearly, this defines a $q$-geometric Langlands correspondence that is a 5d analog of the standard geometric Langlands correspondence for simply-laced $G$! Indeed, when $\beta \to 0$  whence $q \to 1$, this $q$-deformed correspondence reduces to the standard correspondence discussed in $\S$5.2\ref{s5.2}.   
 
 Thus, since the noncommutative deformation of the algebra of holomorphic functions on a space which captures a classical integrable system defines a quantization of the classical integrable system itself, (\ref{cqGL-corr}) therefore relates the quantization of a circle-valued $G$ Hitchin system on $\Sigma_g$ to a circle-valued flat $^LG$-bundle on $\Sigma_g$. 


\bigskip\noindent{\it Quantization of Circle-Valued $G$ Hitchin Systems and Transfer Matrices of a $G$-type XXZ Spin Chain}

In light of the identity (\ref{cq-GL-A-final}) (which also holds for ${\frak g} = \frak {so}(2N)$, as explained thereafter), one can also express (\ref{H=W, g, cqGL}) as
\be
\label{H=Z, g, cqGL}
{\cal H}^{\rm A}_{{\bf I} \times {\bf R}_+}({\cal M}^{{\rm S}^1}_H (G, \Sigma_g))_{{\cal B}^\beta_{c.c.}, \, {\cal B}^\beta_\alpha }  =  \widehat{Z}(U_q(\hat{\frak g})_{\rm crit})_{\Sigma_g}.
\ee
Here, the RHS is a module of  the center $Z(U_q(\hat{\frak g})_{\rm crit})_{\Sigma_g}$ of the quantum affine algebra  $U_q(\hat{\frak g})_{\rm crit}$ at critical level on ${\Sigma_g}$. Hence, since the LHS is the space of states of all $({\cal B}^{\beta}_{c.c.}, {\cal B}^{\beta}_\alpha)$ strings which are themselves modules of ${\cal O}_{\hbar}({\cal M}^{{\rm S}^1}_{{\rm H.S.}} (G, \Sigma_g))$,  it would mean that we have a correspondence between ${\cal O}_{\hbar}({\cal M}^{{\rm S}^1}_{{\rm H.S.}} (G, \Sigma_g))$ and $Z(U_q(\hat{\frak g})_{\rm crit})_{\Sigma_g}$ (where $\hbar \sim \epsilon_2$ and $q = e^{-i \beta \epsilon_2}$).\footnote{The alert reader would have recalled that the algebra ${\cal O}_{\hbar}$ is noncommutative while $Z_{\Sigma_g}$ is commutative. However, there is no inconsistency here, as  over $\Sigma_g$ that $Z_{\Sigma_g}$ is defined on, ${\cal O}_{\hbar}$ is is also commutative (which again can be seen by generalizing to our case, the arguments in~\cite[$\S$11.1]{KW}).}

At any rate, according to~\cite{DE}, one can actually identify $Z(U_q(\hat{\frak g})_{\rm crit})_{{\bf C}^\ast}$ with the algebra ${\cal T}_{\rm xxz}(G, {\bf C}^\ast)$ of polynomials in $\partial_z^{m} \, {\cal T}_i(z)$, where $m \geq 0$; the ${\cal T}_i$'s are commuting transfer matrices of a $G$-type XXZ spin chain with  $U_q(\hat{\frak g})$ symmetry on ${\bf C}^\ast$; and $i = 1, \dots, {\rm rank}(\frak g)$. In other words, since our derivation of (\ref{H=Z, g, cqGL}) also holds for ${\bf C}^\ast$ (which is conformally equivalent to a cylinder) if we just start with the building block in fig.~\ref{fig 1}(b), we would have the following correspondence
  \be
  \label{O = T for XXZ}
  \boxed{{\cal O}_{\hbar}({\cal M}^{{\rm S}^1}_{{\rm H.S.}} (G, {\bf C}^\ast)) \, \, \Longleftrightarrow  \, \, {\cal T}_{\rm xxz}(G, {\bf C}^\ast)}
\ee
which relates the quantization of a circle-valued $G$ Hitchin system on ${\bf C}^\ast$ to the transfer matrices of a $G$-type XXZ spin chain on ${\bf C}^\ast$!


\bigskip\noindent{\it Circle-Valued Hitchin $G$ Systems and $q$-Characters of ${U^{\rm aff}_q}({\frak g})$}

A relevant implication of (\ref{O = T for XXZ}) is the following. Each point $x$ on ${\cal M}^{{\rm S}^1}_{{\rm H.S.}} (G, {\bf C}^\ast)$ defines an arbitrary holomorphic function and therefore generator of ${\cal O}_{\hbar}({\cal M}^{{\rm S}^1}_{{\rm H.S.}} (G, {\bf C}^\ast))$; since the ${\cal T}_i$'s generate ${\cal T}_{\rm xxz}(G, {\bf C}^\ast)$, (\ref{O = T for XXZ}) would mean that to each point $x$, one can associate the ${\cal T}_i$'s. In turn, as the eigenvalue of each ${\cal T}_i$ (acting on the physical states of the $G$-type XXZ spin chain) is given by the $q$-character $\chi_q({V_i})$, where ${V_i}$ is the $i$th fundamental representation of  $U_q(\hat{\frak g}) = {U^{\rm aff}_q}({\frak g})$ (also know as an ``auxiliary space'' in the algebraic Bethe Ansatz)~\cite{FR-ch}, it would mean that we have the correspondence
\be
\label{HS = chi}
\boxed{x \in {\cal M}^{{\rm S}^1}_{{\rm H.S.}} (G, {\bf C}^\ast) \,  \Longleftrightarrow \,  \chi_q({V_i}) = T_i(z), \quad {\rm where}  \, \, \, \, {V_i} \in {\rm Rep} \, [{U^{\rm aff}_q}({\frak g})_{{\bf C}^\ast}],   \, \, \, i = 1, \dots, {\rm rank}(\frak g)} 
\ee
and $T_i(z)$ is a polynomial whose degree depends on $V_i$.

\bigskip\noindent{\it An M-Theoretic Realization of Nekrasov-Pestun-Shatashvili's Results for 5d, ${\cal N} =1$ $G$-Quiver $SU(K_i)$ Gauge Theories}

Let us now realize, via our M-theoretic framework, Nekrasov-Pestun-Shatashvili's result in~\cite{NPS} which relates the moduli space of 5d, ${\cal N} =1$ $G$-quiver $SU(K_i)$ gauge theories to the representation theory of ${U^{\rm aff}_q}({\frak g})$. 

To this end, consider the AGT correspondence for a 5d, $SU(N)$ theory with $N_f = 2N$ fundamental matter discussed in~\cite[$\S$3.2]{5d AGT}, as an illustrative example. In this case, in place of Fig.~\ref{fig 1}(b), we have~\cite[Fig.~2(3)]{5d AGT} (where $n=1$ therein), and according to the RHS of~\cite[eqn.~(3.29)]{5d AGT} and the conformal symmetry of the 2d theory underlying it, it would mean that we can replace ${\bf C}^\ast$ in the above with $\{{\bf R} \times {\bf S}^1 \}_{z_1, z_2}$, a cylinder with two punctures at points $z_{1,2}$ with boundary conditions specified at $\pm \infty$. Therefore, in place of (\ref{HS = chi}), we have
\be
\label{HS = chi for NPS}
x \in {\cal M}^{{\rm S}^1}_{{\rm H.S.}} (G, \{{\bf R} \times {\bf S}^1\}_{z_1, z_2}) \, \, \Longleftrightarrow \, \, \chi_q({V_i}) = T_i(z). 
\ee

Note at this point that over a flat surface such as ${\bf R} \times {\bf S}^1$, one can define a \emph{trivial} circle-bundle. As such, the circle-valued Hitchin equations on $\{{\bf R} \times {\bf S}^1\}_{z_1, z_2}$ can also be regarded as a set of equations on  $\{{ \rm S}^1  \times {\bf R} \times {\bf S}^1\}_{y_1, y_2}$, where ${\rm S}^1$ represents the aforementioned circle, and $y_{1,2}$ are the positions in three-space of the punctures. Since this set of equations must reduce to the ordinary Hitchin equations when the radius of ${\rm S}^1$ goes to zero, i.e. the ordinary Hitchin equations are a dimensional reduction of this set of equations,  it must mean that this set of equations are the  periodic monopole equations on $\{{\rm S}^1  \times {\rm C}_{\rm x} \}_{y_1, y_2}$, where ${\rm C}_{\rm x} = {\bf R} \times {\bf S}^1$. In other words, in place of (\ref{HS = chi for NPS}), we have
\be
\label{mono = chi for NPS}
x \in {\cal M}^{\rm int}_{\rm mono} (G, {\bf k},  \{{\rm S}^1  \times {\rm C}_{\rm x} \}_{y_1, y_2}) \, \, \Longleftrightarrow \, \, \chi_q({V_i}) = T_i(z),
\ee
where ${\cal M}^{\rm int}_{\rm mono} (G, {\bf k}, \{{\rm S}^1  \times {\rm C}_{\rm x} \}_{y_1, y_2})$ captures a classical integrable system of periodic $G$-monopoles of charge $\bf k$ on $\{{\rm S}^1  \times {\rm C}_{\rm x} \}$ with singularities at $y_1$ and $y_2$.  

Since $x$, which is a point in the total space of ${\cal M}^{\rm int}_{\rm mono} (G, {\bf k}, \{{\rm S}^1  \times {\rm C}_{\rm x} \}_{y_1, y_2})$, also defines a point $u$ in its base space ${\frak M}^{G, {\rm C}_{\rm x}, y_1, y_2}_{{{\rm S}^1}\textnormal{-}{\rm mono}, \bf k}$, and since ${\rm rank}(\frak g)$ is equal to the number of vertices in its Dynkin diagram, instead of (\ref{HS = chi}), we would have
\be
\label{NPS for G}
\boxed{u \in {\frak M}^{G, {\rm C}_{\rm x}, y_1, y_2}_{{{\rm S}^1}\textnormal{-}{\rm mono}, \bf k} \,  \Longleftrightarrow \,  \chi_q({V_i}) = T_i(z), \quad {\rm where} \, \, \, {\rm C}_{\rm x} = {\bf R} \times {\bf S}^1,  \, \, \,  {V_i} \in {\rm Rep} \, [{U^{\rm aff}_q}({\frak g})_{\{{\rm C}_{\rm x}\}_{z_1, z_2}}],  \, \, \,  i \in I_{\Gamma}} 
\ee
where $I_{\Gamma}$ denotes the vertices of the Dynkin diagram of $G$. 

Note that ${\bf k} \in H^2({\rm S}^1 \times {\bf S}^1, \pi_1(T)) = \sum_{i \in I_\Gamma} K_i \, \alpha^\vee_i$  (c.f.~\cite[eqn.~(8.17)]{NP}), where $T \subset G$ is a Cartan subgroup, $K_i$ is a positive integer, and $\alpha^\vee_i$ is a simple coroot of $G$. As one might therefore expect, there should be a correspondence between $K_i$ and $T_i$.  In fact, the degree of $T_i$ would be given by $K_i$, as we shall now explain.

First, note that ${\bf k} = {\bf F}_{\rm m}\vert_{{\rm S}^1 \times {\bf S}^1}$, where ${\bf F}_{\rm m}\vert_{{\rm S}^1 \times {\bf S}^1}$ is the restriction of the monopole gauge field strength ${\bf F}_{\rm m}$ on ${\rm S}^1 \times {\bf S}^1 \times {\bf R}$ to ${\rm S}^1 \times {\bf S}^1$. Second, notice that our discussion below (\ref{HS = chi for NPS}) means that we can actually write ${\bf k} = {\bf F}^{{\rm S}^1}_{\rm H}\vert_{{\bf S}^1}$, where ${\bf F}^{{\rm S}^1}_{\rm H}\vert_{{\bf S}^1}$ is a restriction of an ${\rm S}^1$-valued Hitchin gauge field strength ${\bf F}^{{\rm S}^1}_{\rm H}$ on ${\bf S}^1 \times {\bf R}$ to ${\bf S}^1$. Third, from the perspective of an M-theoretic compactification on ${\bf S}^1$ (in (\ref{cqGL-AB-all g}) and (\ref{cqGL-CDG-all g}) with $\Sigma_g$ replaced by ${\rm C}_{\rm x}$), the momentum number around ${\bf S}^1$ would be given by the number of D0-branes along $\bf R$, and the D0-branes, being a charged pointlike object, would in turn source a two-form field strength and therefore ${\bf F}^{{\rm S}^1}_{\rm H}$. Fourth, note that in our M-theoretic realization of the gauge/CFT duality, the momentum number around ${\bf S}^1$ would correspond to the conformal dimension and thus energy level of the chiral CFT on ${\rm C}_{\rm x} = {\bf R} \times {\bf S}^1$ with ${U^{\rm aff}_q}({\frak g})$ symmetry (as recalled below (\ref{4d AGT-TS-EF-identity}) for the 4d case). Last but not least, according to~\cite[eqn.~(2.18)]{H-XXZ}, the energy level of the associated $G$-type XXZ spin chain would depend on $\partial {\rm ln} T_i(z)\vert_{z=1}$; in particular, the higher the degree of $T_i$, the higher the energy level. Altogether, this means that the higher the $K_i$, the higher the degree of $T_i$, and since both are integers,  the degree of $T_i$ ought to be given by $K_i$ (up to some positive integer constant that we can absorb in redefining $\alpha^\vee_i$).    

According to~\cite[$\S$5.1.1--5.1.4]{NPS}\cite{NP}, ${\frak M}^{G, {\rm C}_{\rm x}, y_1, y_2}_{{{\rm S}^1}\textnormal{-}{\rm mono}, \bf k}$ is the moduli space of vacua on the Coulomb  branch of a 5d, ${\cal N}=1$ $G$-quiver $SU(K_i)$ gauge theory with the singularities at $y_i$ determined by its mass and flavor data. Consequently, (\ref{NPS for G}), and the conclusion in the previous paragraph that the degree of $T_i$ is given by $K_i$, are, together, Nekrasov-Pestun-Shatashvili's main result in~\cite[$\S$1.3]{NPS} which relates the moduli space of 5d $G$-quiver gauge theory  to the representation theory of ${U^{\rm aff}_q}({\frak g})$!

Two comments are in order before we end this subsection. Firstly, our above analysis can be trivially generalized to the case of multiple $y_i$'s by considering at the start, a general linear quiver for the 5d theory in the underlying AGT correspondence. Secondly, as (\ref{NPS for G}) is just (a projection to the base space of) (\ref{HS = chi}) with ${\bf C}^\ast \cong {\rm C}_{\rm x}$ replaced by the punctured cylinder $\{{\rm C}_{\rm x}\}_{z_1, z_2}$, we can regard (\ref{HS = chi}) as a ``non-ramified'' version of Nekrasov-Pestun-Shatashvili's main result in~\cite[$\S$1.3]{NPS}.

\newsubsection{A $q$-Geometric Langlands Correspondence for Simply-Laced Kac-Moody Groups}

We would now like to ascertain the $\widehat G$ version of the $q$-geometric Langlands correspondence for simply-laced $G$ in (\ref{cqGL-corr}), where $\widehat G$ is the Kac-Moody generalization of $G$. 

To this end, first note that with regard to our arguments leading up to (\ref{cqGL-corr}), one could, instead of (\ref{SxI-AB-all g}) and (\ref{SxI-CDG-all g}), start with the worldvolumes in~\cite[eqns.~(5.9)]{4d AGT}, whence $\Sigma^{{\rm S}^1}_g$ in (\ref{qGL-AB-all g})--(\ref{cqGL-CDG-all g}) would be replaced by $\{{\bf R} \times {\bf S}^1\}^{{\rm S}^1}$ (with boundary conditions specified at $\pm \infty$), and (\ref{cqGL-corr}) would consequently hold for unpunctured flat ${\bf R} \times {\bf S}^1$ instead of $\Sigma_g$.  In other words, in addition to  (\ref{cqGL-corr}), we also have 
\be
\label{cqGL-corr-flat curve}
 {{{\cal O}_{\hbar}({\cal M}^{{\rm S}^1}_{{\rm H.S.}} (G, {\Sigma})){\textnormal{-}}{\rm{module}}}} \, \, \, \,
  {\Longleftrightarrow} \, \, \, \, { {\textnormal {circle-valued flat}}~{^LG}{\textnormal{-bundle on}}~{{\Sigma}} },
  \ee 
where ${\Sigma} = {\bf R} \times {\bf S}^1$.\footnote{The alert reader may wonder if one can consistently define the Hitchin equations over ${\bf R} \times {\bf S}^1$. The answer is ``yes''. This is because the Hitchin equations would have well-behaved solutions when there are singularities~\cite{ward 2-3}, and ${\bf R} \times {\bf S}^1$ (which in our case, has  boundary conditions  specified at $\pm \infty$) can, in the context of the conformally-invariant Hitchin equations, be regarded as a Riemann surface (a sphere to be exact) with singularities (at the poles).} 

Next, note that it has been established in~\cite{Garland} that nonsingular $\widehat G$-monopoles on a flat three space ${M}_3$ can also be regarded as $G$-instantons on $\hat{{\rm S}}^1 \times { M}_3$, and that as explained below (\ref{HS = chi for NPS}), nonsingular $G$-monopoles on ${ M}_3 = {\rm S}^1 \times {\Sigma}$ correspond to ${\rm S}^1$-valued $G$ Hitchin equations on ${\Sigma}$. In other words, one can interpret the ${\rm S}^1$-valued $\widehat G$ Hitchin equations on ${\Sigma}$ as\emph{ well-behaved} doubly-periodic $G$-instantons on $ {\hat{\rm S}}^1 \times  {\rm S}^1 \times {\Sigma}$, i.e. a consistent $\widehat G$ version of the LHS of (\ref{cqGL-corr-flat curve}) and hence (\ref{cqGL-corr}), would be ``${{{\cal O}_{\hbar}({\cal M}^{{\rm S}^1}_{{\rm H.S.}} ({\widehat G}, {{\Sigma}})){\textnormal{-}}{\rm{module}}}} $''. Since principal bundles on a flat space with Kac-Moody  structure group are also well-defined~\cite{Garland}, it would mean that a consistent $\widehat G$ version of the RHS of  (\ref{cqGL-corr-flat curve}) and hence (\ref{cqGL-corr}), would be ``${ {\textnormal {circle-valued flat}}~{\widehat{^LG}}{\textnormal{-bundle on}}~{{\Sigma}} }$''. 

Thus, in place of (\ref{cqGL-corr}), we would have 
\be
\label{cqGL-KM-corr}
\boxed{{{\cal O}_{\hbar}({\cal M}^{{\rm S}^1}_{{\rm H.S.}} ({\widehat G}, {\Sigma})){\textnormal{-}}{\rm{module}}}}  \,\,\, \Longleftrightarrow \, \, \,  \boxed{ {\textnormal {circle-valued flat}}~{\widehat{^LG}}{\textnormal{-bundle on}}~{\Sigma} }
\ee
where ${\Sigma} = {\bf R} \times {\bf S}^1$. This defines a $\widehat G$ version of the $q$-geometric Langlands correspondence for simply-laced $G$. 

Indeed, when the radius of  $\hat{{\rm S}}^1$ goes to zero whence ${\widehat G} \to G$, (\ref{cqGL-KM-corr}) reduces to the physically consistent correspondence (\ref{cqGL-corr-flat curve}), as expected.

\bigskip\noindent{\it Quantization of Elliptic-Valued $G$ Hitchin Systems and Circle-Valued $\widehat{^LG}$-bundles on a Flat Complex Curve}


We now wish to express the LHS of (\ref{cqGL-KM-corr}) in terms of $G$, as it will soon prove useful to do so. To this end, note that since the ordinary Hitchin equations are a double dimensional reduction of the instanton equation, it would mean that doubly-periodic $G$-instantons on $ {\hat{\rm S}}^1 \times  {\rm S}^1 \times {\Sigma}$ can be regarded as the $({\hat{\rm S}}^1 \times  {\rm S}^1)$-valued $G$ Hitchin equations on ${\Sigma}$; in other words, the $ {\rm S}^1$-valued $\widehat G$ Hitchin equations are the same as the  $({\hat{\rm S}}^1 \times  {\rm S}^1)$-valued $G$ Hitchin equations on  ${\Sigma}$. 
Thus, we can rewrite (\ref{cqGL-KM-corr}) as
\be
\label{cqGL-KM-G-corr}
 \boxed{{{\cal O}_{\hbar}({\cal M}^{ {\hat{\rm S}}^1 \times {\rm S}^1}_{{\rm H.S.}} (G, \Sigma)){\textnormal{-}}{\rm{module}}}} \, \,
  {\Longleftrightarrow} \,  \, \boxed{ {\textnormal {circle-valued flat}}~{\widehat{^LG}}{\textnormal{-bundle on}}~{\Sigma} }
  \ee 
where ${\Sigma} = {\bf R} \times {\bf S}^1$. As desired, the LHS of the correspondence is now expressed solely in terms of $G$. 
 
 Since the noncommutative deformation of the algebra of holomorphic functions on a space which captures a classical integrable system defines a quantization of the classical integrable system itself, (\ref{cqGL-KM-G-corr}) therefore relates the quantization of an elliptic-valued $G$ Hitchin system on $\Sigma$ to a circle-valued flat $\widehat{^LG}$-bundle on $\Sigma$.

As we will see shortly, (\ref{cqGL-KM-G-corr}) actually underlies Nekrasov-Pestun-Shatashvili's main result in~\cite[$\S$1.3]{NPS} for 5d $\widehat G$-quiver gauge theories.

\bigskip\noindent{\it Quantization of Elliptic-Valued $G$ Hitchin Systems and Transfer Matrices of a $\widehat G$-type XXZ Spin Chain}

In light of the fact that a $\widehat {\mathscr G}$-bundle can be obtained from a ${\mathscr G}$-bundle by replacing the underlying Lie algebra $\bf g$ of the latter bundle with its Kac-Moody generalization $\hat {\bf g}$,  it is clear that  just as circle-valued $^LG$-bundles have a correspondence with ${{\cal W}^q_{\rm cl}}(^L\frak g)$, circle-valued $\widehat{^LG}$-bundles would have a correspondence with ${{\cal W}^q_{\rm cl}}(\widehat{^L\frak g})$. Also, just as (\ref{cqGL-corr}) implies the identity (\ref{cq-GL-A-final}) on $\Sigma_g$, (\ref{cqGL-KM-corr}) would imply the identity ${Z(U_q(\hat{\hat{\frak g}})_{\rm crit}) = {\cal W}^q_{\rm cl}(\widehat{^L\frak g})}$ on $\Sigma$, where $U_q(\hat{\hat{\frak g}})$ is the quantum toroidal algebra of $\frak g$. Last but not least, just as we can identify $Z(U_q(\hat{\frak g})_{\rm crit})_{\Sigma}$ with the algebra ${\cal T}_{\rm xxz}(G, \Sigma)$, we can identify $Z(U_q(\hat{\hat{\frak g}})_{\rm crit})_{\Sigma}$ with the algebra ${\cal T}_{\rm xxz}({\widehat G}, \Sigma)$ of polynomials in $\partial_z^{m} \, {\widehat{\cal T}}_i(z)$, where $m \geq 0$; the ${\widehat{\cal T}}_i$'s are commuting transfer matrices of a $\widehat G$-type XXZ spin chain with  $U_q(\hat{\hat{\frak g}})$ symmetry on $\Sigma$; and $i = 0, \dots, {\rm rank}(\frak g)$. In all, together with (\ref{cqGL-KM-G-corr}), it would mean that in place of (\ref{O = T for XXZ}), we now have
 \be
  \label{O = T for G hat-XXZ}
  \boxed{{\cal O}_{\hbar}({\cal M}^{ {\hat{\rm S}}^1 \times {\rm S}^1}_{{\rm H.S.}} ({G}, \Sigma)) \, \, \Longleftrightarrow  \, \, {\cal T}_{\rm xxz}({\widehat G}, \Sigma)}
\ee
which relates the quantization of an elliptic-valued $G$ Hitchin system on $\Sigma$ to the transfer matrices of a $\widehat G$-type XXZ spin chain on $\Sigma$!


\bigskip\noindent{\it Elliptic-Valued Hitchin $G$ Systems and $q$-Characters of ${U^{\rm aff}_q}(\hat{\frak g})$}

Just as (\ref{HS = chi}) is an implication of (\ref{O = T for XXZ}), an implication of (\ref{O = T for G hat-XXZ}) is the following correspondence 
\be
\label{elliptic HS = chi}
\boxed{x \in {\cal M}^{{\hat{\rm S}}^1 \times {\rm S}^1}_{{\rm H.S.}} (G, \Sigma) \,  \Longleftrightarrow \,  \chi_q({{\hat V}_i}) = {\hat T}_i(z), \quad {\rm where}   \, \, \, {{\hat V}_i} \in {\rm Rep} \, [{U^{\rm aff}_q}(\hat{\frak g})_{\Sigma}],   \, \, \, i = 0, \dots, {\rm rank}( {\frak g})} 
\ee
 ${\hat T}_i(z)$ is a polynomial whose degree depends on ${\hat V}_i$, and ${U^{\rm aff}_q}(\hat{\frak g})$ is the quantum toroidal algebra of $\frak g$.

\newpage 
\bigskip\noindent{\it An M-Theoretic Realization of Nekrasov-Pestun-Shatashvili's Results for 5d, ${\cal N} =1$ $\widehat G$-Quiver $SU(K_i)$ Gauge Theories}

Let us now realize, via our M-theoretic framework, Nekrasov-Pestun-Shatashvili's result in~\cite{NPS} which relates the moduli space of 5d, ${\cal N} =1$ $\widehat G$-quiver $SU(K_i)$ gauge theories to the representation theory of ${U^{\rm aff}_q}(\hat {\frak g})$. 

To this end, recall that  doubly-periodic $G$-instantons on $ {\hat{\rm S}}^1 \times  {\rm S}^1 \times \Sigma$ can be regarded as the $({\hat{\rm S}}^1 \times  {\rm S}^1)$-valued $G$ Hitchin equations on $\Sigma$, where $\Sigma = {\bf R} \times {\bf S}^1 = {\rm C}_{\rm x} $. This means that we can also express (\ref{elliptic HS = chi}) as 
\be
\label{rexpress 5d}
x \in {\cal M}^{\rm int}_{\rm inst} (G, {k}, {\hat{\rm S}}^1 \times  {\rm S}^1 \times {\rm C}_{\rm x}) \,  \Longleftrightarrow \,  \chi_q({{\hat V}_i}) = {\hat T}_i(z),
\ee
where ${\cal M}^{\rm int}_{\rm inst} (G, {k}, {\hat{\rm S}}^1 \times  {\rm S}^1 \times {\rm C}_{\rm x})$ captures a classical integrable system of doubly-periodic $G$-instantons of charge $k$ on ${\hat{\rm S}}^1 \times  {\rm S}^1 \times {\rm C}_{\rm x}$.

Since $x$, which is a point in the total space of ${\cal M}^{\rm int}_{\rm inst} (G, { k}, {\hat{\rm S}}^1 \times  {\rm S}^1 \times {\rm C}_{\rm x})$ , also defines a point $u$ in its base space ${\frak M}^{G, {\rm C}_{\rm x}, { k}}_{{{\hat{\rm S}}^1 \times {\rm S}^1}\textnormal{-}{\rm inst}}$, and since ${\rm rank}({\frak g}) + 1$ is equal to the number of vertices of its affine Dynkin diagram, (\ref{elliptic HS = chi}) would also mean that
\be
\label{NPS for hat-G}
\boxed{u \in {\frak M}^{G, {\rm C}_{\rm x}, {k}}_{{{\hat{\rm S}}^1 \times {\rm S}^1}\textnormal{-}{\rm inst}} \,  \Longleftrightarrow \,  \chi_q({{\hat V}_i}) = {\hat T}_i(z), \quad {\rm where} \, \, \, {\rm C}_{\rm x} = {\bf R} \times {\bf S}^1,  \, \, \,  {{\hat V}_i} \in {\rm Rep} \, [{U^{\rm aff}_q}(\hat{\frak g})_{{\rm C}_{\rm x}}],  \, \, \,  i \in {\hat I}_{\Gamma}} 
\ee
where ${\hat I}_{\Gamma}$ denotes the vertices of the affine Dynkin diagram of $G$. 

Like the monopole case in the previous subsection, the degree of ${\hat T}_i$ would depend on $k$. To see this, First, note that ${k} =\int_{{\hat{\rm S}}^1 \times  {\rm S}^1 \times {\rm C}_{\rm x}} {\bf F}_{\rm I} \wedge {\bf F}_{\rm I}$, where ${\bf F}_{\rm I}$ is two-form which can be regarded as an  $({\hat{\rm S}}^1 \times  {\rm S}^1)$-valued gauge field strength on ${\rm C}_{\rm x}$. Second, from the perspective of an M-theoretic compactification on ${\bf S}^1$ (in (\ref{cqGL-AB-all g}) and (\ref{cqGL-CDG-all g}) with $\Sigma_g$ replaced by ${\rm C}_{\rm x}$), the momentum number around ${\bf S}^1$ would be given by the number of D0-branes along $\bf R$, and the D0-branes, being a charged pointlike object, would in turn source a two-form field strength and therefore ${\bf F}_{\rm I}$. Third, note that in our M-theoretic realization of the gauge/CFT duality, the momentum number around ${\bf S}^1$ would correspond to the conformal dimension and thus energy level of the chiral CFT on ${\rm C}_{\rm x} = {\bf R} \times {\bf S}^1$ with ${U^{\rm aff}_q}(\hat {\frak g})$ symmetry (as recalled below (\ref{4d AGT-TS-EF-identity}) for the 4d case). Last but not least, according to~\cite[eqn.~(2.18)]{H-XXZ}, the energy level of the associated $\widehat G$-type XXZ spin chain would depend on the degree of ${\hat T}_i$; in particular, the higher the degree of ${\hat T}_i$, the higher the energy level. Altogether, this means that the higher the $k$, the higher the degree of ${\hat T}_i$, and since both are integers, the degree of ${\hat T}_i$ ought to be some positive integer times $k$.    

According to~\cite[$\S$5.1.1--5.1.4]{NPS}\cite{NP}, ${\frak M}^{G, {\rm C}_{\rm x}, {k}}_{{{\hat{\rm S}}^1 \times {\rm S}^1}\textnormal{-}{\rm inst}}$ is the moduli space of vacua on the Coulomb  branch of a 5d, ${\cal N}=1$ $\widehat G$-quiver $SU(K_i)$ gauge theory with $K_i = k a_i$, where the $a_i$'s are the positively-integered Dynkin labels. Consequently, (\ref{NPS for hat-G}), and the conclusion in the previous paragraph that the degree of ${\hat T}_i$ is given by $k$ times a positive integer, are, together, Nekrasov-Pestun-Shatashvili's main result in~\cite[$\S$1.3]{NPS} which relates the moduli space of the 5d $\widehat G$-quiver gauge theory  to the representation theory of ${U^{\rm aff}_q}(\hat{\frak g})$!

\newsubsection{A $q, v$-Geometric Langlands Correspondence for Simply-Laced Lie Groups}

We would now like to ascertain, for simply-laced Lie groups $G$, what a $q, v$-geometric Langlands correspondence that is related to the classical $q, v$-geometric Langlands duality in (\ref{cqv-GL-A-final}), means. 

To this end, note that in the 6d AGT correspondence behind our derivation of (\ref{cqv-GL-A-final}), the 2d CFT is defined over $\Sigma^{1,2}_{t, 1}$, which is a torus ${\bf S}^1 \times {\bf S}^1_t$ with two punctures at positions $z_{1,2}$~\cite[$\S$5.1]{5d AGT}. Here, ${\bf S}^1$ corresponds to the decompactified fifth circle of radius $\beta \nrightarrow 0$, while ${\bf S}^1_t$ corresponds to the sixth circle formed by gluing the ends of an interval ${\bf I}_t$ of radius $R_6$ much smaller than $\beta$. In other words, we effectively have a \emph{single }decompactification of circles whence like in the 5d AGT correspondence, we ought to associate to each and every point on the compactified Riemann surface $\Sigma^{1,2}_{t, 1} = \Sigma_{1, 2}$, a loop which effectively represents the decompactified circle of radius $\beta$. This is consistent with the fact that it is ${\bf L}{\frak g}_{\Gamma  \, \textrm{aff},1}$ which appears in \emph{both }(\ref{Z6d-SU(N)-matter-vertex operator-map-Zp}) and (\ref{G on Zp}) for the 6d and 5d case (while it is ${\frak g}_{\Gamma  \, \textrm{aff},1}$ which appears in the upper line of (\ref{dia 1}) for the 4d case).  

\bigskip\noindent{\it Quantization of Circle-Valued $G$ Hitchin Systems and  Elliptic-Valued $^LG$-bundles on a Punctured Torus}

Hence, this means that in place of (\ref{cqGL-AB-all g}) and (\ref{cqGL-CDG-all g}), we ought to have 
\be
\label{cqvGL-A-torus punctured}
\underbrace{ {\bf T}^2_{0, \epsilon_2}\times {\bf R}_+ \times   {\bf I} \times {{\Sigma}^{{\rm S}^1}_{1,2}}}_{\textrm{$N$ M5-branes}}
\ee
for the simply-laced $A$ groups, and
\be
\label{cqvGL-CDG-torus punctured}
\underbrace{{\bf T}^2_{0, \epsilon_2} \times {\bf R}_+  \times {\bf I} \times {{\Sigma}^{{\rm S}^1}_{1,2}}}_{\textrm{$N$ M5-branes + OM5-plane}}
\ee
for  the simply-laced $D$ groups, where ${{\Sigma}^{{\rm S}^1}_{1,2}}$ is the Riemann surface $\Sigma_{1,2}$ with an ${\rm S}^1$ loop of radius $\beta$ over every point.   

Consequently, in place of (\ref{H=W, g, cqGL}), we have
\be
\label{H=W, punctured torus, cqvGL}
{\cal H}^{\rm A}_{{\bf I} \times {\bf R}_+}({\cal M}^{{\rm S}^1}_H (G, \Sigma_{1,2}))_{{\cal B}^\beta_{c.c.}, \, {\cal B}^\beta_\alpha }  =  { \widehat{{\cal W}^{q,v}_{\rm cl}}(^L\frak g)_{\Sigma_{1,2}}},
\ee
where $G = SU(N)$ or $SO(2N)$, $q = e^{-i \beta \epsilon_2}$ and $v = e^{-{1\over R_6}}$. Also, in place of (\ref{cat cqGL}), bearing in mind that states on the RHS of (\ref{H=W, punctured torus, cqvGL}) have a projection onto the fifth \emph{and} sixth circle~\cite[$\S$5.1]{5d AGT}, we have
\be
\label{cat cqvGL}
{{\rm C}^{\rm mod}_{{\cal O}_{\hbar}}({\cal M}^{{\rm S}^1}_{{\rm H.S.}} (G, \Sigma_{1,2})) = {\rm M}^{{{\mathbb S}^1 \times{\rm S}^1}}_{^L G}(\Sigma_{1,2})_{\rm flat} },
\ee
where the RHS is a ``module'' of elliptic-valued flat $^LG$-bundles on $\Sigma_{1,2}$, and the ${\mathbb S}^1$ loop has radius $R_6$. 

In turn, in place of (\ref{cqGL-corr}), we have the following correspondence
\be
\label{cqvGL-corr}
 \boxed{{{\cal O}_{\hbar}({\cal M}^{{\rm S}^1}_{{\rm H.S.}} (G, \Sigma_{1,2})){\textnormal{-}}{\rm{module}}}} \, \,
  {\Longleftrightarrow} \,  \, \boxed{ {\textnormal {elliptic-valued flat}}~{^LG}{\textnormal{-bundle on}}~{\Sigma_{1,2}} }
  \ee 
Clearly, this defines a $q,v$-geometric Langlands correspondence that is a 6d analog of the standard geometric Langlands correspondence for simply-laced $G$! Indeed, when $\beta \to 0$ and $R_6 \to 0$  whence $q \to 1$ and $v \to 0$, this $q, v$-deformed correspondence reduces to (a ``ramified'' version of) the standard correspondence discussed in $\S$5.2\ref{s5.2}.   
 
 Thus, since the noncommutative deformation of the algebra of holomorphic functions on a space which captures a classical integrable system defines a quantization of the classical integrable system itself, (\ref{cqvGL-corr}) therefore relates the quantization of a circle-valued $G$ Hitchin system on $\Sigma_{1,2}$ to an elliptic-valued flat $^LG$-bundle on $\Sigma_{1,2}$.

\bigskip\noindent{\it Quantization of Circle-Valued $G$ Hitchin Systems and Transfer Matrices of a $G$-type XYZ Spin Chain}

In light of the identity (\ref{cqv-GL-A-final}) (which also holds for ${\frak g} = \frak {so}(2N)$, as explained thereafter), one can also express (\ref{H=W, punctured torus, cqvGL}) as
\be
\label{H=Z, punctured torus, cqvGL}
{\cal H}^{\rm A}_{{\bf I} \times {\bf R}_+}({\cal M}^{{\rm S}^1}_H (G, \Sigma_{1,2}))_{{\cal B}^\beta_{c.c.}, \, {\cal B}^\beta_\alpha }  =  \widehat{Z}(U_{q,v}(\hat{\frak g})_{\rm crit})_{\Sigma_{1,2}}.
\ee
Here, the RHS is a module of  the center $Z(U_{q, v}(\hat{\frak g})_{\rm crit})_{\Sigma_{1,2}}$ of the elliptic affine algebra $U_{q,v}(\hat{\frak g})_{\rm crit}$ at critical level on $\Sigma_{1,2}$. Hence, since the LHS is the space of states of all $({\cal B}^{\beta}_{c.c.}, {\cal B}^{\beta}_\alpha)$ strings which are themselves modules of ${\cal O}_{\hbar}({\cal M}^{{\rm S}^1}_{{\rm H.S.}} (G, \Sigma_{1,2}))$,  it would mean that we have a correspondence between ${\cal O}_{\hbar}({\cal M}^{{\rm S}^1}_{{\rm H.S.}} (G, \Sigma_{1,2}))$ and $Z(U_{q, v}(\hat{\frak g})_{\rm crit})_{\Sigma_{1,2}}$  (where $\hbar \sim \epsilon_2$ and $q = e^{-i \beta \epsilon_2}$).

At any rate, since the eigenvectors of transfer matrices can be constructed from intertwining operators between modules of the symmetry algebra of the underlying integrable system at critical level, from~\cite[eqn.~(6.1)--(6.3)]{5d AGT}, it is clear that one can identify $Z(U_{q, v}(\hat{\frak g})_{\rm crit})_{\Sigma_{1,2}}$ with the algebra ${\cal T}_{\rm xyz}(G, \Sigma_{1,2})$ of polynomials in $\partial_z^{m} \, {\cal T}_i(z)$, where $m \geq 0$; the ${\cal T}_i$'s are commuting transfer matrices of a $G$-type XYZ spin chain with  $U_{q,v}(\hat{\frak g})$ symmetry on $\Sigma_{1,2}$; and $i = 1, \dots, {\rm rank}(\frak g)$. In other words, we have the following correspondence
  \be
  \label{O = T for XYZ}
  \boxed{{\cal O}_{\hbar}({\cal M}^{{\rm S}^1}_{{\rm H.S.}} (G, \Sigma_{1,2})) \, \, \Longleftrightarrow  \, \, {\cal T}_{\rm xyz}(G, \Sigma_{1,2})}
\ee
which relates the quantization of a circle-valued $G$ Hitchin system on $\Sigma_{1,2}$ to the transfer matrices of a $G$-type XYZ spin chain on $\Sigma_{1,2}$!

Note that when $\beta \to 0$ and $R_6 \to 0$ whence $q \to 1$ and $v \to 0$, the generators of the LHS of (\ref{O = T for XYZ}) would correspond  (according to~\cite{GW}) to commuting differential operators ${\cal D}_{K^{1/2}_{{\frak M}_{1,2}}}$ on $K^{1/2}_{{\frak M}_{1,2}}$, where $K_{{\frak M}_{1,2}}$ is the canonical bundle of ${\frak M}_{1,2}$, the moduli space of $G$-bundles on $\Sigma_{1,2}$, while the generators of the RHS would correspond (according to~\cite{Frenkel-Ram}) to commuting Segal-Sugawara fields ${\cal S}^{1,2}_i$ which define  holomorphic differentials of degree $d_i+1$ on $\Sigma_{1,2}$, where the $d_i$'s are the exponents of $\frak g$. This correspondence between the ${\cal D}_{K^{1/2}_{{\frak M}_{1,2}}}$'s on ${\frak M}_{1,2}$ and the ${\cal S}^{1,2}_i$'s on $\Sigma_{1,2}$ is nothing but (a ``ramified'' version of) the quantized Hitchin map which underlies the (``ramified'' version of the) original formulation by Beilinson-Drinfeld of the standard geometric Langlands correspondence for $G$ (explained via 4d gauge theory and 2d CFT in~\cite{GW}  and~\cite{Frenkel-Ram}, respectively). In other words, (\ref{O = T for XYZ}) is consistent with established results in the 4d limit, as expected.

\bigskip\noindent{\it Circle-Valued Hitchin $G$ Systems and $q,v$-Characters of ${U^{\rm ell}_{q,v}}({\frak g})$}

Let $U_{q,v}(\hat{\frak g}) = {U^{\rm ell}_{q,v}}({\frak g})$ with $q,v$-character $\chi_{q,v}$. Then,  just as (\ref{O = T for XXZ}) implied the correspondence (\ref{HS = chi}),  (\ref{O = T for XYZ}) would imply the correspondence
\be
\label{HS = chi - qv}
\boxed{x \in {\cal M}^{{\rm S}^1}_{{\rm H.S.}} (G, \Sigma_{1,2}) \,  \Longleftrightarrow \,  \chi_{q,v}({V_i}) = T_i(z), \quad {\rm where}  \, \, \, {V_i} \in {\rm Rep} \, [{U^{\rm ell}_{q,v}}({\frak g})_{\Sigma_{1,2}}],   \, \, \, i = 1, \dots, {\rm rank}(\frak g)} 
\ee
and $T_i(z)$ is a polynomial whose degree depends on $V_i$.

\bigskip\noindent{\it An M-Theoretic Realization of Nekrasov-Pestun-Shatashvili's Results for 6d, ${\cal N} =1$ $G$-Quiver $SU(K_i)$ Gauge Theories}

Let us now realize, via our M-theoretic framework, Nekrasov-Pestun-Shatashvili's result in~\cite{NPS} which relates the moduli space of 6d, ${\cal N} =1$ $G$-quiver $SU(K_i)$ gauge theories to the representation theory of ${U^{\rm ell}_{q,v}}({\frak g})$. 

To this end, note that since $\Sigma_{1,2} = \{{\bf S}^1 \times {\bf S}^1_t\}_{z_1, z_2} = \{{\rm C}_{\rm x} \}_{z_1, z_2}$, like  $\{{\bf R} \times {\bf S}^1\}_{z_1, z_2}$ in (\ref{HS = chi for NPS}), is a flat (punctured) surface, from (\ref{HS = chi - qv}), we have
\be
\label{mono = chi for NPS 6d}
x \in {\cal M}^{\rm int}_{\rm mono} (G, {\bf k},  \{{\rm S}^1  \times {\rm C}_{\rm x} \}_{y_1, y_2}) \, \, \Longleftrightarrow \, \, \chi_{q,v}({V_i}) = T_i(z),
\ee
where ${\cal M}^{\rm int}_{\rm mono} (G, {\bf k}, \{{\rm S}^1  \times {\rm C}_{\rm x} \}_{y_1, y_2})$ captures a classical integrable system of periodic $G$-monopoles of charge $\bf k$ on $\{{\rm S}^1  \times {\rm C}_{\rm x} \}$ with singularities at $y_1$ and $y_2$.  

Since $x$, which is a point in the total space of ${\cal M}^{\rm int}_{\rm mono} (G, {\bf k}, \{{\rm S}^1  \times {\rm C}_{\rm x} \}_{y_1, y_2})$, also defines a point $u$ in its base space ${\frak M}^{G, {\rm C}_{\rm x}, y_1, y_2}_{{{\rm S}^1}\textnormal{-}{\rm mono}, \bf k}$, and since ${\rm rank}(\frak g)$ is equal to the number of vertices in its Dynkin diagram, (\ref{HS = chi - qv}) would also mean that
\be
\label{NPS for G 6d}
\boxed{u \in {\frak M}^{G, {\rm C}_{\rm x}, y_1, y_2}_{{{\rm S}^1}\textnormal{-}{\rm mono}, \bf k} \,  \Longleftrightarrow \,  \chi_{q,v}({V_i}) = T_i(z), \quad {\rm where} \, \, \, {\rm C}_{\rm x} = {\bf S}^1 \times {\bf S}^1_t,  \, \, \,  {V_i} \in {\rm Rep} \, [{U^{\rm ell}_{q,v}}({\frak g})_{\{{\rm C}_{\rm x}\}_{z_1, z_2}}],  \, \, \,  i \in I_{\Gamma}} 
\ee
where $I_{\Gamma}$ denotes the vertices of the Dynkin diagram of $G$. 

Note that ${\bf k} \in H^2({\rm S}^1 \times {\bf S}^1, \pi_1(T)) = \sum_{i \in I_\Gamma} K_i \, \alpha^\vee_i$  (c.f.~\cite[eqn.~(8.17)]{NP}), where $T \subset G$ is a Cartan subgroup, $K_i$ is a positive integer, and $\alpha^\vee_i$ is a simple coroot of $G$. As one might therefore expect, there should be a correspondence between $K_i$ and $T_i$.  In fact, repeating verbatim the arguments below (\ref{NPS for G}), whilst bearing in mind that according to~\cite[$\S$8.3.3]{Gaudin}, the energy level of a $G$-type XYZ spin chain would depend on a derivative of $T_i$, we find that the the degree of $T_i$ would be given by $K_i$. 

According to~\cite[$\S$5.1.1--5.1.4]{NPS}\cite{NP}, ${\frak M}^{G, {\rm C}_{\rm x}, y_1, y_2}_{{{\rm S}^1}\textnormal{-}{\rm mono}, \bf k}$ is the moduli space of vacua on the Coulomb  branch of a 6d, ${\cal N}=1$ $G$-quiver $SU(K_i)$ gauge theory with the singularities at $y_i$ determined by its mass and flavor data. Consequently, (\ref{NPS for G 6d}), and the conclusion in the previous paragraph that the degree of $T_i$ is given by $K_i$, are, together, Nekrasov-Pestun-Shatashvili's main result in~\cite[$\S$1.3]{NPS} which relates the moduli space of 6d $G$-quiver gauge theory  to the representation theory of ${U^{\rm ell}_{q,v}}({\frak g})$!

\newsubsection{A $q, v$-Geometric Langlands Correspondence for Simply-Laced Kac-Moody Groups}

We would now like to ascertain the $\widehat G$ version of the $q,v$-geometric Langlands correspondence for simply-laced $G$ in (\ref{cqvGL-corr}), where $\widehat G$ is the Kac-Moody generalization of $G$. 

To this end, first note that with regard to our arguments leading up to (\ref{cqvGL-corr}), one could consider $\Sigma^{1,0}_{t,1} = \Sigma_1$ instead  of $\Sigma^{1,2}_{t, 1} = \Sigma_{1,2}$ in (\ref{cqvGL-A-torus punctured}) and (\ref{cqvGL-CDG-torus punctured}) (which is the same as considering the massless limit of the underlying linear quiver theory). In other words, in addition to  (\ref{cqvGL-corr}), we also have 
\be
\label{cqvGL-corr-flat curve}
 {{{\cal O}_{\hbar}({\cal M}^{{\rm S}^1}_{{\rm H.S.}} (G, {\Sigma_1})){\textnormal{-}}{\rm{module}}}} \, \, \, \,
  {\Longleftrightarrow} \, \, \, \, { {\textnormal {elliptic-valued flat}}~{^LG}{\textnormal{-bundle on}}~{{\Sigma_1}} },
  \ee 
where ${\Sigma_1} = {\bf S}^1 \times {\bf S}^1_t$.\footnote{The alert reader would have noticed that the Hitchin equations on $ {\bf S}^1 \times {\bf S}^1_t = {\bf T}^2$ may have subtleties with reducible solutions (and this is why (\ref{cqGL-corr}) was defined for $g >1$), but since we are working with the circle-valued equations, which as explained earlier, are equivalent to well-behaved monopoles, such subtleties will not affect us.} 

Next, note that according to our explanations below (\ref{cqGL-corr-flat curve}) (which also apply in this case because $\Sigma_1$, like $\Sigma$ therein, is also flat), a consistent $\widehat G$ version of (\ref{cqvGL-corr-flat curve}) would be given by 
\be
\label{cqvGL-KM-corr}
\boxed{{{\cal O}_{\hbar}({\cal M}^{{\rm S}^1}_{{\rm H.S.}} ({\widehat G}, {\Sigma_1})){\textnormal{-}}{\rm{module}}}}  \,\,\, \Longleftrightarrow \, \, \,  \boxed{ {\textnormal {elliptic-valued flat}}~{\widehat{^LG}}{\textnormal{-bundle on}}~{\Sigma_1} }
\ee
where ${\Sigma_1} = {\bf S}^1 \times {\bf S}^1_t$. This defines a $\widehat G$ version of the $q$-geometric Langlands correspondence for simply-laced $G$. 

Indeed, when the radius of  $\hat{{\rm S}}^1$ goes to zero whence ${\widehat G} \to G$, (\ref{cqvGL-KM-corr}) reduces to the physically consistent correspondence (\ref{cqvGL-corr-flat curve}), as expected.

\bigskip\noindent{\it Quantization of Elliptic-Valued $G$ Hitchin Systems and Elliptic-Valued $\widehat{^LG}$-bundles on a Torus}

Just as we could express (\ref{cqGL-KM-corr}) as (\ref{cqGL-KM-G-corr}),  we can express (\ref{cqvGL-KM-corr}) as
\be
\label{cqvGL-KM-G-corr}
 \boxed{{{\cal O}_{\hbar}({\cal M}^{ {\hat{\rm S}}^1 \times {\rm S}^1}_{{\rm H.S.}} (G, \Sigma_1)){\textnormal{-}}{\rm{module}}}} \, \,
  {\Longleftrightarrow} \,  \, \boxed{ {\textnormal {elliptic-valued flat}}~{\widehat{^LG}}{\textnormal{-bundle on}}~{\Sigma} }
  \ee 
where ${\Sigma_1} = {\bf S}^1 \times {\bf S}^1_t$. 
 
Since the noncommutative deformation of the algebra of holomorphic functions on a space which captures a classical integrable system defines a quantization of the classical integrable system itself, (\ref{cqvGL-KM-G-corr}) therefore relates the quantization of an elliptic-valued $G$ Hitchin system on $\Sigma_1$ to an elliptic-valued flat $\widehat{^LG}$-bundle on $\Sigma_1$. 

As we will see shortly, (\ref{cqvGL-KM-G-corr}) actually underlies Nekrasov-Pestun-Shatashvili's main result in~\cite[$\S$1.3]{NPS} for 6d $\widehat G$-quiver gauge theories. 

\newpage

\bigskip\noindent{\it Quantization of Elliptic-Valued $G$ Hitchin Systems and Transfer Matrices of a $\widehat G$-type XYZ Spin Chain}

In light of the fact that a $\widehat {\mathscr G}$-bundle can be obtained from a ${\mathscr G}$-bundle by replacing the underlying Lie algebra $\bf g$ of the latter bundle with its Kac-Moody generalization $\hat {\bf g}$,  it is clear that  just as elliptic-valued $^LG$-bundles have a correspondence with ${{\cal W}^{q,v}_{\rm cl}}(^L\frak g)$, elliptic-valued $\widehat{^LG}$-bundles would have a correspondence with ${{\cal W}^{q,v}_{\rm cl}}(\widehat{^L\frak g})$. Also, just as (\ref{cqvGL-corr}) implies the identity (\ref{cqv-GL-A-final}) on $\Sigma_{1,2}$, (\ref{cqvGL-KM-corr}) would imply the identity ${Z(U_{q,v}(\hat{\hat{\frak g}})_{\rm crit}) = {\cal W}^{q,v}_{\rm cl}(\widehat{^L\frak g})}$ on $\Sigma_1$, where $U_q(\hat{\hat{\frak g}})$ is the elliptic toroidal algebra of $\frak g$. Last but not least, just as we can identify $Z(U_{q,v}(\hat{\frak g})_{\rm crit})_{\Sigma_{1,2}}$ with the algebra ${\cal T}_{\rm xyz}(G, \Sigma_{1,2})$, we can identify $Z(U_{q,v}(\hat{\hat{\frak g}})_{\rm crit})_{\Sigma_1}$ with the algebra ${\cal T}_{\rm xyz}({\widehat G}, \Sigma_1)$ of polynomials in $\partial_z^{m} \, {\widehat{\cal T}}_i(z)$, where $m \geq 0$; the ${\widehat{\cal T}}_i$'s are commuting transfer matrices of a $\widehat G$-type XYZ spin chain with  $U_{q,v}(\hat{\hat{\frak g}})$ symmetry on $\Sigma_1$; and $i = 0, \dots, {\rm rank}(\frak g)$. In all, together with (\ref{cqvGL-KM-G-corr}), it would mean that in place of (\ref{O = T for XYZ}), we now have
 \be
  \label{O = T for G hat-XYZ}
  \boxed{{\cal O}_{\hbar}({\cal M}^{ {\hat{\rm S}}^1 \times {\rm S}^1}_{{\rm H.S.}} ({G}, \Sigma_1)) \, \, \Longleftrightarrow  \, \, {\cal T}_{\rm xyz}({\widehat G}, \Sigma_1)}
\ee
which relates the quantization of an elliptic-valued $G$ Hitchin system on $\Sigma_1$ to the transfer matrices of a $\widehat G$-type XYZ spin chain on $\Sigma_1$!


\bigskip\noindent{\it Elliptic-Valued Hitchin $G$ Systems and $q$-Characters of ${U^{\rm ell}_{q,v}}(\hat{\frak g})$}

Just as (\ref{HS = chi - qv}) is an implication of (\ref{O = T for XYZ}), an implication of (\ref{O = T for G hat-XYZ}) is the following correspondence 
\be
\label{elliptic HS = chi G-hat}
\boxed{x \in {\cal M}^{{\hat{\rm S}}^1 \times {\rm S}^1}_{{\rm H.S.}} (G, \Sigma_1) \,  \Longleftrightarrow \,  \chi_{q,v}({{\hat V}_i}) = {\hat T}_i(z), \quad {\rm where}   \, \, \, {{\hat V}_i} \in {\rm Rep} \, [{U^{\rm ell}_{q,v}}(\hat{\frak g})_{\Sigma_1}],   \, \, \, i = 0, \dots, {\rm rank}( {\frak g})} 
\ee
 ${\hat T}_i(z)$ is a polynomial whose degree depends on ${\hat V}_i$, and ${U^{\rm ell}_{q,v}}(\hat{\frak g})$ is the elliptic toroidal algebra of $\frak g$.

\bigskip\noindent{\it An M-Theoretic Realization of Nekrasov-Pestun-Shatashvili's Results for 6d, ${\cal N} =1$ $\widehat G$-Quiver $SU(K_i)$ Gauge Theories}

Let us now realize, via our M-theoretic framework, Nekrasov-Pestun-Shatashvili's result in~\cite{NPS} which relates the moduli space of 6d, ${\cal N} =1$ $\widehat G$-quiver $SU(K_i)$ gauge theories to the representation theory of ${U^{\rm ell}_{q,v}}(\hat{\frak g})$. 

To this end, recall that  doubly-periodic $G$-instantons on $ {\hat{\rm S}}^1 \times  {\rm S}^1 \times \Sigma_1$ can be regarded as the $({\hat{\rm S}}^1 \times  {\rm S}^1)$-valued $G$ Hitchin equations on $\Sigma_1$, where $\Sigma_1 = {\bf S}^1 \times {\bf S}^1_t = {\rm C}_{\rm x}$. This means that we can also express (\ref{elliptic HS = chi G-hat}) as 
\be
x \in {\cal M}^{\rm int}_{\rm inst} (G, {k}, {\hat{\rm S}}^1 \times  {\rm S}^1 \times {\rm C}_{\rm x}) \,  \Longleftrightarrow \,  \chi_{q,v}({{\hat V}_i}) = {\hat T}_i(z),
\ee
where ${\cal M}^{\rm int}_{\rm inst} (G, {k}, {\hat{\rm S}}^1 \times  {\rm S}^1 \times {\rm C}_{\rm x})$ captures a classical integrable system of doubly-periodic $G$-instantons of charge $k$ on ${\hat{\rm S}}^1 \times  {\rm S}^1 \times {\rm C}_{\rm x}$.

Since $x$, which is a point in the total space of ${\cal M}^{\rm int}_{\rm inst} (G, { k}, {\hat{\rm S}}^1 \times  {\rm S}^1 \times {\rm C}_{\rm x})$ , also defines a point $u$ in its base space ${\frak M}^{G, {\rm C}_{\rm x}, { k}}_{{{\hat{\rm S}}^1 \times {\rm S}^1}\textnormal{-}{\rm inst}}$, and since ${\rm rank}({\frak g}) + 1$ is equal to the number of vertices of its affine Dynkin diagram, (\ref{elliptic HS = chi G-hat}) would also mean that
\be
\label{NPS for hat-G 6d}
\boxed{u \in {\frak M}^{G, {\rm C}_{\rm x}, {k}}_{{{\hat{\rm S}}^1 \times {\rm S}^1}\textnormal{-}{\rm inst}} \,  \Longleftrightarrow \,  \chi_{q,v}({{\hat V}_i}) = {\hat T}_i(z), \quad {\rm where} \, \, \, {\rm C}_{\rm x} = {\bf S}^1 \times {\bf S}^1_t,  \, \, \,  {{\hat V}_i} \in {\rm Rep} \, [{U^{\rm ell}_{q,v}}(\hat{\frak g})_{{\rm C}_{\rm x}}],  \, \, \,  i \in {\hat I}_{\Gamma}} 
\ee
where ${\hat I}_{\Gamma}$ denotes the vertices of the affine Dynkin diagram of $G$. 

Like the monopole case of the previous subsection, the degree of ${\hat T}_i$ would depend on $k$. Repeating our explanations below (\ref{NPS for hat-G}) with ${\rm C}_{\rm x} = {\bf S}^1 \times {\bf S}^1_t$ and ${U^{\rm aff}_q}(\hat {\frak g})$ therein replaced by ${U^{\rm ell}_{q,v}}(\hat{\frak g})$ (whence we have a $\widehat G$-type XYZ spin chain on ${\bf S}^1 \times {\bf S}^1_t$), we find that the degree of ${\hat T}_i$ ought to be some positive integer times $k$.    

According to~\cite[$\S$5.1.1--5.1.4]{NPS}\cite{NP}, ${\frak M}^{G, {\rm C}_{\rm x}, {k}}_{{{\hat{\rm S}}^1 \times {\rm S}^1}\textnormal{-}{\rm inst}}$ is the moduli space of vacua on the Coulomb  branch of a 6d, ${\cal N}=1$ $\widehat G$-quiver $SU(K_i)$ gauge theory with $K_i = k a_i$, where the $a_i$'s are the positively-integered Dynkin labels. Consequently, (\ref{NPS for hat-G 6d}), and the conclusion in the previous paragraph that the degree of ${\hat T}_i$ is given by $k$ times a positive integer, are, together, Nekrasov-Pestun-Shatashvili's main result in~\cite[$\S$1.3]{NPS} which relates the moduli space of 6d $\widehat G$-quiver gauge theory  to the representation theory of ${U^{\rm ell}_{q,v}}(\hat{\frak g})$!

\end{document}